\documentclass{article}
\usepackage[a4paper,
            bindingoffset=0.2in,
            left=0.6in,
            right=0.6in,
            top=1in,
            bottom=1in,
            footskip=.25in]{geometry}

\usepackage[utf8]{inputenc} \usepackage[english]{babel}

\usepackage{jcappub}
\usepackage{mathtools} \usepackage{amsfonts}
\usepackage{amssymb} \usepackage{braket}
\usepackage{graphicx} \usepackage{bigstrut}
\usepackage{multirow} \usepackage{geometry}
\usepackage{ltablex} 
\usepackage{natbib}
\usepackage{fancyhdr}
\setcitestyle{square,numbers}

\usepackage[page]{appendix}

\usepackage{url}  \usepackage{csquotes} \usepackage{ragged2e}

\usepackage{afterpage} \usepackage{upgreek}
\usepackage[usenames,dvipsnames]{xcolor}  
\usepackage{subfigure} \usepackage{float}
\usepackage{slashed} \usepackage{vmargin}
\usepackage{pdfpages}
\usepackage{makecell} 

\allowdisplaybreaks
\numberwithin{equation}{section} \numberwithin{table}{section} \numberwithin{figure}{section}
\setpapersize{A4} \setmarginsrb {2.5cm} {1.5cm} {2.5cm} {1.5cm} {0pt} {1cm} {0pt} {2cm}
\usepackage{empheq}
\definecolor{myblue}{rgb}{.8, .8, 1}
\definecolor{mygray}{rgb}{.9, .9, .9}
\newlength\mytemplen
\newsavebox\mytempbox

\makeatletter
\newcommand\mybluebox{%
    \@ifnextchar[
       {\@mybluebox}%
       {\@mybluebox[0pt]}}

\def\@mybluebox[#1]{%
    \@ifnextchar[
       {\@@mybluebox[#1]}%
       {\@@mybluebox[#1][0pt]}}

\def\@@mybluebox[#1][#2]#3{
    \sbox\mytempbox{#3}%
    \mytemplen\ht\mytempbox
    \advance\mytemplen #1\relax
    \ht\mytempbox\mytemplen
    \mytemplen\dp\mytempbox
    \advance\mytemplen #2\relax
    \dp\mytempbox\mytemplen
    \colorbox{mygray}{\hspace{1em}\usebox{\mytempbox}\hspace{1em}}}

\makeatother

\usepackage[subfigure]{tocloft} 

\renewcommand{\arraystretch}{1.2} \usepackage{wrapfig}

\usepackage{soul}

\title{Tests of BSM Higgs interactions \\ by the combination $(\kappa_V^2-\kappa_{2V})$ \\ in HHjj production at LHC}  
\author{D. Domenech,}
\author{G. García-Mir,}
\author{M. Herrero}
\affiliation{Departamento de F\'{\i}sica Te\'orica and Instituto de F\'{\i}sica Te\'orica, IFT-UAM/CSIC,\\
Universidad Aut\'onoma de Madrid, Cantoblanco, 28049 Madrid,  Spain}
\emailAdd{daniel.domenech@uam.es}
\emailAdd{guillermo.garciamir@estudiante.uam.es}
\emailAdd{maria.herrero@uam.es}

\abstract{In this work, we study the potential of analyzing the particular combination of $\kappa$ modifiers given by $\kappa_V^2-\kappa_{2V}$ to disentangle Beyond Standard Model Higgs signals 
in double Higgs production with two extra light jets at LHC. We use the Higgs Effective Field Theory approach, HEFT, for the gauge invariant description of these two $\kappa_V$ and $\kappa_{2V}$ parameters representing the BSM interactions of the Higgs particle with the electroweak gauge bosons,  $HVV$ and $HHVV$ respectively. We illustrate the bonus of studying this particular combination $\kappa_V^2-\kappa_{2V}$ focusing in just one process,  $HHjj$, instead of studying $\kappa_V$ and $\kappa_{2V}$ separately using various processes.  For the detailed analysis here, we focus on the particular 
final state $H(\to \gamma \gamma) H (\to b \bar b)jj'$, with two photons, two b-jets and two light jets,  which we analyze fully including parton showering,  fragmentation, hadronization and detector effects, in both signal and main background.  We find that the sensitivity to  $\kappa_V^2-\kappa_{2V}$ could be notably improved at the HL-LHC by the proper cuts on the final state $b \bar b \gamma \gamma jj'$ that select optimally the events with VBF-topology and $HH$-topology.   In particular,  we propose strategies based on a good isolation of $\gamma \gamma$ and b-jet  pairs, and focus specially on specific variables like $\eta_{\gamma \gamma}$ and $P^T_{\gamma \gamma}$ where the signal events distributions may efficiently discriminate between BSM signal  versus SM background. We will show here how the high transversality of these two pairs in the signal (inherited from the high transversality of the $H$'s) is correlated with the non-vanishing value of this $(\kappa_V^2-\kappa_{2V})$ combination. }

\begin{document}
\begin{flushright}
	IFT-UAM/CSIC-25-80 
\end{flushright}
\maketitle

\section{Introduction}

At present,  due to the absence of any experimental signal of new particles or resonances beyond the Standard Model (SM) particle content,   the main searches of Beyond Standard Model (BSM) physics are focused  on the study  of anomalous couplings of the SM particles and their phenomenological consequences at colliders.  For that purpose,  
the use of $\kappa$ coupling modifiers to analyze  BSM Higgs physics  at  the Large Hadron Collider (LHC) has been widely explored 
in the past years.  For summaries in joint reports  see,  for instance,  
Refs. \cite{LHCHiggsCrossSectionWorkingGroup:2013rie, Dainese:2019rgk, ATLAS:2019mfr, Cepeda:2019klc,Mlynarikova:2023bvx}.  
Due to their relevance in connection with the Higgs mechanism of mass generation,  among the most frequently studied BSM couplings are those describing new Higgs interactions with the electroweak (EW) gauge bosons $V=W^\pm, Z$.  These BSM Higgs couplings are usually parametrized in terms of two modifiers $\kappa_V$ and $\kappa_{2V}$,  which refer to  deviations from the  $HVV$ and $HHVV$ SM couplings,  respectively.  Both ATLAS and CMS collaborations have performed multiple searches at LHC,  looking at  different channels,  and  have  provided a set of bounds on these $\kappa$ parameters.  A summary of the expected reach to the $\kappa$ parameters in the future phase at LHC with $\sqrt{s}=14$ TeV and high luminosity (HL-LHC) can be found in the recent joint paper by the ATLAS and CMS collaborations,   Ref.\cite{ ATL-PHYS-PUB-2025-018}. 

On the computational side,  the $\kappa$ framework is not the common tool among the Theoretical Physics community,  because in general this framework does not provide a full gauge invariant theory. In particular,  to study BSM Higgs interactions,   the usual  framework adopted is that of  Effective Field Theories (EFTs),  which allow us to  access these anomalous couplings under a gauge  invariant and model-independent approach.  
On the other hand,  the great advantage of EFTs is that they encode the low-energy limit of the particular high-energy \enquote{ultraviolet} (UV) physics model that one could be interested in.  EFTs are usually built with effective Lagrangians by means of a tower of $SU(3)_C \times SU(2)_L \times U(1)_Y$  gauge invariant operators,  with factors in front suppressed by inverse powers of the high energy scale or heavy masses associated to the UV theory.  These operators are built exclusively with the SM fields,  and the coefficients in front  (usually called Wilson coefficients) are the only ones that encode information about the underlying UV physics.  In this respect,  it is worth  remarking that the usual procedure  of setting the specific values for these coefficients by integrating out the heavy modes of the UV theory may also lead to interesting correlations among those coefficients.  Then,  EFTs are designed specifically to predict observable quantities  being measured at colliders,  with the hope of finally decoding the underlying physics operating at much higher energy scales.  If the  mentioned correlations can be tested at colliders,  this will be a powerful tool to find hints of the new UV physics.  

Currently, two of the most popular gauge invariant EFTs for describing BSM Higgs physics are the Standard Model Effective Field Theory (SMEFT) and the Higgs Effective Field Theory (HEFT) (for EFTs reviews see,  for instance,  Refs.\cite{Brivio:2017vri,  Dobado:2019fxe}).  These two EFTs have linear and non-linear approaches,  respectively,  for the implementation of the Higgs Mechanism by means of the four scalar fields needed for the electroweak symmetry breaking 
$SU(2)_L \times U(1)_Y \to U(1)_{\rm em}$,  i.e. ,  the three would-be Goldstone bosons (GBs) and the Higgs field.  In the SMEFT,  these four fields are placed within a doublet,  as in the SM,  whereas in the HEFT,  the Higgs field  is a singlet and the three GBs are placed in a non-linear representation,  usually in an exponential one.  
The present work is carried out using the HEFT,  and,  in particular,  the HEFT framework defined in 
\cite{PhysRevD.102.075040, PhysRevD.104.075013, PhysRevD.106.073008, PhysRevD.106.115027}.  The reason behind this choice is twofold.  First,  it provides an immediate connection to the $\kappa$ framework.  In particular,  the Leading Order HEFT Lagrangian (LO-HEFT) with chiral dimension 2 contains two effective coefficients,  usually named $a$ and $b$,  describing the effective interactions of our interest here,   $HVV$ and $HHVV$,  which can be identified directly with $\kappa_V$ and $\kappa_{2V}$,  respectively.  As a consequence,  the experimental tests of these two HEFT effective coefficients,  $a$ and $b$,  can be taken from those performed on their corresponding  $\kappa$'s.   One can  always obtain the SM predictions from the general HEFT predictions  by simply setting $a$ and $b$ to the reference values $a=1$ and $b=1$.   Besides,  since the HEFT is a fully gauge invariant theory,  as said above,  it provides gauge invariant predictions for all scattering amplitudes of phenomenological interest,  which is not the case if one uses just a scaling of SM couplings by the  $\kappa$'s.  Second, and most importantly,   due to the Higgs being a singlet within the HEFT,  these two parameters $a$ and $b$ are independent and,  therefore,  provide generally uncorrelated $HVV$ and $HHVV$ interactions.  This is in contrast to the SM case,  where the two interaction vertices are correlated by $v V_{HHVV}=V_{HVV}$,  with $v=246$ GeV,  and  this relation appears as a consequence of $H$ being a component of the SM doublet.   Going BSM,  other different correlations between $a$ and $b$ and,  therefore,  between the $HVV$ and $HHVV$ interactions at low energies,  are expected to appear. 
For instance,  considering small deviations from the SM values,  $\Delta a \equiv 1 - a$ and $\Delta b \equiv  1 - b$,  in the case of BSM theories  that incorporate the Higgs boson as a component of a SU(2) doublet,   correlations between $a$ and $b$ appear.  This is the case of the SMEFT,  where these correlations among $a$ and $b$ translate into a correlation between  $\Delta a$ and $\Delta b$ given specifically  by $4\Delta a = \Delta b$ (see,  for instance,  \cite{PhysRevD.106.115027,  Domenech:2025gmn}).  In contrast,  for a two-Higgs-doublet model (2HDM),  once the heavy Higgs modes $H^\pm$,  $H^0$ and $A$  are integrated out,   a different correlation between $a$ and $b$  leading to $2\Delta a = - \Delta b$ appears (see \cite{Arco:2023sac}).   Thus,  there is a clear motivation to explore whether such correlations can be checked experimentally at future colliders.   
 
The aim of the present work is to discuss the phenomenological implications of  BSM $HVV$ and $HHVV$ couplings within the HEFT,  specifically in the double Higgs production with two light jets at the LHC, i.e,  in $pp \to HHjj$.  Since our main purpose here is to explore the expected sensitivity  at the LHC to the previously mentioned correlations via the particular combination 
$(\kappa_V^2 -\kappa_{2V})$,   we will focus our study on the $HHjj$ events with vector boson fusion (VBF) topology.   This $HHjj$ production with VBF topology is known to be the second-largest  $HH$ production mode at LHC,  behind gluon-gluon fusion,  and has the advantage of  allowing us access to the two $\kappa_V$ and $\kappa_{2V}$ parameters together in just one process.  The relevant remark for the present work is that this dependence on $\kappa_V$ and $\kappa_{2V}$ appears in $HHjj$ via VBF precisely through this particular combination $(\kappa_V^2 -\kappa_{2V})$.  This can be easily understood from the large energy behaviour of the dominant helicity scattering amplitude for the subprocess of $HH$ production from the fusion of two electroweak (EW) gauge bosons, $VV \to HH$ with $VV=WW, ZZ$.  This dominant subprocess is known to be that of the longitudinal modes,  i.e.,  $V_LV_L \to HH$.   Specifically,   at large energies compared to the boson masses,  $\sqrt{s}\gg m_H, m_V$,  this amplitude ${\cal A}(V_LV_L \to HH)$ within the HEFT behaves as $\simeq  -(a^2-b) s/v^2$,  with $v$ the EW vacuum expectation value (see,  for instance,   \cite{Domenech:2025gmn},  \cite{D_vila_2024}).  And it is this dependence what is approximately transmitted to the full process , $pp \to HHjj$.  Furthermore,  it has been shown in  \cite{D_vila_2024} that also the differential cross section with respect to the final Higgs pseudo-rapidity $\eta_H$ manifests the relevance of this particular combination.  Specifically,  $d\sigma(VV \to HH)/d\eta_H$ clearly shows a central peak  at $\eta_H=0$ when $(\kappa_V^2 -\kappa_{2V}) \neq 0$ and,  in addition,  the height of this peak is correlated with the size of the $(\kappa_V^2 -\kappa_{2V})$ combination,  indicating that the final Higgs bosons in $VV \to HH$ within the HEFT are produced with high transversality as compared to the SM case.  

The final objective of this work is  to explore the sensitivity at LHC to this $(\kappa_V^2 -\kappa_{2V})$ combination,  both in the total cross section 
 and  also in specific differential cross-section distributions with respect to variables that properly show
the best sensitivity to this particular combination.  There exist some previous studies focusing on the role of this specific combination.  For $e^+e^-$ colliders we refer to  \cite{D_vila_2024, Gonzalez-Lopez:2021aa},  and for $pp$ colliders an important work making reference to this issue was done some time ago in \cite{Contino_2010}.
In this work,  we will focus on the access to $(\kappa_V^2 -\kappa_{2V})$ in $pp$ collisions considering  the particular final state $b\overline{b}\gamma\gamma jj$,  having  two b-jets, two photons and two light jets.  Thus,  the full process for the present study of BSM signals is $pp \to HHjj \to b\overline{b}\gamma\gamma jj$,  where the  final Higgs bosons decay to $H \to b \bar b$ and $H \to \gamma \gamma$,  respectively.  For this purpose,  we will first show the HEFT predictions at the parton-level for $pp \to HHjj$,   for both the total cross section and the most relevant differential cross sections,  and then we will move to a more realistic computation for the full process $pp \to b \bar b \gamma \gamma jj$ including showering, hadronization,  clustering,  and detector level effects.   The events simulations in this work are produced with MG5 (aMC@NLO version-3.5.1 Montecarlo event generator \cite{Alwall_2014}),  where we have implemented  our HEFT model defined in Refs. \cite{PhysRevD.102.075040, PhysRevD.104.075013, PhysRevD.106.073008, PhysRevD.106.115027},   and all the final events will be filtered with PYTHIA \cite{Sj_strand_2015} and DELPHES  \cite{Favereau:2014aa} to incorporate the showering, hadronization,  clustering,  and detector level effects.  Specifically,   for the detector,  we choose CMS.   The Phase-2 upgraded CMS detector  will be simulated using DELPHES  for a fast and parametric detector simulation \cite{delphes_card_idea}.   For the predictions at the future HL-LHC,  we use $\sqrt{s}=14$ TeV and the high luminosity option with integrated luminosity of 3000 ${\rm fb}^{-1}$.   For  the partons involved in the proton-proton collisions,  we use the set of parton distribution functions (PDFs) NNPDF2.3 \cite{Ball2013290}.  The final part of the work is devoted to showing the improvement one can get on the sensitivity to this $(\kappa_V^2 -\kappa_{2V})$ combination at HL-LHC by means of the proper selection of variables and cuts on the final state, that are good discriminants between the signal and main background.  In this last study,  we will also propose several strategies focusing on the variables of the final $\gamma \gamma$ and $b$-jets pairs.  

The outline of the present work is as follows.  Section \ref{section:1} presents a short review of the $HVV$ and $HHVV$ effective interactions within HEFT and their relation to the $\kappa$ formalism.  In section \ref{section:2}  we recall the role of $(\kappa_V^2-\kappa_{2V})$ in  Higgs pair production from VBF.  
In section \ref{section:3} we present the details of our analysis of $HHjj$ production within the HEFT at proton-proton collisions,  including our selection strategies for both the total $HHjj$ event rates and the relevant distributions.  Section  \ref{section:4} is devoted to the analysis and the prospects for the full process 
$pp \to HH jj' \to  b\overline{b}\gamma\gamma jj'$ at LHC,  including all the realistic effects from showering, hadronization,  clustering,  and detector in both the total cross section and the most relevant differential cross sections.  Section \ref{section:5} presents our final results and strategies for the improvement of the sensitivity to $(\kappa_V^2-\kappa_{2V})$  in the future HL-LHC.  The last section summarizes our conclusions.

\section{$HVV$ and $HHVV$ interactions within HEFT:  relation to the $\kappa$-formalism} \label{section:1}

In this section,  we review the main aspects of BSM Higgs physics within HEFT, and its relation to the $\kappa$-formalism.

The HEFT, also called EChL (Electroweak Chiral Lagrangian) in the literature,  is built from a gauge principle based on the requirement of  EW gauge invariance $SU(2)_L\times U(1)_Y$.    It has an additional  global symmetry,  called EW chiral symmetry, $SU(2)_L \times SU(2)_R$ (resembling the chiral symmetry of low energy QCD), which is spontaneously broken to the diagonal subgroup $SU(2)_{L+R}$ also called the custodial symmetry group.  

In analogy to Chiral Perturbation Theory (ChPT)  for low energy QCD,   and by means of a unitary matrix $U$, 
the GBs of the EW theory are embedded using an exponential representation. The $U$ matrix transforms linearly as a bi-doublet under chiral rotations $SU(2)_L \times SU(2)_R, U' = LUR^\dagger$, whereas the GBs transform non-linearly, being this later a feature which is characteristic of chiral Lagrangians.  The GBs fields in the EW case are normalized by the EW symmetry breaking  scale, $v$ = 246 GeV,  in contrast to ChPT,  where the GBs are normalized by the pion decay constant $f_\pi=94$ MeV.  The counting rules of HEFT are similar to the counting rules in ChPT,  and are common to all EFTs based on chiral Lagrangians. These rules order the tower of effective operators in the effective Lagrangian by their chiral dimension,  starting with chiral dimension 2, then chiral dimension 4, etc.  The chiral dimension ordering sets also the predictions of these EFTs in terms of an expansion in powers of momentum.   The typical energy scale setting the validity of the expansion in powers of momentum  within the HEFT is set by $4 \pi v \simeq 3100 \, {\rm GeV}$.  This is also in close analogy to the typical energy scale setting the validity of the expansion in powers of momentum within ChPT which is given by  $4 \pi f_\pi  \simeq 1.2 \,  {\rm GeV}$. 
The introduction of the SM gauge bosons in the HEFT is usually performed by means of the covariant derivative.  All these features and pieces of the HEFT are  summarized  in the following.

The relevant effective bosonic operators for the present work are contained in the Leading Order (LO)  HEFT Lagrangian with chiral dimension 2.  Therefore,  here we restrict ourselves to this bosonic LO-HEFT Lagrangian. In the following,  we recall this LO-HEFT Lagrangian for arbitrary covariant $R_\xi$ gauges which was introduced in a series of works  \cite{PhysRevD.102.075040, PhysRevD.104.075013, PhysRevD.106.073008, PhysRevD.106.115027}.
  We use here the same notation as in those works: 
        \begin{eqnarray}
        {\cal L}_{\rm LO}^{\rm HEFT} \, = \, \frac{v^2}{4}  \left(1 + 2 a \frac{H}{v} + b \frac{H^2}{v^2}+\dots \right)
        \text{Tr}[D_{\mu} U^{\dagger} D^{\mu} U] + \frac12 \partial_{\mu} H \partial^{\mu} H - V(H) \nonumber \\
        - \frac{1}{2 g^2} \text{Tr}[\hat{W}_{\mu \nu} \hat{W}^{\mu \nu}] - \frac{1}{2 g'^2} \text{Tr}[\hat{B}_{\mu \nu} \hat{B}^{\mu \nu}] + {\cal L}_{GF} + {\cal L}_{FP}.
        \label{eqn: leading}
    \end{eqnarray} \vspace{0em}
    
 The relevant fields and quantities appearing in this Lagrangian are the following.  
$H$ is the Higgs boson field , which is introduced in the HEFT as a singlet field, in contrast to the SM or the SMEFT,  where it is introduced inside a component of the usual doublet $\Phi$.  The $U$ field is a $2\times2$ matrix which in the exponential representation is given by: 
\begin{equation}
        U \, = \, \exp \left( i \frac{{\omega_i} {\tau_i}}{v} \right),
    \end{equation}  
   that contains the three GB fields ${\omega}_i$ ($i=1,2,3$) in a non-linear representation of the $SU(2)$ symmetry group,  and $\tau_i$ are the three Pauli matrices.  The EW covariant derivative of this $U$ field 
   is given by:
   \begin{equation}
        D_{\mu} U \, = \, \partial_{\mu} U - i \hat{W}_{\mu} U + i U \hat{B}_{\mu},
    \end{equation} 
 that contains the EW gauge fields with a hat,   $\hat{W}_{\mu}$ and $\hat{B}_{\mu}$,  which are defined in terms of the usual EW fields,  $W^{1,2,3}_\mu$ and $B_\mu$,  
    and the gauge couplings $g$ and $g'$  by:
\begin{equation}
\hat{W}_{\mu} = \frac{g}{2} W^i_{\mu} \tau^i \, ,\,  \hat{B}_{\mu} = \frac{g'}{2} B_{\mu} \tau^3 \,.
\end{equation}

    The corresponding EW field strength tensors are given respectively  by:
       \begin{equation}
        \hat{W}_{\mu \nu} \, = \, \partial_{\mu} \hat{W}_{\nu} - \partial_{\nu} \hat{W}_{\mu} - i [\hat{W}_{\mu}, \hat{W}_{\nu}], \hspace{8mm} \hat{B}_{\mu \nu} \, = \, \partial_{\mu} \hat{B}_{\nu} - \partial_{\nu} \hat{B}_{\mu}.
    \end{equation}

$ {\cal L}_{GF}$  and $ {\cal L}_{FP}$ are the proper gauge fixing and Fadeev-Popov terms for the $R_\xi$ covariant gauges (for details, see the previously quoted references).  Notice that we use here the same sign conventions as in Ref.\cite{Domenech:2025gmn},  such that the FRs for $a=b=\kappa_3=\kappa_4=1$ coincide (also in sign) with the SM FRs set in \textsc{MG5}.

    The physical gauge fields are then given by:
\begin{equation}
W_{\mu}^\pm = \frac{1}{\sqrt{2}}(W_{\mu}^1 \mp i W_{\mu}^2) \,,\quad
Z_{\mu} = c_W W_{\mu}^3 - s_W B_{\mu} \,,\quad
A_{\mu} = s_W W_{\mu}^3 + c_W B_{\mu} \,,
\label{eq-gaugetophys}
\end{equation}
where we have used the short notation $s_W=\sin \theta_W$ and $c_W=\cos \theta_W$,  with $\theta_W$ the weak angle.

$V(H)$ is the Higgs potential within the HEFT, which includes the modified triple and quartic  Higgs self-interactions:
    \begin{equation}
        V(H) \, = \, \frac12 m_H^2 H^2 + \kappa_3 \lambda v H^3 + \kappa_4 \frac{\lambda}{4} H^4.
    \end{equation}
Notice that the relations among the physical boson masses and couplings within the LO-HEFT are as in the SM.   Concretely,  $m_W= g v/2$,  $m_Z=m_W/\cos \theta_W$ and $m_H^2= 2 \lambda v^2$,  where $v=(\sqrt{2} G_F)^{-1/2}=246\,\,{ \rm GeV}$.  
Notice also that because $H$ is a singlet within the HEFT,   the Higgs  interactions with the gauge bosons are encoded  by generic polynomials $\mathcal{F}_i (H)$ with completely arbitrary coefficients.  In particular,  the Higgs interactions with the EW gauge bosons are encoded in ${\cal F}(H)$, 
which includes the parameters $a$ and $b$ of our interest in this work:
\begin{equation} \label{eq:2_07}
 \mathcal{F}({H}) = \left[ 1 + 2a\frac{H}{v} + b\frac{H^2}{v^2} ... \right]. 
\end{equation} 
For the explicit computations in the following sections,  we will simplify the HEFT model assuming $\kappa_3=\kappa_4=1$ and leave only the two relevant LO-HEFT parameters,  $a$ and $b$,  describing the new Higgs effective couplings.  The interaction vertices we are interested in,   $HVV$ and $HHVV$ ,  are present in the first term of Eq. \ref{eqn: leading}, which incorporates the covariant derivatives of the matrix $U$.  From the expansion of $U$,   Feynman rules (FRs) for the $HVV$ and $HHVV$ \text{\small ($V = W, Z$)} and other vertices within the HEFT are obtained in general gauges.  In particular, to get them in the unitary gauge one 
takes $U=1$ and rotates the bosonic fields to the physical basis.  These rules are similar to those from the SM,  except for the $a$ and $b$ effective coefficients, which in the SM are equal to 1.  In this present work,  there is no need to consider higher-order terms of $\mathcal{F}(H)$. 
\begin{table}[!t] \small \centerline{
\begin{tabular}{|c|c|c|c|}
\hline
$i\Gamma^{\rm SM}_{H W_\mu W_\nu}=i g m_W g_{\mu\nu}$  & 
$i \Gamma^{\rm SM}_{H Z_\mu Z_\nu}= i\frac{g}{\text{cos   }\theta_W}m_Z g_{\mu\nu}$   & 
$i\Gamma^{\rm SM}_{H H W_\mu W_\nu}=i\frac{g^2}{2} g_{\mu\nu}$     & 
$i\Gamma^{\rm SM}_{H H Z_\mu Z_\nu}=i\frac{g^2}{2\text{cos   }\theta^2_W} g_{\mu\nu}$                                             
\\ [5pt] \hline
$i\Gamma^{\rm HEFT}_{H W_\mu W_\nu}=i\textcolor{Red}{a}gm_W g_{\mu\nu}$ & 
$i \Gamma^{\rm HEFT}_{H Z_\mu Z_\nu}= i\textcolor{Red}{a}\frac{g}{\text{cos   }\theta_W}m_Z g_{\mu\nu}$ & 
$i\Gamma^{\rm HEFT}_{H H W_\mu W_\nu}=i\textcolor{Cerulean}{b}\frac{g^2}{2} g_{\mu\nu}$& 
$i\Gamma^{\rm HEFT}_{H H Z_\mu Z_\nu}=i\textcolor{Cerulean}{b}\frac{g^2}{2\text{cos   }\theta^2_W} g_{\mu\nu}$ 
\\ [5pt] \hline
\end{tabular}} \caption{Examples of Higgs-EW gauge boson interaction's Feynman rules within SM and HEFT frameworks in the unitary gauge.  Here,  $g_{\mu \nu}$  denotes the metric tensor. The HEFT coefficients $a$ (in red) and $b$ (in blue) can be replaced by $\kappa_V$ and $\kappa_{2V}$ respectively.  Other HEFT interactions,  as those not shown in the dots of Eq. \ref{eqn: leading} like,  for instance,  $HHHVV$ interactions which involve the HEFT parameter $c$,  are not present in the SM (hence,  not included either in the $\kappa$-formalism).}
\label{table:01_01}
\end{table}
We summarize  in  Table \ref{table:01_01} the set of relevant HEFT Feynman Rules (FR). We have also included the SM Feynman rules for comparison.  It is worth noting that all LO-HEFT Feynman Rules preserve the original SM tensor structure,  and this is the reason why the HEFT $a$ and $b$ coefficients can be directly identified with the $\kappa_V$ and $\kappa_{2V}$ modifiers,  respectively.
 
 The present data from LHC set important constraints on these $\kappa$ parameters,  which can be translated directly into constraints on the LO-HEFT parameters $a=\kappa_V$,  $b=\kappa_{2V}$ and $\kappa_3=\kappa_\lambda$.  The constraints on $\kappa_4$  are very weak still.  
 For the present work, we select the following set of constraints to the values of the Higgs-gauge couplings and Higgs self-coupling: 
 \begin{equation}
\label{ATLASconstraints}
{\rm ATLAS} \, \, (95\% CL) \,\,: \, \,\kappa_V \in  [0.99,1.11]  \,\,,  \,\,\kappa_{2V} \in [0.6,1.5] \,\,,  \,\, \kappa_\lambda \in [-1.2,7.2] , 
\end{equation}
\noindent taken from \cite{ATLAS_main} and \cite{PhysRevLett.133.101801},  and 
\begin{equation}
\label{CMSconstraints}
{\rm CMS} \, \, \, (95\% CL) \,\,: \, \,\kappa_V \in  [0.97 , 1.09] \,\,,  \,\,\kappa_{2V} \in [0.6, 1.4] \,\,,  \,\, \kappa_\lambda \in [-1.2, 7.5] , 
\end{equation}
\noindent taken from  \cite{CMS_main} and \cite{CMS:2024awa}  . \par 
In a recent paper,  \cite{ATL-PHYS-PUB-2025-018},  a summary by ATLAS and CMS collaborations is presented,   with references to some improved constraints  and the prospective improvements for the future operation of HL-LHC.
For the main focus of the present paper, it is important to remark that there is no constraint being obtained directly from a unique process on the combination parameter of our interest here, $(\kappa_V^2-\kappa_{2V})$. 


\section{The role of $(\kappa_V^2-\kappa_{2V})$ in  Higgs pair production from VBF} \label{section:2}

In this section,  we  explain shortly why  the specific combination of $\kappa$ modifiers given by $\kappa_V^2-\kappa_{2V}$  plays an important role in  Higgs pair production at colliders.  For this,  we first tell on
the main aspects of the scattering amplitude of the subprocess $VV \to HH$,  which is involved at colliders within the LO-HEFT framework introduced before,  and then discuss their  implications on the cross-section and differential cross-sections.  We consider here the two interesting cases,  with $VV$ being either $WW$ or $ZZ$. The most relevant subprocess for the $pp$ collisions, explored in the next section, is known to be $WW \to HH$, both for the SM and  HEFT cases. The double Higgs production rates coming from $ZZ$ fusion at the LHC are known to be much smaller than those from $WW$ fusion,  due to the reduced luminosity  for radiating $ZZ$ pairs from the $pp$  compared to radiating $WW$ pairs. However, to be complete in the forthcoming estimates for the LHC in the next section,  all the channels will be taken into account here.

\begin{figure}[!h] \centerline{ \includegraphics[width=1.0\textwidth]{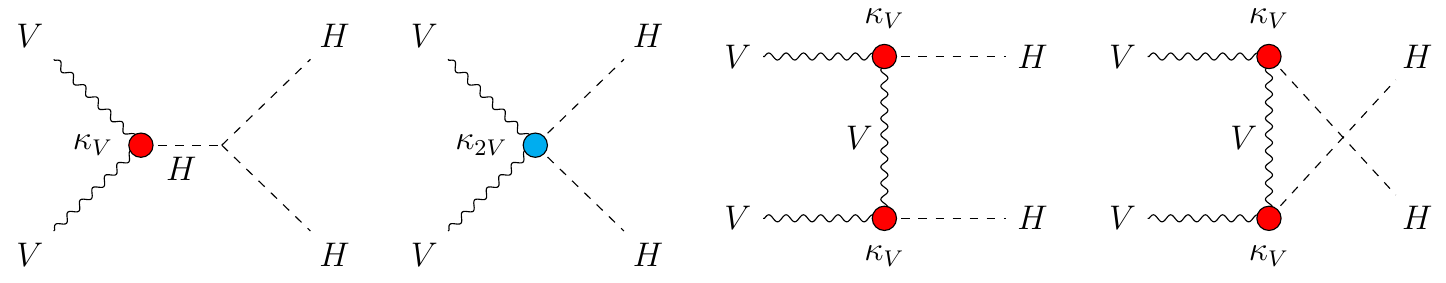} }
\caption{Feynman diagrams contributing to $VV \to HH$  (with  $VV=WW, ZZ$) within the \textcolor{orange}{LO-}HEFT in the Unitary gauge.  From left to right, diagrams for the s-channel, contact c-channel, t-channel and u-channel.  Effective vertices proportional to \textcolor{Red}{$\kappa_V=a$} and \textcolor{Cerulean}{$\kappa_{2V}=b$} are colored in \textcolor{Red}{red} and \textcolor{Cerulean}{blue}, respectively.} 
\label{fig:02_01} \end{figure} \vspace{-1em}

As we said before,  the considered BSM theory only involves modifications of the interaction vertices between the scalar and massive gauge bosons. At $pp$ colliders, we expect the leading contribution to the partonic $q_1q_2 \rightarrow q_3q_4HH$ BSM signal to come from the vector boson  $VV \rightarrow HH$ \text{\small ($VV = WW, ZZ$)} scattering subprocess. Other channels not mediated by VBF are known to be subdominant (see, for instance,  Ref. \cite{ARGANDA2019114687}).  Compared to the production of a Higgs boson pair through a lepton initial state, hadron collisions involve a wider variety of possible initial partons (we refer the reader to appendix \ref{anexo} for the complete list of topologies). Both neutral and charged vector bosons have to be taken into consideration when determining the double Higgs production amplitude through VBF. 

All diagrams contributing to the tree-level amplitude of the $VBF$ subprocess in the unitary gauge are included in Figure \ref{fig:02_01}.  Amplitudes, contrary to observables such as the cross-section, provide us with information on all kinematics of an event. Since the amplitude is not expected to be modified by a certain gauge choice, due to gauge invariance, we choose here the simplest unitary gauge, reached by setting $U=1$ in the HEFT Lagrangian, which avoids involving any diagram with GB internal lines. With this particular section, we aim to illustrate the sensitivity of this subprocess to the particular combination of our interest $a^2 - b$, or equivalently $\kappa_V^2 - \kappa_{2V}$, in terms of $\kappa$ modifiers. 

The individual contribution of each diagram to the scattering amplitude of $V(p_1)V(p_2) \to H(p_3)H(p_4)$,  corresponding to $s$-channel,  contact $c$-channel,  $t$-channel,  and $u$-channel,  respectively,  can be written as:

\minipage{0.4\textwidth} \small
\begin{align} \label{eq:2_01} \mathcal{A}_s & = a\frac{3g^2\delta}{2m_W}\frac{m_H^2}{s - m_H^2} ( \epsilon_1 \cdot  \epsilon_2 ) \\ \label{eq:2_02} \mathcal{A}_c & = b\frac{g^2\beta}{2} ( \epsilon_1 \cdot \epsilon_2 )
\end{align} \endminipage \hfill
\minipage{0.55\textwidth} \small
\begin{align} \label{eq:2_03} \mathcal{A}_t & = a^2g^2\frac{\delta^2}{t - m_V^2} \big( (\epsilon_1 \cdot \epsilon_2) + \frac{(\epsilon_1 \cdot  p_3 )
(\epsilon_2 \cdot  p_4)}{m_V^2} \big) \\ \label{eq:2_04} \mathcal{A}_u & = a^2 g^2\frac{\delta^2}{u - m_V^2} \big( (\epsilon_1 \cdot  \epsilon_2) + \frac{(\epsilon_1 \cdot p_4) (\epsilon_2 \cdot  p_3)}{m_V^2}\big) \end{align} \endminipage \vspace{-1em}

\begingroup \small
\begin{align} \notag \text{where }\delta = \begin{cases} ~m_W &\text{V = W$^\pm$} \\ \frac{m_Z}{cos\theta_W} &\text{V = Z} \end{cases} \text{and } \beta = \begin{cases} ~~~1 &\text{V = W$^\pm$} \\ \frac{1}{cos^2\theta_W} &\text{V = Z} \end{cases} ~~~~~~~~~~~~~~~~~~~ \end{align} \endgroup

Since the FR have the same tensor dependence independently of the considered vector boson $W^\pm, Z$, with their only difference being present in the coupling constants and the masses, a general term for each diagram's amplitude can be obtained. $\delta$ and $\beta$ parameters are introduced to particularize results for $VV=WW$ or $VV=ZZ$ initiated  processes. 
$\epsilon_{1,2}$ identify the two initial massive $V$ bosons polarization vectors,   and p$_{3,4}$ identify the two final Higgs' 4-momenta. 

The parameter $\kappa_{2V}$ is exclusively accessible through the contact diagram, justifying requiring Higgs boson pair production processes by VBF to be accessible. Effectively, $\kappa_{2V}$ contributes directly to the scattering amplitude, marking a qualitative and quantitative physical difference from $\kappa_V$. The rest of the diagrams involve $\kappa_V$, with order 1 in the s-channel diagram and order 2 in the t-channel and u-channel diagrams. The contribution of $\kappa_V$ to the amplitude is expected to be of order $\kappa_V^2$. These last statements are directly related to the behavior of the amplitude at high energies, as it has already been shown in \cite{D_vila_2024}.  At high energies, the dominant helicity  amplitude is the one for the longitudinal modes,  which at high energies grows with the VBF subprocess energy. This can be understood by means of the Equivalence Theorem (ET) \cite{Cornwall:1974km, Vayonakis:1976vz, Lee:1977eg} that states the equality of scattering amplitudes at high energies  if one replaces the external longitudinal gauge bosons by the corresponding GBs.  In the present case,  the ET implies in particular: ${\cal A}(W^+_LW^-_L \to HH) \simeq {\cal A}(\omega^+ \omega^- \to HH)$ for $\sqrt{s} \gg m_W$.  This can be explicitly shown by comparing our next results (Equation \ref{eq:2_06}) with the results, for instance,  in \cite{Delgado_2014},  where they use the ET. 
On the other hand, a complete analysis going beyond the leading order considered here of the scattering amplitude for  the full $WW \to HH$ process,  including all the relevant HEFT operators up to next to leading order,  and with the explicit comparison of the different polarization channels can be found in \cite{PhysRevD.106.115027}.

Next, we shortly review the computations of the LO-HEFT amplitude for the longitudinal gauge bosons, $\mathcal{A}^L = \mathcal{A}\left( V_L V_L \rightarrow HH \right)$.  Here we are interested in the high energy limit,  i.e. ,  for  $\sqrt{s} \gg m_V, m_H$.   For this exercise,  we use the Center of Mass (CM) frame, and write the  momenta of the Higgs bosons and polarization vectors for the longitudinal modes in terms of the CM energy $\sqrt{s}$, the particle masses,  and the CM scattering angle $\theta$:
\begingroup \small
\begin{align} \label{eq:2_05}  & p_{3,4} = \left( \frac{\sqrt{s}}{2}, \pm \text{sin}\theta \frac{\sqrt{s-4m_H^2}}{2}, 0 , \pm \text{cos}\theta \frac{\sqrt{s-4m_H^2}}{2} 	\right),  \\ \notag
 & \epsilon^{L}_{1,2} = \left( \frac{\sqrt{s-4m^2_V}}{2m_V}, 0, 0, \frac{\pm \sqrt{s}}{2m_V} \right). 
\end{align} \endgroup 
Then, considering the high energy limit,  $\sqrt{s} \gg m_V, m_H$,  in the previous amplitudes,  taking 
$\epsilon_{1,2} = \epsilon^{L}_{1,2}$,  and using the following high energy limit for the Mandelstam variables:
 \begingroup \small
\begin{align} \notag t \simeq - \frac{s}{2}\left(1-\text{cos}\theta \right) \text{, } u \simeq  - \frac{s}{2}\left(1+\text{cos}\theta \right) 
\end{align} \endgroup 
 we get the leading terms  of the amplitude with respect to the $s$ variable,  which we separate into the  various contributions from the various diagrams:
 \begin{align} \mathcal{A}^L = \mathcal{A}^L_s + \mathcal{A}^L_c + \mathcal{A}^L_t + \mathcal{A}^L_u \end{align}
where:

\vspace{-1em} 
\minipage{0.4\textwidth} \small
\begin{align} \notag \mathcal{A}^L_s & = \mathcal{O}(s^0) \\ \notag \mathcal{A}^L_c & =  b\frac{g^2\beta}{4m^2_V}s  + \mathcal{O}(s^0) \end{align} \endminipage \hfill
\minipage{0.6\textwidth} \small
\begin{align} \notag \mathcal{A}^L_t & = -a^2g^2\frac{\delta^2}{8m^4_V} \left( 1 - \text{cos}\theta \right)s + O(s^0) \\  \notag \mathcal{A}^L_u & = -a^2 g^2\frac{\delta^2}{8m^4_V} \left( 1 + \text{cos}\theta \right)s + O(s^0) \end{align} \endminipage 

The total amplitude at high energies,  $\sqrt{s} \gg m_H, m_V$,   is then the following:
\begingroup 
\begin{align} \label{eq:2_06}
\mathcal{A}^L = -\left(a^2 - b\right)\frac{g^2\beta}{8m_V^2}s + \mathcal{O}(s^0) ~~~~~~~~~
\end{align} \endgroup 
where we see explicitly  the announced combination of HEFT coefficients $(a^2-b)$. 
Therefore, the high energy behavior of the $VBF$ scattering amplitude for longitudinal gauge bosons exhibits the linear dependence on the $\kappa$ framework  combination $(\kappa_V^2 - \kappa_{2V})$.  In the particular case of the SM, i.e. ,  for $\kappa_V,  \kappa_{2V}$  = 1,  the linear term of order $s$  vanishes and the resulting SM amplitude is of ${\cal O}(s^0)$.  Following from the expression of $\mathcal{A}^L$, this vanishing is not exclusive to the SM,  thus expecting similar high energy behaviour for the set of HEFT models whose $\kappa$ parameters are such that $(\kappa_V^2-\kappa_V)=0$,  or equivalently for BSM physics where the 
$\kappa$'s are related by  $\kappa_V^2$ = $\kappa_{2V}$.  This particular behaviour will help us in identifying BSM scenarios where this relation is not fulfilled, thus their experimental signals will be better distinguishable from the SM ones.  Our main objective here is precisely to focus on these particular BSM scenarios with $(\kappa_V^2-\kappa_V) \neq 0$, and explore which kind of experimental measurements and types of observables could be more sensitive to this particular combination.   Notice that here and from now on in this work,  for shortness,  whenever we refer to 'HEFT models', 'HEFT scenarios',  'BSM scenarios within HEFT',  etc,  we simply mean HEFT with specific choices for the HEFT parameters/coefficients,  $a$, $b$, etc. 

\begin{figure}[!ht] \leftskip1em
\minipage{0.46\textwidth} \centerline{ \includegraphics[width=.95\textwidth]{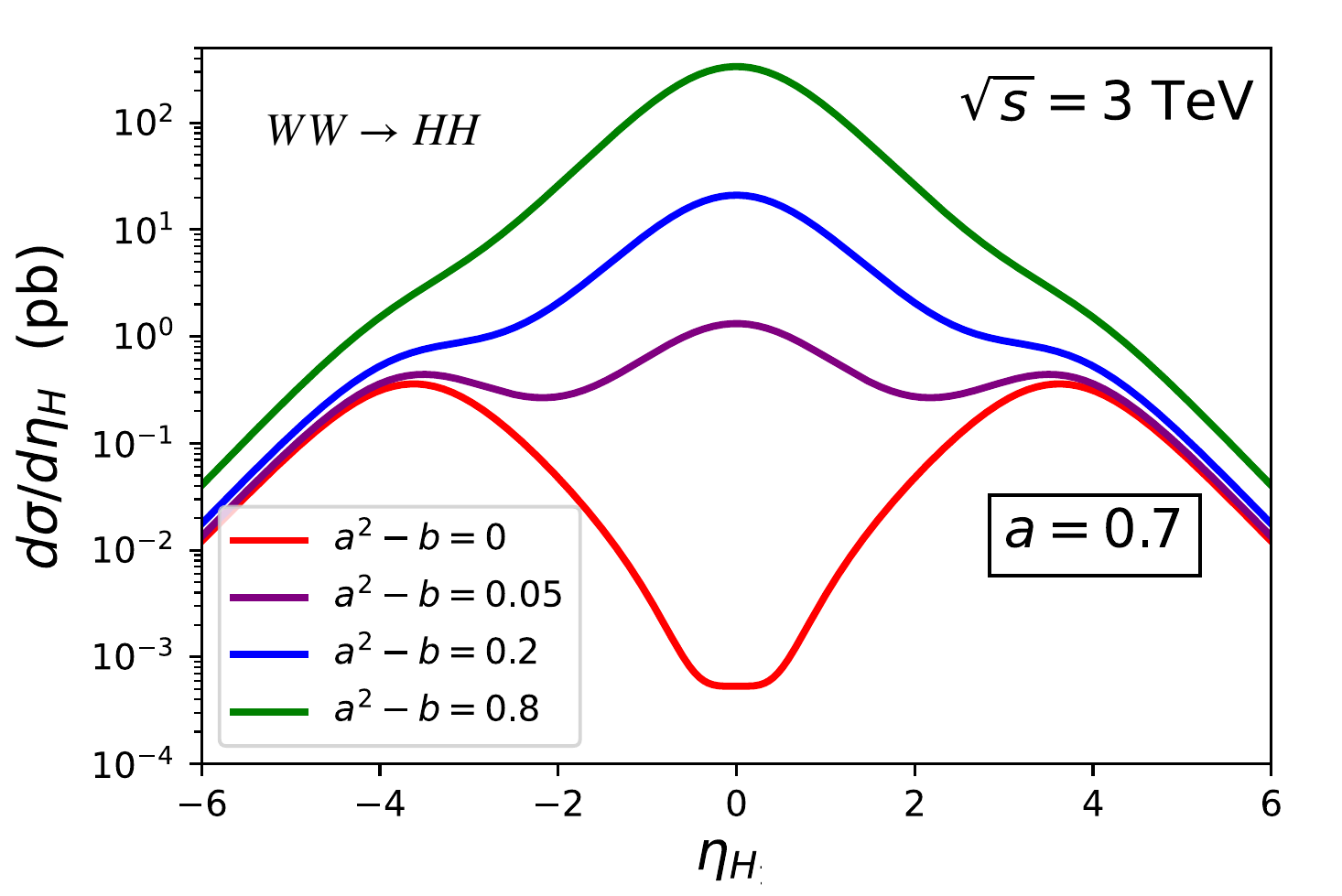}} \endminipage \hspace{2em}
\minipage{0.46\textwidth} \centerline{ \includegraphics[width=.95\textwidth]{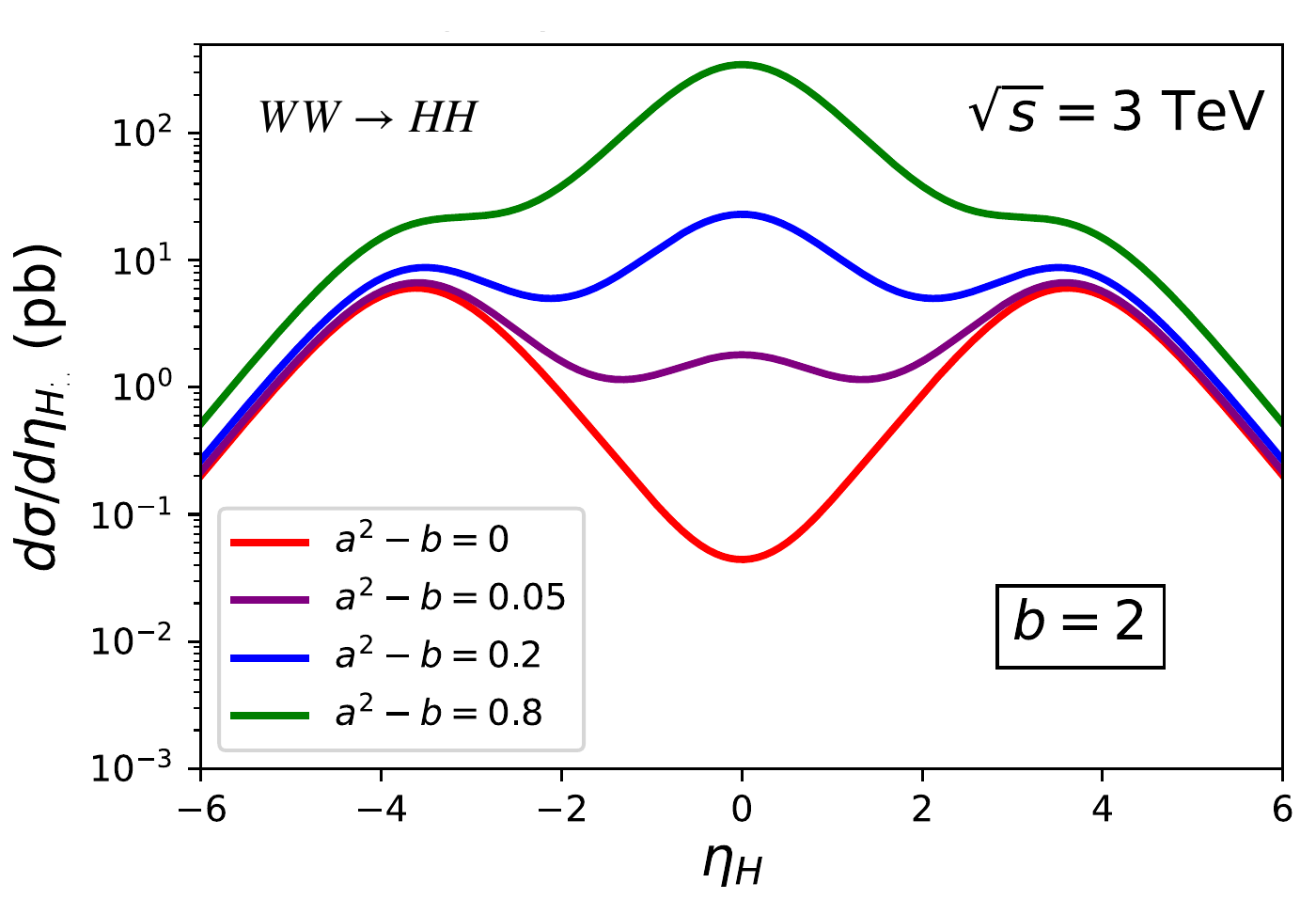}} \endminipage
\caption{ Predictions within the HEFT for the differential cross section of $WW \to HH$  with respect to the final Higgs pseudo-rapidity,  $\eta_H$.  The colored  lines are for different choices of the combination $(a^2-b)$  defining some examples of  BSM scenarios.  We take $(a^2-b)=$ 0 (red line),  0.05 (purple line),  0.2 (blue line) and 0.8 (green line).  Here the total center-of-mass energy is fixed to $\sqrt{s}=3$ TeV.  The plot on the left is for $a=0.7$,  and the plot on the right is for $b=2$.  } \label{fig:etaH} \end{figure}

As in Ref.  \cite{D_vila_2024},  we focus here on the search for particular differential cross-sections to be measured at future colliders which are especially sensitive to the value of $(\kappa_V^2-\kappa_V) \neq 0$.   Given the above shown behaviour of the amplitudes with the scattering angle $\theta$,  it is obvious that the most sensitive differential cross sections to this combination are those with respect to the angular variables of the final Higgs bosons or their decay products.  Concretely,  some examples are:  $d \sigma /d (\cos \theta)$,  $d \sigma /d \eta_H$,  and $d \sigma /d p^T_H$,  where $\sigma$ refers to the unpolarized cross section, and $\eta_H$ and $p^T_H$ are the Higgs pseudo-rapidity and Higgs transverse momentum,  respectively.  

The high sensitivity to $(a^2-b)=(\kappa_V^2-\kappa_{2V})$ in  $d \sigma(VV \to HH) /d \eta_H$ can be seen analitically as follows.  First,  considering the relation between $\eta_H$ and $\theta$  given by: 
 \begin{equation}
        \eta_{H} \, = \, - \log \left( \tan (\theta/2) \right), 
        \label{eqn: eta}
    \end{equation}
    and the corresponding relation among the differential cross sections given by,
    \begin{equation}
                \frac{\text{d} \sigma}{\text{d} \eta_{H}}=(\sin^2\theta)   \frac{\text{d} \sigma}{\text{d} \cos\theta} \,.
      \label{eqn: d_eta-versus-d_cos}
    \end{equation}
    where,
   \begin{equation}
        \frac{\text{d} \sigma}{\text{d} \cos\theta} \, = \,  \frac{1}{64 \pi s} \frac{\sqrt{s - 4m_H^2}}{\sqrt{s - 4m_W^2}}|\bar{\mathcal{A}}|^2,
        \label{eqn: diff_amp}
    \end{equation}
    and  the average in  $|\bar{\mathcal{A}}|^2$ is over the $3 \times 3$ polarization combinations of the initial $V$'s.  
 The factor $1/2$ due to the two identical final Higgs bosons is also included.  Since  the total  unpolarized cross section of $VV \to HH$ is  dominated at large energy by the $V_L V_L \to HH$ component,  ${\cal A}^L$,  then one finds out the announced behaviour of $d \sigma /d \eta_H$ having  a maximum at the central pseudo-rapidity value,  $\eta_H=0$ (or $\theta=\pi/2$),  with the  height of the central peak growing with $|a^2-b|^2$  as expected from Eq.\ref{eq:2_06}.
 In particular,   for $WW \to HH$,   the above commented  sensitivity to $(a^2-b)$  in $d \sigma /d \eta_H$  is illustrated  in Fig.\ref{fig:etaH}.  
 Other examples can be found in Ref.  \cite{D_vila_2024}.   We see that the higher the value of $(a^2-b)$ is,  the more transversely  the Higgs boson is produced,  since the central peak at $\eta_H=0$  is more pronounced.  We wish to emphasize that this behaviour of the final $H$ showing a high transversality is genuine of HEFT models with  $(\kappa_V^2-\kappa_V) \neq 0$ and the height  of the central peak is correlated with this combination value,   and it is not much dependent on the separate values of $\kappa_V$ and $\kappa_{2V}$.  This can be clearly seen comparing the two plots in Fig.  \ref{fig:etaH} that assume different values for 
the two involved $\kappa$'s, or equivalently, different values  for $a$ and $b$,  and both manifest the highest central peak at the highest value of $a^2-b$,  here fixed to 0.8.  

Therefore,  from a collider perspective, these HEFT models will provide BSM signals where the Higgs bosons (or their decay products) are produced highly transverse, i.e. , with angular distances closer to the plane transverse to the beam axis.  We refer the reader to Ref. \cite{D_vila_2024},  where more examples are  included and  a more detailed analysis of the subprocess differential cross-sections is performed. 
The consequences of the previously commented sensitivity of the described subprocess to the combination $(a^2-b)$  will be translated to the context of the real scattering process.  This was explored in Ref. \cite{D_vila_2024} for the case of $e^+e^-$ colliders.  In the present work,  we will explore this sensitivity in the case of hadronic $pp$ colliders.   Finding the proper differential cross section distributions showing this sensitivity for the full process 
$pp \to HH jj$ is indeed one of our main objectives in this work.  This issue will be studied in the following sections.

\section{Exploring $HHjj'$ production in $pp$ collisions within HEFT} \label{section:3}

In this section, the HEFT predictions for $HHjj'$ production from $pp$ collisions are presented and analyzed.  First we evaluate the integrated cross-section for $pp \rightarrow HH jj'$ with VBF topology as a function of $a$,  $b$,  and the combination $a^2-b$.  Then,  we introduce the main phenomenology of Higgs boson pair angular kinematics resulting from  BSM within the HEFT framework,  and study the most interesting effects on  the proper distributions with these angular variables as a function of this same $a^2-b$ combination.  
Throughout this section, only the parton-level signals produced using \textsc{MG5} \cite{Alwall_2014} simulations have been considered.  For definiteness,   we use the following notation/definition for the kinematical variables of the final state $HHjj'$  (composed of four objects,  two Higgs particles  and two jets) which are the usual ones in the lab-frame for $pp$ collisions :  
$\eta_A$ (or $\eta^{A}$)  is the pseudo-rapidity of final object $A$,  $P_{T,A}$ (or $P^T_A$) is the transverse momentum of final object $A$,  
$\Delta \eta_{AB}$ (or $ \Delta \eta_{A,B}$) is the distance in pseudo-rapidity between object $A$ and object $B$,  $\Delta \Phi_{AB}$ is the distance in azimutal angle between object $A$ and object $B$,  $\Delta R_{AB}$ (or $\Delta R_{A,B}$) is  defined in terms of 
$\Delta \eta_{AB}$ and  $\Delta \Phi_{AB}$ as $\Delta R_{AB}=\sqrt{(\Delta \eta_{AB})^2+ (\Delta \Phi_{AB})^2}$,  $M_{AB}$ 
(or  $m_{AB}$) is the invariant mass of the joint system $AB$ formed by the two objects $A$ and $B$.  

\subsection{Vector Boson Fusion (VBF): Events selection strategy} \label{subsection:3_01}

\begin{figure}[!ht] \leftskip1em
\minipage{0.45\textwidth} \centerline{ \includegraphics[width=.95\textwidth]{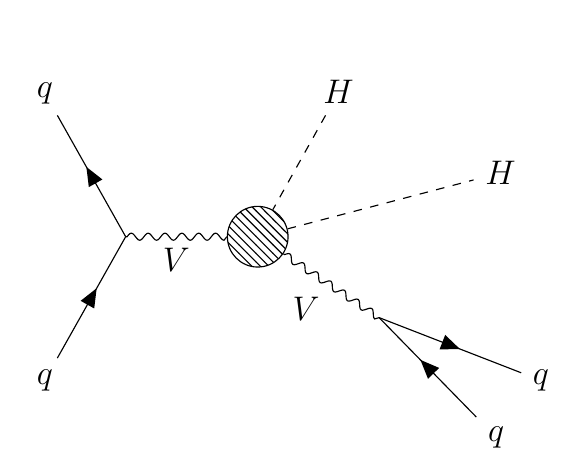}} \endminipage \hspace{2em}
\minipage{0.45\textwidth} \centerline{ \includegraphics[width=.95\textwidth]{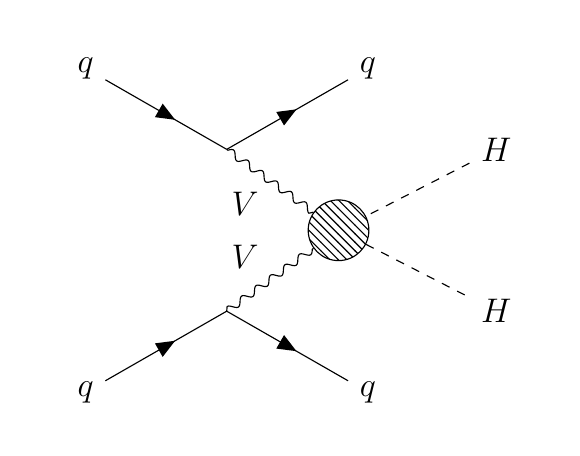}} \endminipage
\caption{Schematic of the two different topology classes involved in the EW Higgs boson pair production in $pp$ collisions,  the associated production (double Higgs-strahlung) [left] and the vector boson fusion (VBF) [right].  Here, $V$ refers generically to either $W^\pm$ or $Z$ and $q$ denotes either quarks $q$ or anti-quarks $\overline{q}$.} \label{fig:04_00} \end{figure}

Our first step here is the selection of $HHjj'$ events with VBF topology from the complete set of possible topologies in the partonic $q_1q_2 \rightarrow HH q_3q_4$ process,  which are schematically represented in figure \ref{fig:04_00}.   Notice that the initial and final quarks $q$ refer generically to either quarks or anti-quarks.  The full set of Feynman diagrams for the case $q_1 \overline{q}_2 \to HH q_3 \overline{q}_4$ is displayed in Fig. \ref{fig:A} in the Appendix.  Recall also that the particular  diagrams including the subprocess $VV \to HH$ cannot be just extracted by hand,  since excluding the other diagrams from the sample simulation would break gauge invariance.  However,  it is well known that events with VBF topology can be efficiently selected, preserving gauge invariance by the proper cuts on the two final light jets $jj'$,  which are typically emitted in narrow cones and in opposite hemispheres, being easily differentiated from non-VBF  jet pairs.  Both light jets in the final state acquire a moderate absolute pseudo-rapidity,  typically in the range $2 \leq |\eta_j| \leq 5$,  and a notable pseudo-rapidity difference between themselves.  This is true not only for the SM case but also for all BSM cases within the HEFT \cite{PhysRevD.86.036011, szleper2015higgsbosonphysicsww, Delgado:2017aa, Gon_alves_2018, ARGANDA2019114687, PhysRevD.98.114016}. 

\vspace{1em} \begin{figure}[!h] 
\includegraphics[width=1.0\textwidth]{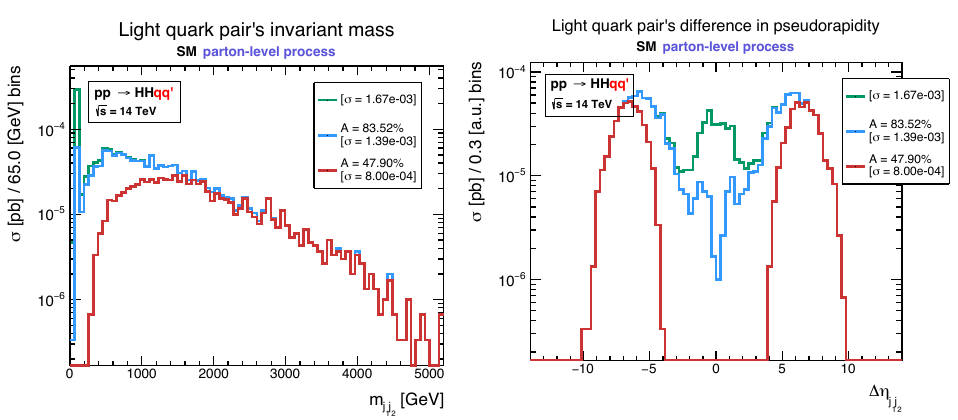} 
\caption{Parton-level SM differential cross section of the double Higgs production with respect to the light quark pair's invariant mass [left],  and to the light quarks pseudo-rapidity difference [right], for different kinematic cuts applied to the event's phase space. In \textcolor{green}{green}, the initial signal. In \textcolor{blue}{blue}, reduced to events with $\eta_{j_1} \times \eta_{j_2} < 0 $, with an acceptance of 83.52 $\%$. In \textcolor{red}{red}, reduced to events with $\eta_{j_1} \times \eta_{j_2} < 0 $, as well as $\left| \eta_{j} \right| > 2$, with an acceptance of 47.90 $\%$.} \label{fig:04_01} \end{figure} 

\begin{figure}[!h] 
\includegraphics[width=1.0\textwidth]{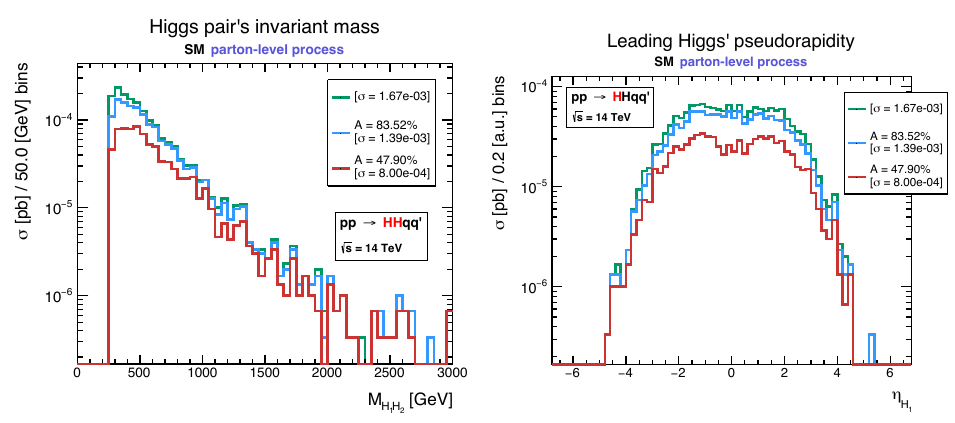}
\caption{Parton-level SM differential cross section of the double Higgs production with respect to the Higgs boson pair's invariant mass [left],  and to the leading Higgs pseudo-rapidity [right], for different kinematic cuts applied to the event's phase space. In \textcolor{green}{green}, the initial signal. In \textcolor{blue}{blue}, reduced to events with $ \eta_{j_1} \times \eta_{j_2} < 0 $, with an acceptance of 83.52 \%. In \textcolor{red}{red}, reduced to events with $ \eta_{j_1} \times \eta_{j_2} < 0 $, as well as $\left| \eta_{j} \right| > 2$, with an acceptance of 47.90 \%.} \label{fig:04_02} \end{figure} \vspace{.5em}

\begin{table}[!h] \footnotesize \vspace{1em} \centerline{
\begin{tabular}{|c|c|c|c|c|c|}
\hline
\textcolor{green}{$S_0$} & $P_{T, j} > 20.0$ GeV & $\left| \eta_j \right| < 5.0$ & $\Delta R_{j,j'} > 0.4$ &                                       &                 \\ \hline
\textcolor{blue}{$S_1$}  & $P_{T, j} > 20.0$ GeV & $\left| \eta_j \right| < 5.0$ & $\Delta R_{j,j'} > 0.4$ & $\eta_{j} \times \eta_{j'}    < 0$ &                 \\ \hline
\textcolor{red}{$S_2$}   & $P_{T, j} > 20.0$ GeV & $\left| \eta_j \right| < 5.0$ & $\Delta R_{j,j'} > 0.4$ & $\eta_{j} \times \eta_{j'}    < 0$ & $ \left| \eta_j \right| > 2 $ \\ \hline
\end{tabular}} \caption{Summary of the kinematic cuts,  for $pp \to HH jj'$,   applied to SM predictions in Figures \ref{fig:04_01} and \ref{fig:04_02}.} \vspace{0em}
\label{table:04_02}
\end{table}

Figures \ref{fig:04_01} and \ref{fig:04_02} show the predictions for the SM differential cross-section of the $pp \rightarrow HH jj'$ process at the partonic level with respect to several kinematic variables.  In these figures the final jets $j$ and $j'$ are also referred to as $j_1$ and $j_2$ respectively.  Figure \ref{fig:04_01} includes, from left to right, the differential cross-section with respect to the invariant mass of the jet pair,  and the jets' pseudo-rapidity difference. Figure \ref{fig:04_02} shows, from left to right, the differential cross-section with respect to the Higgs boson pair's invariant mass,  and the leading Higgs boson's pseudo-rapidity. All signals are generated with the following set of basic cuts: 

\vspace{-2.5em} \bgroup \small \begin{align}  P_{T, j} > 20.0 \text{GeV},~~~ \left| \eta_j \right| < 5.0,~~~ \Delta R_{jj'}  > 0.4 \end{align} \egroup \vspace{-2.7em}

Green distributions only involve said basic cuts. The invariant mass distribution in green from Figure \ref{fig:04_01} shows two maxima,  approximately at $m_{jj'} \approx 500$ GeV and $m_{jj'} \approx 0$. The latter is directly related to resonant diagrams, and we expect this maximum to be present at either $m_W = 80$ or $m_Z = 91$ GeV. In the pseudo-rapidity difference distribution three maxima can be found,  one around $\left|\Delta \eta_{jj'}\right| \approx 0$ and the other two at $\left|\Delta \eta_{jj'}\right| \approx 6$. The first maximum is related to topologies that generate two collimated jets.  Said kinematics regions, i.e. $m_{jj'} \approx 0$ and $\left|\Delta \eta_{jj'}\right| \approx 0$, do not correspond to $VBF$ events.

In order to isolate the VBF events, we will first check the results of applying two kinematic cuts to the pseudo-rapidity of the jets: 

\vspace{-2.7em} \bgroup \small \begin{align} \label{eq:04_02} \eta_{j_1} \times \eta_{j_2} < 0, ~~~~~~~~~~ \left| \eta_{j} \right| > 2 \end{align} \egroup \vspace{-3em}

Blue distributions present in Figures \ref{fig:04_01} and \ref{fig:04_02} correspond to the parton-level signal filtered for events whose jets pseudo-rapidity have opposite signs $\eta_{j_1} \times \eta_{j_2} < 0$, that is, evolve in opposed directions with respect to the beam's axis. For the case of the jet pair invariant mass distribution, the blue signal shows a reduction of almost one order of magnitude in the number of resonant events. In the central region of the jet pseudo-rapidity difference distribution, the maximum is now a minimum. Red distributions correspond to the parton-level signal filtered for both criteria present in equation \ref{eq:04_02}. The differential cross-section with respect to the jet pseudo-rapidity difference perfectly fits a $ 4 < \left|\Delta\eta_{jj'}\right| < 10$ distribution, resulting from the final individual $2 < \left| \eta_{j} \right| < 5$ selection cut. The jet pair invariant mass distributions do not include a visible resonant contribution, and the non-resonant maximum has moved closer to $m_{jj'} \approx 1000$ GeV. 

From Figure \ref{fig:04_02},  we conclude that all differential cross-section distributions with respect to the Higgs boson's kinematics, for either the initial or the filtered signal, have the same exact profile. The only visible difference is in the scale, related to the number of  events resulting from each selection process. The kinematics of the generated Higgs boson pair are not affected by applying both kinematic cuts. 

The acceptance for the SM case results of imposing both kinematic cuts, defined as the quotient of the filtered events over the total number of events, $A = \frac{N_{cut}}{N_{orig.}}$, is $A = 47.90\%$.  Thus, for this  SM scenario,  the events with VBF topology contribute to $50\%$ of the total number of events.  Extending this type of analysis to all BSM scenarios within the HEFT framework considered in this work , we find an acceptance for this event selection with VBF topology ranging  between  $50\%$ and  $75\%$ in agreement with  \cite{ARGANDA2019114687}.  

In summary,  the final selection strategy applied to all parton-level signals generated using \textsc{MG5} in order to isolate the events with VBF topology,  for both the SM and HEFT predictions of the $pp\rightarrow HHjj'$ process,  is composed of the following 5 kinematic cuts:

\vspace{.5em} \begin{table}[!h] \small \centering{
\begin{tabular}{ccccc}
\hline
\multicolumn{5}{|c|}{Event   selection}                                                                                                                                                                                                                                                \\ \hline
                                                       &                                        &                                                        &                                                       &                                                                     \\  [-13pt] \cline{1-1} \cline{3-5} 
\multicolumn{1}{|c|}{\multirow{2}{*}{VBF topology}}    & \multicolumn{1}{c|}{\multirow{2}{*}{}} & \multicolumn{1}{c|}{$P_{T, j} > 20.0$ GeV}             & \multicolumn{1}{c|}{$2 < \left| \eta_j \right| < 5$}             & \multicolumn{1}{c|}{$\eta_{j} \times \eta_{j'}    < 0$}          \\ \cline{3-5} 
\multicolumn{1}{|c|}{}                                 & \multicolumn{1}{c|}{}                  & \multicolumn{1}{c|}{$m_{jj'} > 500$ GeV}                & \multicolumn{1}{c|}{$\Delta R_{jj'} > 0.4$}            & \multicolumn{1}{c|}{}                  \\ \cline{1-1} \cline{3-5}                                                                                                                                                                                                                                                                                                  
\end{tabular}} 	\caption{Event selection strategy for VBF topologies, used for theoretical predictions for the $pp\rightarrow HHjj'$ process.} \label{table:00_00}
\end{table} \vspace{-.5em}

The selection strategy described in Table \ref{table:00_00} will be referred to as the VBF selection strategy. The strategy involves all kinematics cuts discussed throughout this subsection, either basic and jet pseudo-rapidity cuts, as well as cuts to the invariant mass of the light jet pair, $m_{jj'} > 500$ GeV, to further isolate the VBF configuration.  Indeed, this selection strategy will also be presented for the recalculated prediction,  including the $b\overline{b}\gamma\gamma j j'$ final state.

\subsection{HEFT total cross section,  $\sigma(pp \to HHjj')$,  versus $(\kappa_V^2-\kappa_{2V})$} \label{subsection:3_02}

In this section, we present the predictions of the total cross section for the hadronic process,  $\sigma(pp \to HHjj')$,  within the context of the HEFT, and analyze the results as a function of the combination of the $\kappa$  parameters of our interest, given by $(\kappa_V^2-\kappa_{2V})$.  We first analyze the sensitivity to the HEFT parameters $a$ and $b$ separately,  and then show the relevance of combining  them particularly into $(a^2-b)$.  We expect the sensitivity to particular HEFT models of the process $pp \rightarrow HHjj'$ to inherit most of the properties we discussed in section \ref{section:2} for the $VV \rightarrow HH$ subprocess.  

For this analysis, we simplify the parton constituents of the protons to include just the quarks and antiquarks of two lighetest generations and the gluons.  For the process of our interest, $pp \rightarrow HHjj'$, the relevant partonic subprocess, $q_1q_2 \rightarrow HH q_3q_4$, is folded with the proper Parton Distribution Function (PDF's) for the quarks and antiquarks inside the proton, chosen to be NNPDF2.3 \cite{Ball2013290}. The simulations make use of a diagonal Cabibbo-Kobayashi-Maskawa (CKM) matrix,  which is a very good approximation for the present computation.  For this subsection and the following, Subsection \ref{subsection:3_03}, the VBF selection strategy detailed in Table \ref{table:00_00} has been applied.

\vspace{-.5em} \begin{figure}[!h] 

\centerline{ \includegraphics[width=0.55\textwidth]{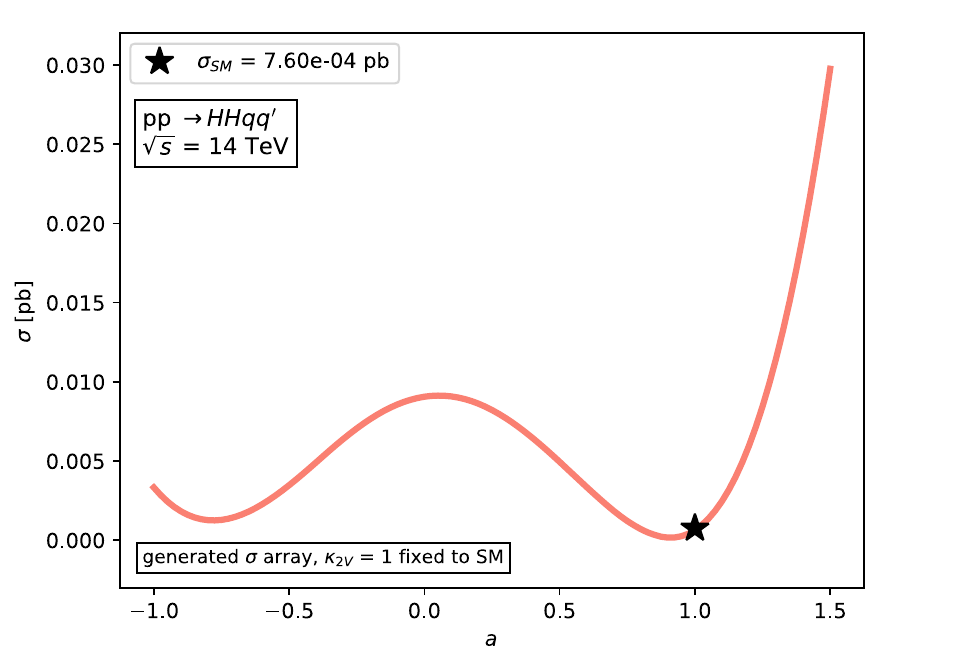} \hspace{-1em}
\includegraphics[width=0.55\textwidth]{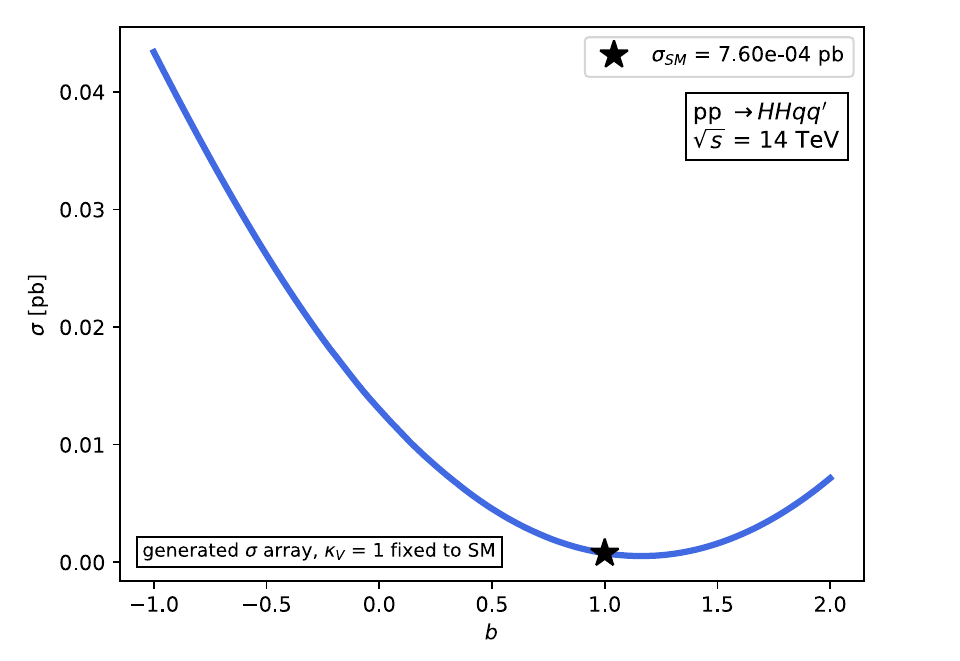} }

\caption{HEFT integrated cross-section of the $pp \to HHjj'$ process, $\sigma(pp \to HHjj')$, as a function of each of the two relevant parameters, either $\kappa_V=a$ [left] or $\kappa_{2V}=b$ [right].  For each graph, the parameter that is not explicit,  has been fixed to the SM value, either $\kappa_{2V} = 1$ [left] or $\kappa_{V} = 1$ [right]. The intervals considered here are $ -1 < \kappa_V < 1.5 $ and $ -1 <  \kappa_{2V} < 2 $.  Notice, however,  that the subintervals allowed by data in Eqs. \ref{ATLASconstraints} and \ref{CMSconstraints} are indeed much smaller.} 
 
\label{fig:04_03} \end{figure} \vspace{0em}

Despite the present constraints for both $\kappa_V$ and $\kappa_{2V}$,  and for illustrative purposes,  we have considered analyzing the sensitivity of the Higgs boson pair production cross-section for a wider set of HEFT models, with $\kappa$ parameters comprising the following values: 

\vspace{-2.5em} \bgroup \small \begin{align} \label{eq:04_03}
-1.0 < \kappa_V < 1.5, ~~~~~~~~~~~~~ -1.0 < \kappa_{2V} < 2.0 0 \end{align} \egroup \vspace{-3em}

Figure \ref{fig:04_03} shows the integrated cross-section for the $pp \rightarrow HHjj'$ process as a function of either $a=\kappa_V$ or $b=\kappa_{2V}$.  Specifically, the function on the left illustrates the sensitivity of the process to a particular set of HEFT models for which the value of $\kappa_{2V}$ is fixed to the SM, $\kappa_{2V} = 1$, and $\kappa_V$ is left as a free parameter. The analogue for $\kappa_{2V}$ is represented on the right. 

The total cross-section with respect to $\kappa_V$ exhibits two local minima in the considered range of values. For the cross-section with respect to $\kappa_{2V}$, there is only a minimum point. The number of local minima reflects the order of the insertions each coefficient has in the amplitude of the VBF subprocess, discussed in section \ref{section:2}. A quadratic dependence with a given $\kappa$ in the amplitude reflects two insertions whereas a linear dependence reflects single insertions. This aspect will be relevant in future discussions throughout this subsection.  Following  from the $\kappa_V$ insertion being of order two, the integrated cross-section with respect to $\kappa_V$ exhibits an axis of symmetry at $\kappa_V \approx 0$, again associated with the leading term of the high energy expansion in the VBF amplitude. However, the local maximum does not appear exactly at $\kappa_V = 0$,  and both local minima are not localized  at the exact values of $\kappa_V = \pm 1$,  but they are slightly shifted.  Furthermore, the cross-section present at the minima at $\kappa_V < 0$ is slightly greater than the predicted value for the minima at $\kappa_V > 0$.  These slight deviations from the simple behaviour found for the subprocess $A(V_L V_L \to HH)$ at high energies in eq. \ref{eq:2_06}  reflect simply the effect of the small  contributions from the next order, ${\cal O}(s^0)$,  in the large energy expansion of this amplitude that are obviously entering in the full process computation.  

\begin{figure}[!t] 

\centerline{ \includegraphics[width=0.68\textwidth]{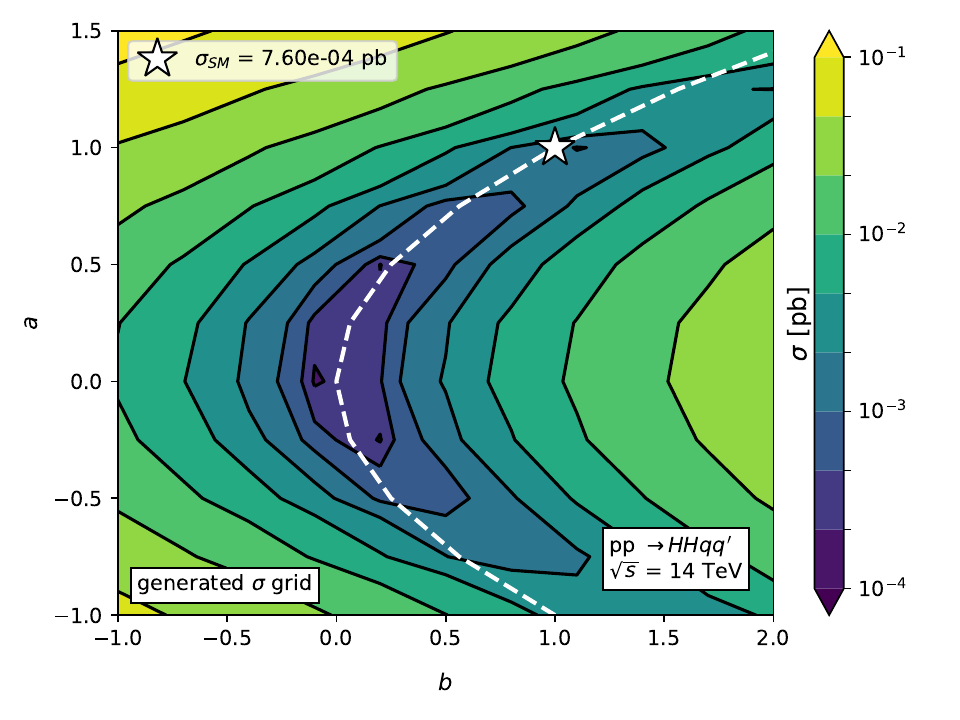} }

\caption{Contour lines of HEFT cross-section predictions for the $HH$ production, $\sigma(pp \to HHjj')$, in the $(\kappa_{2V},\kappa_{V})=(b,a)$ plane. The grid consists of 121 simulations with different $\kappa_V$ and $\kappa_{2V}$ pairs within the intervals $ -1 < \kappa_V < 1.5 $ and $ -1 < \kappa_{2V} < 2 $. The point  $\kappa_V = \kappa_{2V} = 1$ is indicated with a star and corresponds to the standard model, whose cross-section is specified in the legend. The white dashed line corresponds to all points following a $\kappa_V^2 - \kappa_{2V} = 0$ combination.} 

\label{fig:04_04} \end{figure}

\begin{figure}[!b] 

\centerline{ \includegraphics[width=0.68\textwidth]{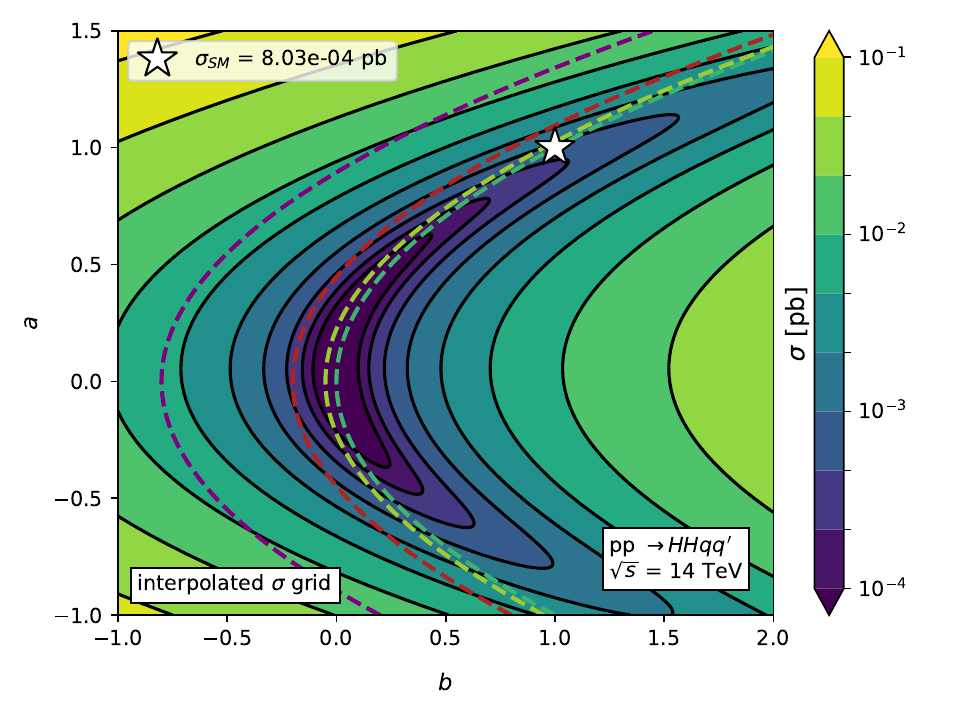} }

\caption{Interpolated two variable function for the integrated cross-section of the contour lines of $\sigma(pp \to HHjj')$ in the $(\kappa_{2V},\kappa_{V})=(b,a)$ plane. The considered intervals are $ -1 < \kappa_V < 1.5 $ and $ -1 < \kappa_{2V} < 2 $. The point $\kappa_V = \kappa_{2V}  = 1$ is indicated with a star and corresponds to the standard model, whose cross-section is specified in the legend. The dashed lines correspond to all points following $\kappa_V^2 - \kappa_{2V} = 0.00$ [\textcolor{ForestGreen}{dark green}], $\kappa_V^2 - \kappa_{2V} = 0.05$ [\textcolor{SpringGreen}{light green}], $\kappa_V^2 - \kappa_{2V} = 0.20$ [\textcolor{BrickRed}{red}] and $\kappa_V^2 - \kappa_{2V} = 0.80$ [\textcolor{RedViolet}{violet}] correlations. } 

\label{fig:04_05} \end{figure}

On top of both distributions in Figure \ref{fig:04_03}, the cross-section for the particular set of $\kappa$ values that correspond to the SM is indicated with a starred mark.  We can see  in both distributions that the SM prediction is quite close to the cross-section minimum.  Thus,  we expect a limited sensitivity in the $pp \rightarrow HHjj'$ process to the set of HEFT models with small departures from $\kappa_V = 1$ and $\kappa_{2V} = 1$.  This sensitivity is slightly greater for the $\kappa_V$ distribution,  as a result of the minimum having a bigger curvature. 

The previous graphs offer us an insight into the role that each $\kappa$ parameter has on the total cross-section of the process.  However, as we mentioned in the introduction, one of the main benefits of using the HEFT formalism is the treatment of the $a$ and $b$ coefficients as independent parameters, in contrast to other EFTs such as the SMEFT.  Figures \ref{fig:04_04} and \ref{fig:04_05} show the integrated cross-section of the $pp \rightarrow HHjj'$ process with respect to both $\kappa_V$ and $\kappa_{2V}$ parameters together,  as contour maps over the ($\kappa_{2V}$, $\kappa_{V}$) parameter space.  Both maps are obtained from the same calculations,  but Figure \ref{fig:04_04} shows the total cross-section obtained from exact parton-level simulations whereas Figure \ref{fig:04_05} shows the approximate results from an interpolated function with a 100 times increase in the resolution. 
This interpolation  is performed via the fitting of the set of cross-sections from each parton-level simulation to a polynomial in two variables, namely $\kappa_V \equiv a$ and $\kappa_{2V} \equiv b$. The polynomial is formed as an expansion in powers of $a$ and $b$ up to $a^4$ and $b^2$ as follows: 
\bgroup \begin{align} \label{eq:04_04}  \sigma_{pp\rightarrow HHjj'} (a, b) = C_{a^4}a^4 + C_{a^3}a^3 + C_{a^2}a^2 + C_{a^2b}a^2b + C_{ab}ab + C_{b^2}b^2 \end{align} \egroup
Notice that other potential  combinations of coefficients have not been considered since they do not result from any  interference from the different diagrams discussed in the amplitude calculation of Section \ref{section:2}. 
The resulting cross-section interpolated function is described by the following coefficients: 

\vspace{-2.7em} \bgroup \begin{alignat}{2} \label{eq:04_05}
C_{a^4} &= ~~0.01662248 \text{ pb}, ~~~ C_{a^3} =& ~-0.00414373 \text{ pb}, ~~~ C_{a^2} &= 0.00051408 \text{ pb}  \\ \notag
C_{a^2b} &= -0.02407124 \text{ pb}, ~~~ C_{ab} =& ~0.00261345 \text{ pb}, ~~~ C_{b^2} &= 0.00926828 \text{ pb} \, . \end{alignat} \egroup  \vspace{-2.7em}

Since the coefficients $C_{a^2b}$, $C_{ab}$,  and $C_{a^3}$ are not negligible, we expect our process to be sensitive not only to the absolute value of $a$ and $b$ coefficients, but also to their individual and relative sign. The direction in which we move along the ($\kappa_{2V}$, $\kappa_{V}$) parameter space will play an important role in our analysis of the sensitivity to the $\kappa_{V}^2-\kappa_{2V}$ combination.  The main conclusion  from the comparison of these two contour plots in Figs. \ref{fig:04_04} and \ref{fig:04_05} is that this interpolating function works pretty well and provides quite accurate predictions for the total rates of the $pp \to HHjj$ (with VBF topology)  process  within the HEFT models. 
We have also checked that this same function with the coefficients scaled down by the factor ${\rm BR}(H \to b \bar b) \times {\rm BR}(H \to \gamma \gamma)$  also works pretty well for the extrapolated events in the forthcoming analysis where we study the HEFT signal events with final state $b \bar b \gamma \gamma jj$. 

The most important outcome from our study  of the total cross section $\sigma(pp \to HHjj)$ with VBF topology in these two figures,  Figs. \ref{fig:04_04} and \ref{fig:04_05},  is the relevance of the value of the combination $\kappa_{V}^2-\kappa_{2V}$.  The larger this value is, the larger is the departure of the HEFT cross-section with respect to the SM prediction, given by:
 \bgroup \small \begin{align} \label{eq:04_06}
\sigma^{SM, a=b=1}_{pp\rightarrow HHjj'} = 8.033\times10^{-4} pb \,. \end{align}
\egroup 
From a quantitative standpoint, the least sensitive region to test is the central darker region, $-0.4 < \kappa_V <  0.6, -0.2 < \kappa_{2V} <  0.4$,   which is near the $\kappa_{V}^2 - \kappa_{2V} = 0$ contour line.   For the particular cases studied here,  we see that the largest values of the cross-section are reached for the largest values considered of $a^2-b$ and the smallest cross-section values are reached for the smallest values of  $a^2-b$ considered,   i.e.,  those  close to the contour line $a^2-b=0$.  
In particular,  the highest cross-sections, and thus sensitivities, are obtained near the point $(a, b) = (1.5, -1.0)$,  with $a^2-b=3.25$ and with a calculated value of $\sigma^{a=1.5, b=-1.0} = 0.132$ pb, more than two orders of magnitude greater than the SM cross-section. 
Therefore, we can already conclude that the largest sensitivity to the $\kappa$'s is reached in the regions of the plane distant from the $(\kappa_V^2-\kappa_{2V})=0$ line (dashed white line in Figure \ref{fig:04_04}).  Thus,  going to a dedicated study of the sensitivity to BSM physics via the combination 
$(\kappa_V^2-\kappa_{2V})$ is clearly motivated. 

\begin{figure}[!t] 
\centerline{ \includegraphics[width=0.52\textwidth]{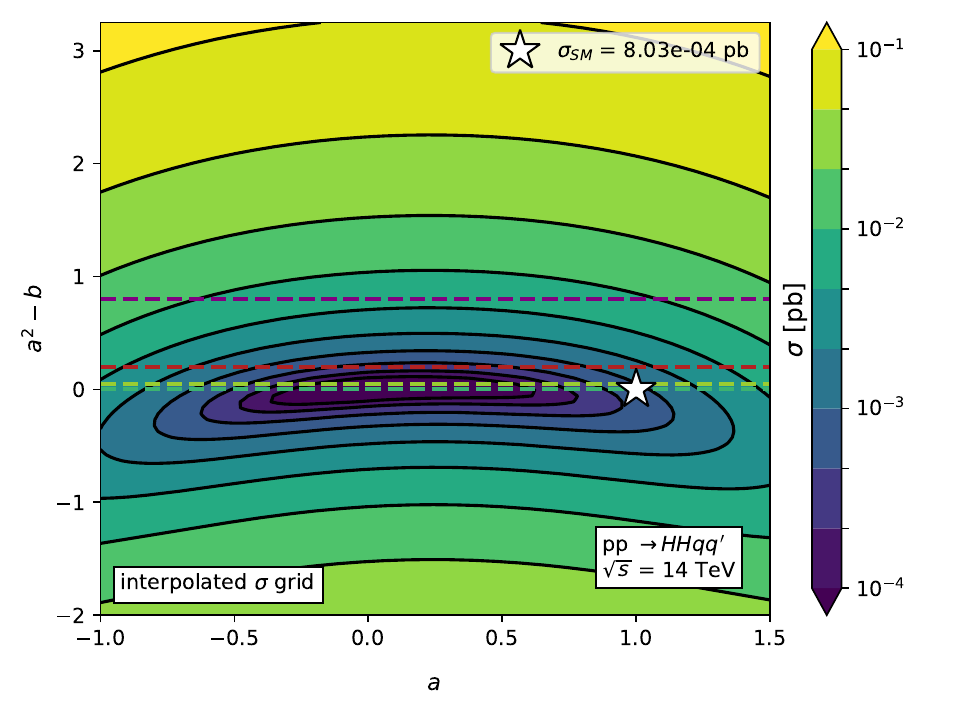} 
\includegraphics[width=0.52\textwidth]{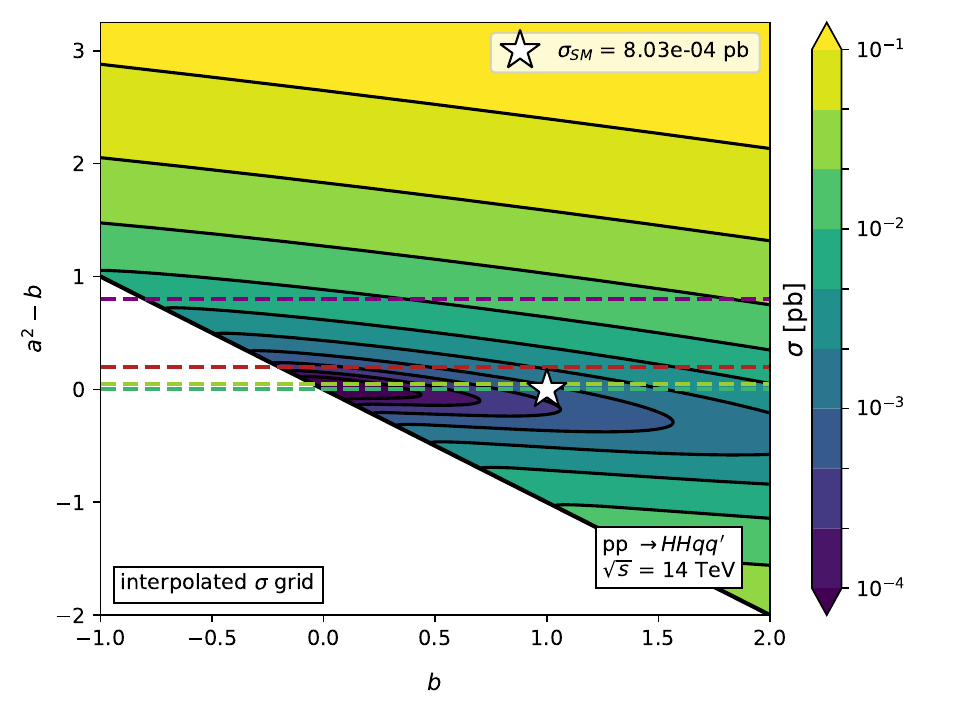} }
\caption{Interpolated two variable function of the contour lines of $\sigma(pp\to HHjj')$ in the $(\kappa_{V},\kappa^2_{V} - \kappa_{2V}) = (a,a^2 - b)$ plane [left] or $(\kappa_{2V},\kappa^2_{V} - \kappa_{2V}) = (b,a^2 - b)$ plane [right]. The considered intervals are $ -1 < \kappa_V < 1.5 $, $ -1 < \kappa_{2V} < 2 $ and $ -2 < \kappa^2_V -\kappa_{2V} < 3.25 $,  respectively. The SM point  is indicated with a star.  The dashed lines correspond to all points following $\kappa^2_V -\kappa_{2V} = 0.00$ [\textcolor{ForestGreen}{dark green}], $\kappa^2_V -\kappa_{2V} = 0.05$ [\textcolor{SpringGreen}{light green}], $\kappa^2_V -\kappa_{2V} = 0.20$ [\textcolor{BrickRed}{red}] and $\kappa^2_V -\kappa_{2V} = 0.80$ [\textcolor{RedViolet}{violet}] correlation parameters.  The white region present in graph B corresponds to values of $\kappa_V^2 - \kappa_{2V}$ and $\kappa_{2V}$ for which $\kappa_{V}$ results imaginary. }
\label{fig:04_06} 
\end{figure}

The most clear test that this $\kappa^2_V -\kappa_{2V}$ combination is the most relevant parameter is provided in Fig. \ref{fig:04_06},  where the contourlines of the cross section,  $\sigma(pp \to HHjj')$,  are shown differently than in the previous figures.  Here,  the cross-section predictions are done with the interpolating function and are displayed in the  $(\kappa_{V},\kappa^2_{V} - \kappa_{2V}) = (a,a^2 - b)$ plane on the left plot,  and on the $(\kappa_{2V},\kappa^2_{V} - \kappa_{2V}) = (b,a^2 - b)$ plane in the right plot.  The ranges considered in this plot for $\kappa_V$,  $\kappa_{2V}$,   and  $\kappa_{V}^2-\kappa_{2V}$  are $ -1 < \kappa_V < 1.5 $, $ -1 < \kappa_{2V} < 2 $,  and $ -2 < \kappa^2_V -\kappa_{2V} < 3.25 $,  respectively.  Several dashed lines with particular values of $a^2-b$ are also included for reference.
Notice that in the right plot,  there exists a white region where there are no predictions because it corresponds to values of $\kappa_V^2 - \kappa_{2V}$ and $\kappa_{2V}$ for which $\kappa_{V}$ results imaginary.  The most important feature in these two plots is the shape found for the contourlines that are nearly flat   along either $\kappa_V$ or $\kappa_{2V}$ axis,  indicating that there is a weaker sensitivity to these two parameters when considered separately and a stronger sensitivity  when combined via $\kappa_V^2 - \kappa_{2V}$.

Therefore,  we conclude from this subsection that the particular combination of the HEFT $a$ and $b$ parameters given by $(a^2-b)$, or, equivalently, by $(\kappa_{V}^2-\kappa_{2V})$, is indeed the most relevant one to study the implications of BSM Higgs physics in the total rates of $HHjj'$ production at LHC.  
In the next subsection,  we will show that this relevance of the combination $(a^2-b)$ also appears in the differential cross section with respect to the pseudo-rapidity of the final Higgs bosons.  

\subsection{HEFT  differential cross section,  $d\sigma(pp \to HHjj')/d \eta_{H}$,  versus  $(\kappa^2_V - \kappa_{2V})$} \label{subsection:3_03}
\begin{figure}[!t] 
\centerline{ \includegraphics[width=0.96\textwidth]{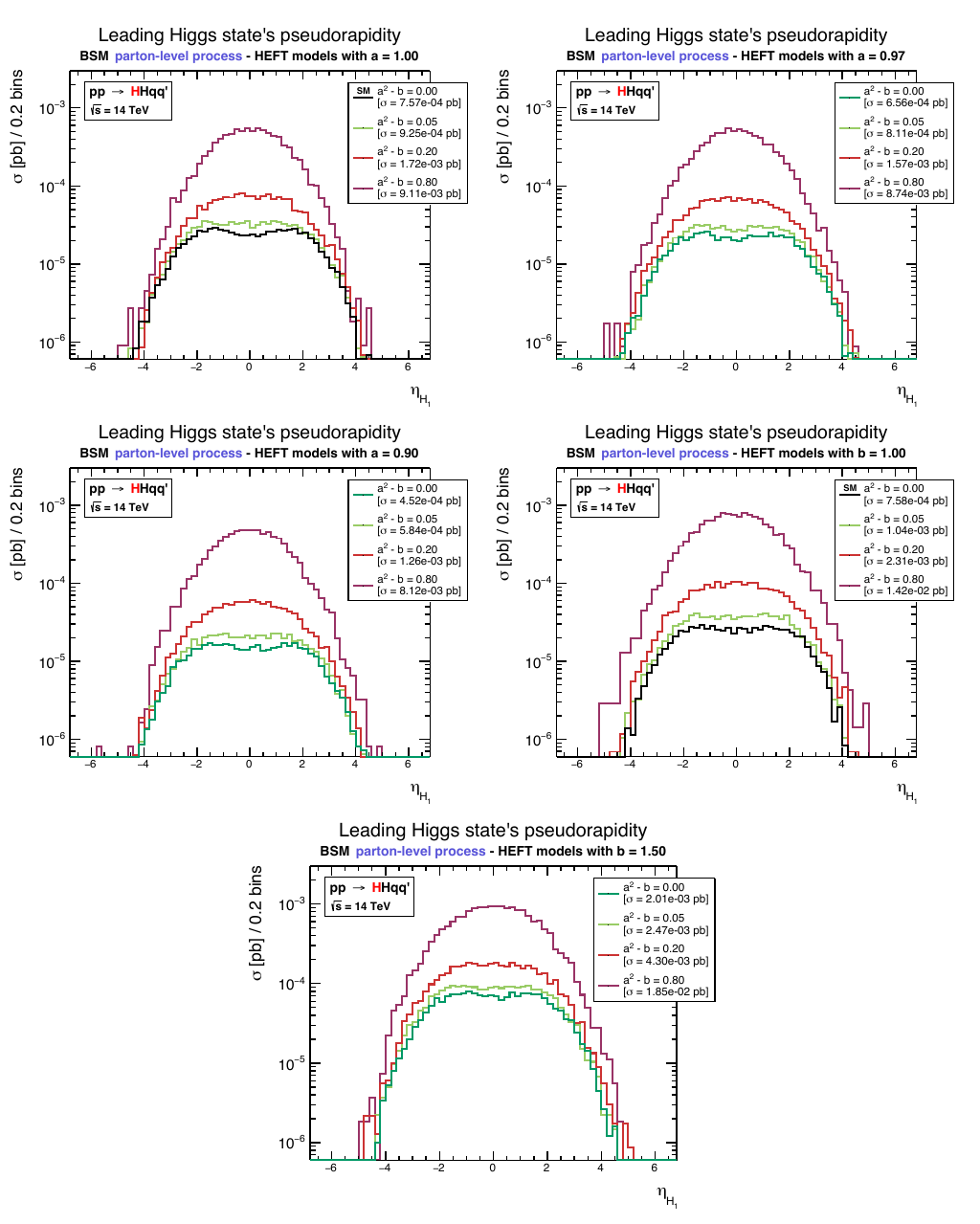}}
\caption{Parton level HEFT predictions for the differential cross-section of the $pp \to HHjj'$ process with respect to the pseudo-rapidity of the Higgs boson leading in $P_T$, $d\sigma(pp\to HHjj') / d\eta_{H_1}$, for different values of $a$ and $b$ parameters.  Each line's color corresponds to a given value of the combination $(a^2-b)$:  
$(a^2-b) = 0.00$ [\textcolor{ForestGreen}{dark green}], $(a^2-b) = 0.05$ [\textcolor{SpringGreen}{light green}], $(a^2-b) = 0.20$ [\textcolor{BrickRed}{red}] and $(a^2-b) = 0.80$ [\textcolor{RedViolet}{violet}]. The parameter fixed for each graph, either $a=\kappa_V$ or $b=\kappa_{2V}$, is indicated in the title of each plot. Specifically, from top to bottom, left to right,  the choices are  $a$ = 1.00,  $a$ = 0.97,  $a$ = 0.9,  $b$ = 1.00,  $b$ = 1.50.  Distributions in black correspond to SM ($a=b=1$) predictions. } 
\label{fig:04_07} 
\end{figure}  
In this subsection, we  explore the HEFT predictions for some differential cross-sections that may enhance the sensitivity to the combination $(\kappa^2_V - \kappa_{2V})$  of our interest here.  In particular,  we focus on the differential cross section with respect to  
the Higgs boson's pseudo-rapidity $\eta_{H}$,  which turns out to be an interesting variable in this study, given the expected high transversality of the final Higgs bosons,  as discussed in section \ref{section:2}. 

In order to study the angular phenomenology of our process with respect to the aforementioned combination of $\kappa_V$ and $\kappa_{2V}$, we simulated a sample consisting of 20 signals obtained using HEFT models with the particular combinations of the $\kappa$-parameters $\kappa^2_V - \kappa_{2V} = 0.00, 0.05, 0.20 \text{ and } 0.80$. 

Figure \ref{fig:04_07} shows the predicted differential cross-section within HEFT of the $pp\rightarrow HHjj'$ signals with respect to the pseudo-rapidity $\eta_{H_1}$ of the Higgs boson leading in $P_T$.  Each of the four lines included within each graph are obtained using a HEFT model with a different value for the $\kappa^2_V - \kappa_{2V}$ combination.  Signals in dark green correspond to $\kappa^2_V - \kappa_{2V} = 0.00$, in light green to $\kappa^2_V - \kappa_{2V} = 0.05$, in red to $\kappa^2_V - \kappa_{2V} = 0.20$,  and in violet to $\kappa^2_V - \kappa_{2V} = 0.80$. The title of each graph also indicates explicitly the value of the $a$ or $b$ parameter fixed for all four lines. Predictions obtained using the values $a = b = 1$ corresponding to the SM case are shown in black.   

We put our focus first on the first plot ($\kappa_V = 1.00$) and  the fourth plot ($\kappa_{2V} = 1.00$) since both plots include the SM prediction and we can compare this reference case with the HEFT predictions.  Using HEFT models with increasing values for the combination parameter $\kappa^2_V - \kappa_{2V}$,  the BSM signals deviate from the SM twin-peak distribution to a central distribution with respect to the Higgs pseudo-rapidity.  These twin-peak distributions with maxima at $\left| \eta_{H}\right| \approx 2$ characteristic of the SM are associated typically with the VBF topology.  Moreover, we clearly see that the height of the central peak around $\eta_H=0$ in the HEFT predictions grows with the value of $(\kappa^2_V - \kappa_{2V})$,  as expected from the commented behaviour of the subprocess amplitude $\mathcal{A}_L$ with this combination of HEFT parameters. 

The fact that all plots In  Fig.  \ref{fig:04_07} show a similar pattern for the four displayed lines is what precisely demonstrates that this pattern is fixed  by the value of $a^2-b$ and not by the separate values of $a$ and $b$.   For instance,  the lines for 
$(a^2-b)=0$ correspond to the predictions for the SM,  with $a=1$ and $b=1$,  and also correspond to the predictions for the BSM cases with $a \neq 1$ and $b \neq 1$ but following the relation $(a^2-b)=0$.  The main conclusion from these plots,  which is important for the objective of this work,  is  that the most relevant parameter to describe BSM signals in the process of our interest here,  $pp \to HH jj$,  is 
$(a^2-b)$ rather than $a$ and $b$ separately. 
In  Fig.  \ref{fig:04_07},   we also see that  all HEFT predictions  with a null value for the combination parameter $\kappa^2_V - \kappa_{2V} = 0.00$ have the same twin-peak $\eta_{H_1}$ shape as that of the SM.  In consequence,  the difference between HEFT models in the angular phenomenology of the Higgs bosons is mainly determined by this combination parameter.  


Therefore,  exploring the differential cross section with respect to the Higgs boson pseudo-rapidity and looking for the HEFT  enhancements in the surrounding region close to the central peaks at $\eta_H=0$ seems to offer a good discriminant to disentangle the BSM physics with the highest sensitivity via the 
combination parameter $(\kappa^2_V - \kappa_{2V}) \neq 0$.  As we will see in the following sections,  these same features will manifest when moving to the realistic final state $pp \to HHjj  \to b \bar b\gamma\gamma jj' $,  even after including showering and detector effects.

\section{Detector-level HEFT prospects for  $pp \to HHjj  \to b \bar b\gamma\gamma jj' $} \label{section:4}
In the following,  we show the HEFT predictions for the total cross section and differential cross sections as a function of $(\kappa_V^2-\kappa_{2V})$,  once showering, hadronization and detector level effects are taken into account, considering the realistic final state with two b-jets, two light jets,  and two photons.   The total process that will be studied  here is $pp \to HHjj  \to b \bar b\gamma\gamma jj' $,  which is produced from the previously studied $HHjj$ case after the two final Higgs bosons decay to $H \to b \bar b$ and $H \to \gamma \gamma$,  respectively.  

For  a realistic study,  jets at the final state are properly defined considering not only parton showering and hadronization phenomena, but also the clustering techniques leading to the definition of the jets from the results of the hadron fragmentation. Thus, instead of the partonic level process,  $pp \to b \bar b \gamma \gamma q_1 q_2$,   one must consider $pp \to j_bj_{\overline{b}} \gamma \gamma jj'$, with $j_bj_{\overline{b}}$ the two \enquote{true}  $b$-jets and $\gamma \gamma$ the two \enquote{true} photons from the $HH$ pair decay, and $jj'$ the two \enquote{true} light jets characterizing the VBF topology. Parton showering, hadronization,  and the jet clustering process are implemented on top of the original \textsc{MG5} parton-level simulations using \textsc{Pythia}8.235 \cite{Sj_strand_2015}. In particular, the algorithm used to perform the clustering process is the anti-k$_t$ \cite{Cacciari_2008} algorithm, implemented using the FastJet \cite{Cacciari_2012} package. As stated in the introduction, the Phase-2 upgraded CMS detector \cite{delphes_card_idea} will be simulated using \textsc{Delphes} \cite{Favereau:2014aa} for a fast and parametric detector simulation. The result of the implementation of \textsc{Pythia} constitutes an intermediate step used in combination with \textsc{Delphes} to produce detector-level simulated signals.  Regarding the notation for the kinematical variables of the final state in this section,  with six identified objects which are now two b-jets,  $j_{b}$ and $j_{\bar b}$,  two light jets,  $ j$ and $j'$,  and two photons,  we use again the usual definitions for pp collisions in the lab-frame:  
$\eta_A$ (or $\eta^{A}$)  is the pseudo-rapidity of final object $A$,  $P_{T,A}$ (or $P^T_A$) is the transverse momentum of final object $A$,  
$\Delta \eta_{AB}$ (or $ \Delta \eta_{A,B}$) is the distance in pseudo-rapidity between object $A$ and object $B$,  $\Delta \Phi_{AB}$ is the distance in azimutal angle between object $A$ and object $B$,  $\Delta R_{AB}$ (or $\Delta R_{A,B}$) is  defined in terms of 
$\Delta \eta_{AB}$ and  $\Delta \Phi_{AB}$ as $\Delta R_{AB}=\sqrt{(\Delta \eta_{AB})^2+ (\Delta \Phi_{AB})^2}$,  $M_{AB}$ 
(or  $m_{AB}$) is the invariant mass of the joint system $AB$ formed by the two objects $A$ and $B$. 

\subsection{Detector-level SM prospects for the  $b\overline{b}\gamma\gamma jj'$ final state} \label{subsection:4_05}

As a reference case,  we first analyze the efficiency loss when moving from parton level to detector level in the SM case (remember that the SM  can be considered as a particular  HEFT case with $a=b=1$).   Concretely,  we start with the SM sample that we have generated at the parton level,  as already outlined in 
table \ref{table:00_00}.  This selected sample is then passed through PYTHIA and DELPHES imposing additional selection criteria,  as specified in table \ref{table:SM-efficiency}.

\vspace{.5em} \begin{table}[!h] \footnotesize \centerline{
\begin{tabular}{clccc}
\hline
\multicolumn{5}{|c|}{Detector-level event selection strategy for the SM case}                                                                                                                                                                                                                                            \\ \hline
\multicolumn{5}{c}{}                                                                                                                                                                                                                                                                                     \\ [-10pt] \cline{5-5} 
\multicolumn{4}{c|}{}                                                                                                                                                                                                                                     & \multicolumn{1}{c|}{SM Selection efficiencies ($E_i$)}              \\  \cline{5-5} 
\multicolumn{5}{c}{}                                                                                                                                                                                                                                                                                     \\ [-10pt] \cline{1-2} \cline{4-5} 
\multicolumn{2}{|c|}{\multirow{4}{*}{\makecell{\\\\VBF topology}}}    & \multicolumn{1}{c|}{\multirow{4}{*}{}} & \multicolumn{1}{c|}{\makecell{At least two light jets with: $P_{T, j} >   20$ GeV, $2 < \left| \eta_j \right| < 5$}}
& \multicolumn{1}{c|}{\multirow{2}{*}{\makecell{\\$66\%$}}} \\ \cline{4-4}
\multicolumn{2}{|c|}{}                                   & \multicolumn{1}{c|}{}                  & \multicolumn{1}{c|}{\makecell{Selected VBF jet pair $jj'$ maximizes\\the $\Delta R$ between light jets}}                                                                       & \multicolumn{1}{c|}{}                        \\ \cline{4-5} 
\multicolumn{2}{|c|}{}                                   & \multicolumn{1}{c|}{}                  & \multicolumn{1}{c|}{\makecell{VBF jets with: $\eta_j\times\eta_{j'} < 0$}}                                                                                              & \multicolumn{1}{c|}{$91\%$}                  \\ \cline{4-5} 
\multicolumn{2}{|c|}{}                                   & \multicolumn{1}{c|}{}                  & \multicolumn{1}{c|}{\makecell{VBF jet pair with: $m_{jj'} > 500$ GeV}}                                                                                           & \multicolumn{1}{c|}{$98\%$}                  \\ \cline{1-2} \cline{4-5} 
\multicolumn{5}{c}{}                                                                                                                                                                                                                                                                                     \\ [-10pt] \cline{1-2} \cline{4-5} 
\multicolumn{2}{|c|}{\multirow{4}{*}{\makecell{\\\\\\$HH$ topology}}} & \multicolumn{1}{c|}{}                  & \multicolumn{1}{c|}{\makecell{At least two loosely identified photons\\ with $P_{T, \gamma} > 18$ GeV, $\left| \eta_\gamma \right|   < 2.5$}}                                            & \multicolumn{1}{c|}{\multirow{2}{*}{\makecell{\\$86\%$}}} \\ \cline{4-4}
\multicolumn{2}{|c|}{}                                   & \multicolumn{1}{c|}{\multirow{3}{*}{}} & \multicolumn{1}{c|}{\makecell{Selected photon pair $\gamma\gamma$ maximizes\\proximity to $m_{\gamma\gamma} = m_{H} = 125$ GeV}}                                               & \multicolumn{1}{c|}{}                        \\ \cline{4-5} 
\multicolumn{2}{|c|}{}                                   & \multicolumn{1}{c|}{}                  & \multicolumn{1}{c|}{\makecell{At least two loosely b-tagged jets\\with $P_{T, j_b} > 25$ GeV, $\left| \eta_{j_b} \right| < 2.5$}}                                     & \multicolumn{1}{c|}{\multirow{2}{*}{\makecell{\\$81\%$}}} \\ \cline{4-4}
\multicolumn{2}{|c|}{}                                   & \multicolumn{1}{c|}{}                  & \multicolumn{1}{c|}{\makecell{Selected $b$-jet pair $j_bj_{\overline{b}}$ maximizes\\proximity to $m_{j_bj_{\overline{b}}} = m_{H} = 125$ GeV}}                                                      & \multicolumn{1}{c|}{}                        \\ \cline{1-2} \cline{4-5} 
\multicolumn{5}{l}{}                                                                          \\ [-8pt] \hline \multicolumn{4}{|c|}{Total selection efficiency $\Pi^iE_i$}                                                                                                                                                                                                          & \multicolumn{1}{c|}{$36\%$}                  \\ \hline
\end{tabular}} \caption{Detector-level event selection strategy for the SM case with $b\overline{b}\gamma\gamma jj'$  final state} 
 \label{table:SM-efficiency}
\end{table}
The selection efficiency used to evaluate each individual criterion $i$ is defined as:
\begin{equation}
 \label{eq:4_B} 
E_i = N_{i ~ \text{detector-level}} ~/~ N_{\text{parton-level}} 
\end{equation}
where $N_{i ~ \text{detector-level}}$ denotes the number of events resulting from the detector-level selection criterion $i$ applied to the initial $N_{\text{parton-level}}$ parton-level events. 
Table \ref{table:SM-efficiency} shows the selection efficiencies resulting from the application of each selection criterion to the SM sample.   It is worth noting that, for consistency reasons, the detector-level selection strategy for the VBF topology is similar to the event selection present at the parton level, outlined in table \ref{table:00_00}.  For the photon and the b-tagged jet definitions at detector level,  we have considered a \enquote{loose} working point from the preestablished \textsc{Delphes} options.  According to \cite{CMS:2018ccd} this \enquote{loose} working point is defined as that presenting an efficiency in the identification/tagging  of a photon of $90\%$  and a b-jet of $90\%$. 
The combination of all selection efficiencies is then defined as the product of the individual efficiencies  $\Pi^iE_i$,  which is also shown in this table.  

Some comments are in order for the various selection  criteria.   The selected events in the first criterion are required to have a minimum of two light jets with $P_{T,j} > 20$ GeV and $2 < \left| \eta_j  \right| < 5$.   This criterion also includes the requirement that the selected two light jets maximize the distance $\Delta R$ between them.  
The resulting selection efficiency is  $E_1 = 66\%$ for the SM signal (and close to  $E_1 \approx 65\%$ for all studied BSM signals).  This criterion is the one contributing the most to the selection efficiency loss. The selected VBF jet pair is required to have an invariant mass greater than $m_{jj'} > 500$ GeV and both light jets to have a pseudo-rapidity with opposed sign. These two requirements have a general selection efficiency close to  $E_2 \approx E_3 \approx 95\%$. The combination of the selection efficiencies from the detector-level VBF topology selection is, in general, of approximately a $E_1 \times E_2 \times E_3 \approx 60\%$ and dominates the selection efficiency loss at detector-level.  Aside from the VBF selection process, the selected events are required to have a pair of reconstructed photons and $b$-jets satisfying the following set of criteria. The selected events are required to have a minimum of 2 b-tagged jets with $P_T > 25$ GeV and $\left| \eta \right| < 2.5$ and a minimum of two photons with $P_T > 18$ GeV and $\left| \eta \right| < 2.5$.  For all studied signals, the selection efficiency of both sets of criteria is  close to a $E_4 \approx E_5 \approx 80\%$ respectively.  In summary,  For the particular case of the SM sample, 
the application of the selection strategy in Table \ref{table:SM-efficiency} at detector level results in a combined selection efficiency value of $(\Pi^iE_i)^{SM} = 36\%$, that is, a reduction of approximately 1/3 of the predicted amount of events produced at parton level.  It is important to note that the value of $\Pi^iE_i$ is not yet fully optimized at this point.  We will investigate later a better optimized choice of selection criteria when studying together the HEFT signal and the background  events in the last section. 

Regarding the efficiency loss in the SM  differential cross-section distributions we have studied several kinematic variables of interest for the final state,  $b\overline{b}\gamma\gamma jj'$.  In particular, we have considered:   invariant mass of the photon pair $M_{\gamma \gamma}$,    invariant mass of the $b$-jet pair  $M_{b \bar b}$,  invariant mass of the light jet pair $M_{j_1j_2}$,  pseudo-rapidity of the light jet with $P_T$  leading $\eta_{j1}^{\rm leading}$,   pseudo-rapidity difference between both light jets $\Delta \eta_{j1,j2}$,  invariant mass of the system with two photons and two $b$-jets $M_{\gamma \gamma b \bar b}$,  transverse momentum of the photon pair $P_{T, \gamma \gamma}$,   pseudo-rapidity of the photon pair  $\eta_{\gamma \gamma}$,  transverse momentum of the $b$-jet  pair, $P_{T, b \bar b}$,  and pseudo-rapidity of the $b$-jet pair  $\eta_{b \bar b}$.  As a general conclusion, we have checked that most of the distributions do not show relevant changes in their shape, comparing them to the original parton-level distributions.   In particular,  in the cases of angular variables there is no appreciable change in the filtered distributions after PYTHIA and DELPHES respect to the parton-level ones.  This is clearly the case of the distributions in pseudo rapidity and in transverse momentum variables.  The most relevant change occurs in the distributions with  $M_{\gamma \gamma}$ and $M_{b \bar b}$,  as expected.  The efficiency loss  introduced by the hadronization, showering and detector effects are therefore noticed at the total cross section level and not so much at the differential cross section level.   A similar conclusion will show up in the HEFT predictions  as will be presented  in the next  subsection.  

\subsection{Detector-level BSM prospects within the HEFT for the  $b\overline{b}\gamma\gamma jj'$ final state} \label{subsection:4_05}
In this subsection,  we analyze the efficiency loss when moving from parton level to detector level in the HEFT case with different values for the $a$ and $b$ coefficients other than those of the SM case, i.e. ,  for $a \neq 1$ and/or $b \neq 1$.   We  focus our analysis on the HEFT predictions for the various corresponding values  of the combination of our interest $a^2-b$.   As in the previous case,  we start with the HEFT sample that we have generated at the parton-level,  as already outlined in 
table \ref{table:00_00}.  This selected HEFT sample is then passed through PYTHIA and DELPHES imposing additional selection criteria,  as already specified for the SM in table \ref{table:SM-efficiency}.  We summarize the HEFT results in tables  \ref{table:04}  and \ref{table:07}.  These tables include the HEFT predictions for the parton-level integrated total cross section,  $\sigma(b\overline{b}\gamma\gamma jj')$,  the total efficiency loss 
$\Pi^iE_i$,    and the total number of events $N$ using a prospected integrated luminosity of 3000 fb$^{-1}$ as one of the benchmarks for the HL-LHC operation \cite{CMS-PAS-FTR-21-004}.

\begin{table}[!t] \footnotesize 

\begin{tabular}{ccccccccc}
\cline{2-9}
\multicolumn{1}{c|}{}               & \multicolumn{8}{c|}{$a =   1.00$}                                                                                                                                                                                                                                                                                                          \\ \hline
\multicolumn{1}{|c|}{$a^2 - b$}     & \multicolumn{1}{c|}{$-1.00$}             & \multicolumn{1}{c|}{$-0.80$}             & \multicolumn{1}{c|}{$0.00 \text{ \footnotesize (SM)}$}              & \multicolumn{1}{c|}{$0.05$}              & \multicolumn{1}{c|}{$0.20$}              & \multicolumn{1}{c|}{$0.80$}              & \multicolumn{1}{c|}{$1.00$}              & \multicolumn{1}{c|}{$1.50$}              \\ \hline
\multicolumn{1}{|c|}{$\sigma [pb]$} & \multicolumn{1}{c|}{1,25$\times10^{-5}$} & \multicolumn{1}{c|}{7,66$\times10^{-6}$} & \multicolumn{1}{c|}{9,54$\times10^{-7}$} & \multicolumn{1}{c|}{1,22$\times10^{-6}$} & \multicolumn{1}{c|}{2,49$\times10^{-6}$} & \multicolumn{1}{c|}{1,49$\times10^{-5}$} & \multicolumn{1}{c|}{2,17$\times10^{-5}$} & \multicolumn{1}{c|}{4,39$\times10^{-5}$} \\ \hline
\multicolumn{9}{c}{}                                                                                                                                                                                                                                                                                                                                                                        \\ [-8pt] \hline
\multicolumn{1}{|c|}{$\Pi^iE_i$}    & \multicolumn{1}{c|}{44\%}                & \multicolumn{1}{c|}{44\%}                & \multicolumn{1}{c|}{36\%}                & \multicolumn{1}{c|}{36\%}                & \multicolumn{1}{c|}{39\%}                & \multicolumn{1}{c|}{41\%}                & \multicolumn{1}{c|}{41\%}                & \multicolumn{1}{c|}{43\%}                \\ \hline
\multicolumn{1}{|c|}{$N$}           & \multicolumn{1}{c|}{16}             & \multicolumn{1}{c|}{10}             & \multicolumn{1}{c|}{1.0}             & \multicolumn{1}{c|}{1.3}             & \multicolumn{1}{c|}{2.9}             & \multicolumn{1}{c|}{18}             & \multicolumn{1}{c|}{27}             & \multicolumn{1}{c|}{56}             \\ \hline
\multicolumn{9}{c}{}                                                                                                                                                                                                                                                                                                                                                                        \\ \cline{2-9} 
\multicolumn{1}{c|}{}               & \multicolumn{4}{c|}{$a = 0.97$}                                                                                                                                & \multicolumn{4}{c|}{$a = 0.90$}                                                                                                                                \\ \hline
\multicolumn{1}{|c|}{$a^2 - b$}     & \multicolumn{1}{c|}{$0.00$}              & \multicolumn{1}{c|}{$0.05$}              & \multicolumn{1}{c|}{$0.20$}              & \multicolumn{1}{c|}{$0.80$}              & \multicolumn{1}{c|}{$0.00$}              & \multicolumn{1}{c|}{$0.05$}              & \multicolumn{1}{c|}{$0.20$}              & \multicolumn{1}{c|}{$0.80$}              \\ \hline
\multicolumn{1}{|c|}{$\sigma [pb]$} & \multicolumn{1}{c|}{8,12$\times10^{-7}$} & \multicolumn{1}{c|}{1,06$\times10^{-6}$} & \multicolumn{1}{c|}{2,28$\times10^{-6}$} & \multicolumn{1}{c|}{1,44$\times10^{-5}$} & \multicolumn{1}{c|}{5,55$\times10^{-7}$} & \multicolumn{1}{c|}{7,55$\times10^{-7}$} & \multicolumn{1}{c|}{1,85$\times10^{-6}$} & \multicolumn{1}{c|}{1,34$\times10^{-5}$} \\ \hline
\multicolumn{9}{c}{}                                                                                                                                                                                                                                                                                                                                                                        \\ [-8pt] \hline
\multicolumn{1}{|c|}{$\Pi^iE_i$}    & \multicolumn{1}{c|}{36\%}                & \multicolumn{1}{c|}{36\%}                & \multicolumn{1}{c|}{38\%}                & \multicolumn{1}{c|}{41\%}                & \multicolumn{1}{c|}{35\%}                & \multicolumn{1}{c|}{37\%}                & \multicolumn{1}{c|}{38\%}                & \multicolumn{1}{c|}{42\%}                \\ \hline
\multicolumn{1}{|c|}{$N$}           & \multicolumn{1}{c|}{0.87}             & \multicolumn{1}{c|}{1.1}             & \multicolumn{1}{c|}{2.6}             & \multicolumn{1}{c|}{18}             & \multicolumn{1}{c|}{0.58}             & \multicolumn{1}{c|}{0.84}             & \multicolumn{1}{c|}{2.1}             & \multicolumn{1}{c|}{17}             \\ \hline
\multicolumn{9}{c}{}  \end{tabular} 

\begin{tabular}{ccccccccc}                                                                                                                                                                                                                                                                                                                                                                 \cline{2-9} 
\multicolumn{1}{c|}{}               & \multicolumn{4}{c|}{$b = 1.00$}                                                                                                                             & \multicolumn{4}{c|}{$b = 1.50$}                                                                                                                             \\ \hline
\multicolumn{1}{|c|}{$a^2 - b$}     & \multicolumn{1}{c|}{$0.00 \text{ \footnotesize (SM)}$}              & \multicolumn{1}{c|}{$0.05$}              & \multicolumn{1}{c|}{$0.20$}              & \multicolumn{1}{c|}{$0.80$}              & \multicolumn{1}{c|}{$0.00$}              & \multicolumn{1}{c|}{$0.05$}              & \multicolumn{1}{c|}{$0.20$}              & \multicolumn{1}{c|}{$0.80$}              \\ \hline
\multicolumn{1}{|c|}{$\sigma [pb]$} & \multicolumn{1}{c|}{9,47$\times10^{-7}$} & \multicolumn{1}{c|}{1,36$\times10^{-6}$} & \multicolumn{1}{c|}{3,34$\times10^{-6}$} & \multicolumn{1}{c|}{2,24$\times10^{-5}$} & \multicolumn{1}{c|}{2,64$\times10^{-6}$} & \multicolumn{1}{c|}{3,33$\times10^{-6}$} & \multicolumn{1}{c|}{6,07$\times10^{-6}$} & \multicolumn{1}{c|}{2,89$\times10^{-5}$} \\ \hline
\multicolumn{9}{c}{}                                                                                                                                                                                                                                                                                                                                                                        \\ [-8pt] \hline
\multicolumn{1}{|c|}{$\Pi^iE_i$}    & \multicolumn{1}{c|}{35\%}                & \multicolumn{1}{c|}{35\%}                & \multicolumn{1}{c|}{39\%}                & \multicolumn{1}{c|}{40\%}                & \multicolumn{1}{c|}{36\%}                & \multicolumn{1}{c|}{36\%}                & \multicolumn{1}{c|}{36\%}                & \multicolumn{1}{c|}{40\%}                \\ \hline
\multicolumn{1}{|c|}{$N$}           & \multicolumn{1}{c|}{1.0}             & \multicolumn{1}{c|}{1.4}             & \multicolumn{1}{c|}{3.9}             & \multicolumn{1}{c|}{27}             & \multicolumn{1}{c|}{2.8}             & \multicolumn{1}{c|}{3.6}             & \multicolumn{1}{c|}{6.6}             & \multicolumn{1}{c|}{34}             \\ \hline
\end{tabular}

\caption{
HEFT predictions for the final state  $b\overline{b}\gamma\gamma jj'$  for various selections of the HEFT parameters $a$,  $b$ and the combination $a^2-b$.  We include  the integrated parton-level cross-section,  $\sigma(b\overline{b}\gamma\gamma jj')$,  the combined selection efficiencies value $\Pi^iE_i$ and the expected number of events $N$ at HL-LHC.} \label{table:04}
\end{table}

Table \ref{table:04} gathers the combined selection efficiency values and the predicted integrated number of events for each of the analyzed BSM signals. Four extra BSM signals obtained from HEFT $a = 1$ have been included for reasons of completeness, which feature high positive or negative values of the parameter combination $\kappa_V^2 - \kappa_{2V}$. 

Quantitatively,  the studied BSM signals show combined selection efficiencies values in the range  $35\% < \Pi^iE_i < 44\%$,  therefore better than in the SM case.  Higher values are obtained from HEFT signals with a higher absolute value for the $\kappa^2_V -\kappa_{2V}$ combination. We conclude that BSM signals corresponding to HEFT models with a greater $\kappa^2_V -\kappa_{2V}$ combination are predicted to experience a smaller selection efficiency loss compared to the SM framework.  This is in agreement  with previous analyses within HEFT for the $HHjj$ production at LHC,  which concluded that the contribution of the VBF topologies to the total cross-section of the process is increased for HEFT signals with greater $\kappa^2_V -\kappa_{2V}$ values \cite{ARGANDA2019114687}. 

Using a projected integrated luminosity of 3000 fb$^{-1}$ as one of the benchmarks for the HL-LHC operation \cite{CMS-PAS-FTR-21-004}, the signal with the smallest cross-section, $\sigma^{a=0.90, a^2 - b = 0.00} =  1.92\times10^{-7}$ pb, translates into a prediction of $N^{a=0.90, a^2 - b = 0.00} \approx 0.58$ events. The signal corresponding to the SM, $\sigma^{a=1.00, a^2 - b = 0.00} =  3.41\times10^{-7}$ pb, translates into a prediction of $N^{a=1.00, a^2 - b = 0.00} \approx 1.0$ events. The signal with the greatest cross-section, $\sigma^{a=1.00, a^2 - b = 1.50} =  1.88\times10^{-5}$ pb, translates into a prediction of $N^{a=1.00, a^2 - b = 1.50} \approx 56$ events. BSM signals calculated from HEFT models using either $\kappa^2_V - \kappa_{2V} = 0.00 \text{ or } 0.05$ do not predict, for small deviations from $(\kappa_{2V}, \kappa_V) = (1,1)$, a noticeable production of $HH$ pair during the HL-LHC operation. We conclude that BSM signals which predict the observation of the $HHjj$ production at the future HL-LHC operation are associated with the set of HEFT models with $\kappa_V$ and $\kappa_{2V}$ values far from the $\kappa_V^2 - \kappa_{2V} = 0.00, 0.05$ contour lines within the ($\kappa_{2V}$, $\kappa_{V}$) parameter space, where the SM scenario is present. Taking into account the tight constraints on the $\kappa_V$ value, the upper section of Table \ref{table:04} provides the most relevant HEFT predictions among the selected BSM points for using this particular channel to constrain further the value of $\kappa_{2V}$, assuming  $\kappa_V \approx 1.00$. 

\begin{table}[!t] \footnotesize \centering{

\begin{tabular}{ccccc}
\cline{2-5}
\multicolumn{1}{c|}{}               & \multicolumn{4}{c|}{$a = 1.00$}                                                                                                                                           \\ \hline
\multicolumn{1}{|c|}{$b$}           & \multicolumn{1}{c|}{$3.00$}              & \multicolumn{1}{c|}{$2.00$}              & \multicolumn{1}{c|}{$0.00$}              & \multicolumn{1}{c|}{$-1.00$}             \\ \hline
\multicolumn{1}{|c|}{$a^2 - b$}     & \multicolumn{1}{c|}{$-2.00$}             & \multicolumn{1}{c|}{$-1.00$}             & \multicolumn{1}{c|}{$1.00$}              & \multicolumn{1}{c|}{$2.00$}              \\ \hline
\multicolumn{1}{|c|}{$\sigma [pb]$} & \multicolumn{1}{c|}{5,61$\times10^{-5}$} & \multicolumn{1}{c|}{1,25$\times10^{-5}$} & \multicolumn{1}{c|}{2,17$\times10^{-5}$} & \multicolumn{1}{c|}{7,43$\times10^{-5}$} \\ \hline
\multicolumn{5}{c}{}                                                                                                                                                                                            
\\ [-8pt] \hline
\multicolumn{1}{|c|}{$\Pi^iE_i$}    & \multicolumn{1}{c|}{44\%}                & \multicolumn{1}{c|}{44\%}                & \multicolumn{1}{c|}{41\%}                & \multicolumn{1}{c|}{43\%}                \\ \hline
\multicolumn{1}{|c|}{$N$}           & \multicolumn{1}{c|}{74}             & \multicolumn{1}{c|}{16}             & \multicolumn{1}{c|}{27}             & \multicolumn{1}{c|}{95}             \\ \hline
\end{tabular}

} \caption{HEFT predictions for the final state  $b\overline{b}\gamma\gamma jj'$  as in table \ref{table:04} but now considering other coefficients leading to larger values for the relevant combination parameter $a^2-b$ ranging from -2 to 2. } \label{table:07}
\end{table}

Finally,  Table \ref{table:07} also shows the HEFT predictions for the final state  $b\overline{b}\gamma\gamma jj'$  as in table \ref{table:04},  but now considering other coefficients $a$ and $b$ leading to larger values for the relevant combination parameter $a^2-b$ ranging from -2 to 2.   We clearly see that all the predictions for the total cross-section,   the total efficiency and the total number of events grow with the value of $a^2-b$,  as expected.  The cross-section grows up to near $10^{-4}$ pb.  The total efficiency grows up to $44 \%$.  The number of predicted $b\overline{b}\gamma\gamma jj'$ events at HL-LHC ranges between 10 and 100 for all BSM signals.  As a result, we expect the future HL-LHC operation to be capable of greatly reducing the experimental constraints on the combination $\kappa_V^2-\kappa_{2V}$.

\begin{figure}[!t] 
\centerline{ \includegraphics[width=0.99\textwidth]{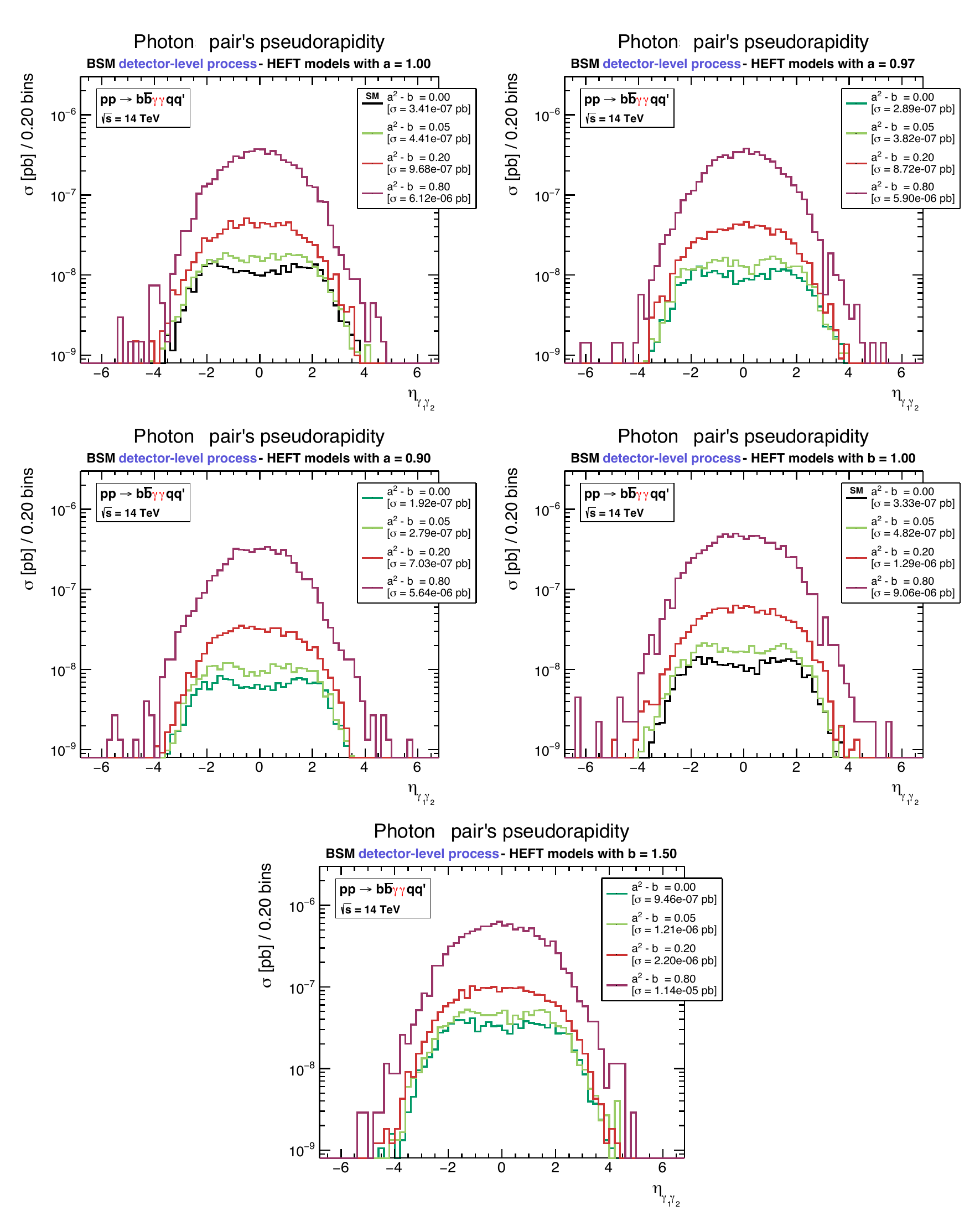}}
\caption{HEFT detector-level predictions for the differential cross-section of the $pp \to HHjj' \to b\overline{b}\gamma\gamma j j'$ process with respect to the photon pair pseudo-rapidity, $d\sigma(pp\to b\overline{b}\gamma\gamma j j') / d\eta_{\gamma\gamma}$, for different values of $a$ and $b$ parameters. Each line's color corresponds to a given value of the combination $(a^2-b)$:  
$(a^2-b) = 0.00$ [\textcolor{ForestGreen}{dark green}], $(a^2-b) = 0.05$ [\textcolor{SpringGreen}{light green}], $(a^2-b) = 0.20$ [\textcolor{BrickRed}{red}] and $(a^2-b) = 0.80$ [\textcolor{RedViolet}{violet}]. The parameter fixed for each graph, either $a=\kappa_V$ or $b=\kappa_{2V}$, is indicated in the title of each plot. Specifically, from top to bottom, left to right, the choices are  $a$ = 1.00,  $a$ = 0.97,  $a$ = 0.9,  $b$ = 1.00,  $b$ = 1.50.  Distributions in black correspond to SM ($a=b=1$) predictions. } \label{fig:06_21} \end{figure}

\begin{figure}[!t] 
\centerline{ \includegraphics[width=0.99\textwidth]{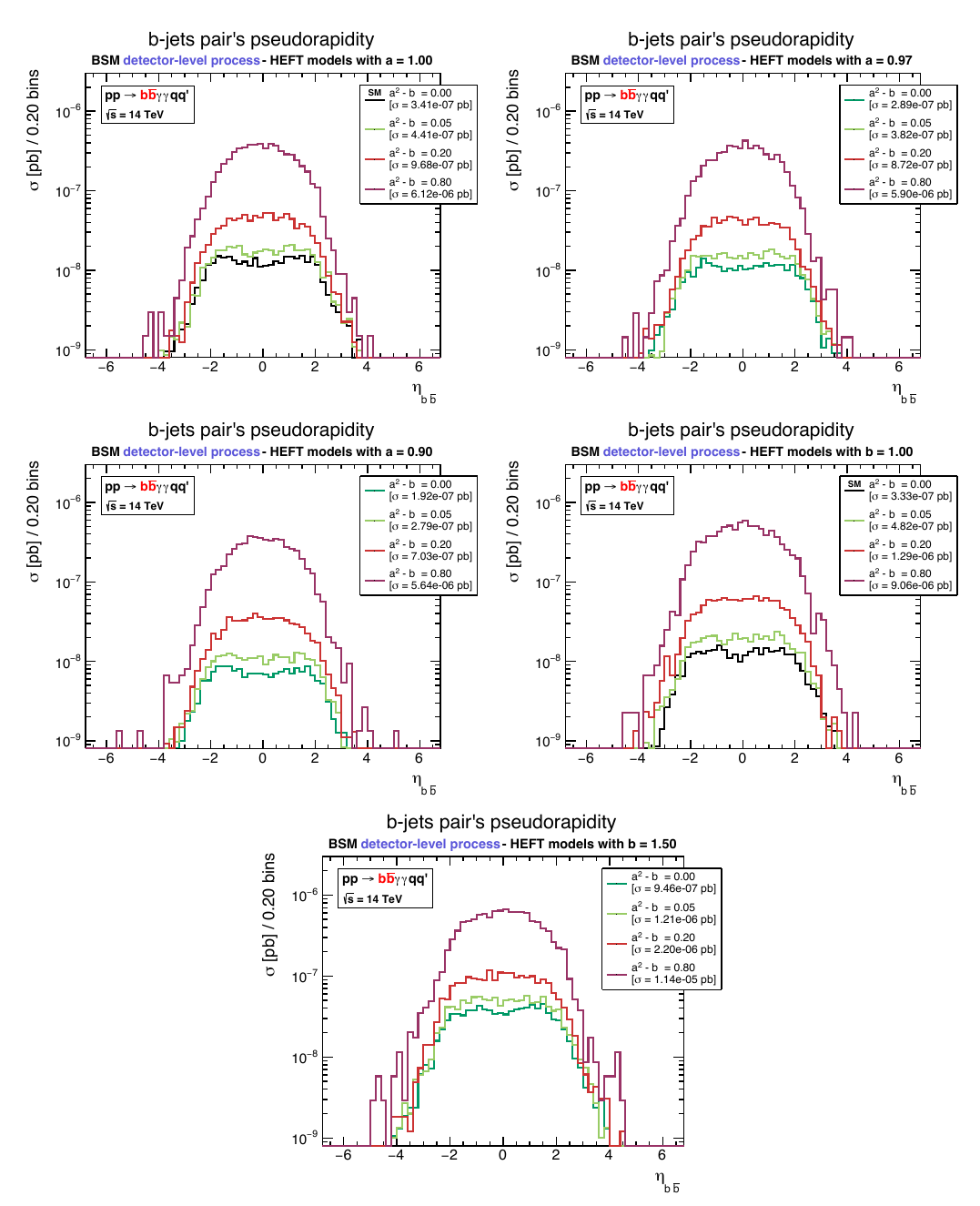}}
\caption{HEFT detector-level predictions for the differential cross-section of the $pp \to HHjj' \to b\overline{b}\gamma\gamma j j'$ process with respect to the $b$-jet pair pseudo-rapidity, $d\sigma(pp\to b\overline{b}\gamma\gamma j j') / d\eta_{b\overline{b}}$,  for different values of $a$ and $b$ parameters. Each line's color corresponds to a given value of the combination $(a^2-b)$:  
$(a^2-b) = 0.00$ [\textcolor{ForestGreen}{dark green}], $(a^2-b) = 0.05$ [\textcolor{SpringGreen}{light green}], $(a^2-b) = 0.20$ [\textcolor{BrickRed}{red}] and $(a^2-b) = 0.80$ [\textcolor{RedViolet}{violet}]. The parameter fixed for each graph, either $a=\kappa_V$ or $b=\kappa_{2V}$, is indicated in the title of each plot. Specifically, from top to bottom, left to right,  the choices are  $a$ = 1.00,  $a$ = 0.97,  $a$ = 0.9,  $b$ = 1.00,  $b$ = 1.50.  Distributions in black correspond to SM ($a=b=1$) predictions.} \label{fig:06_22} \end{figure}

\begin{figure}[!t] 
\centerline{ \includegraphics[width=0.99\textwidth]{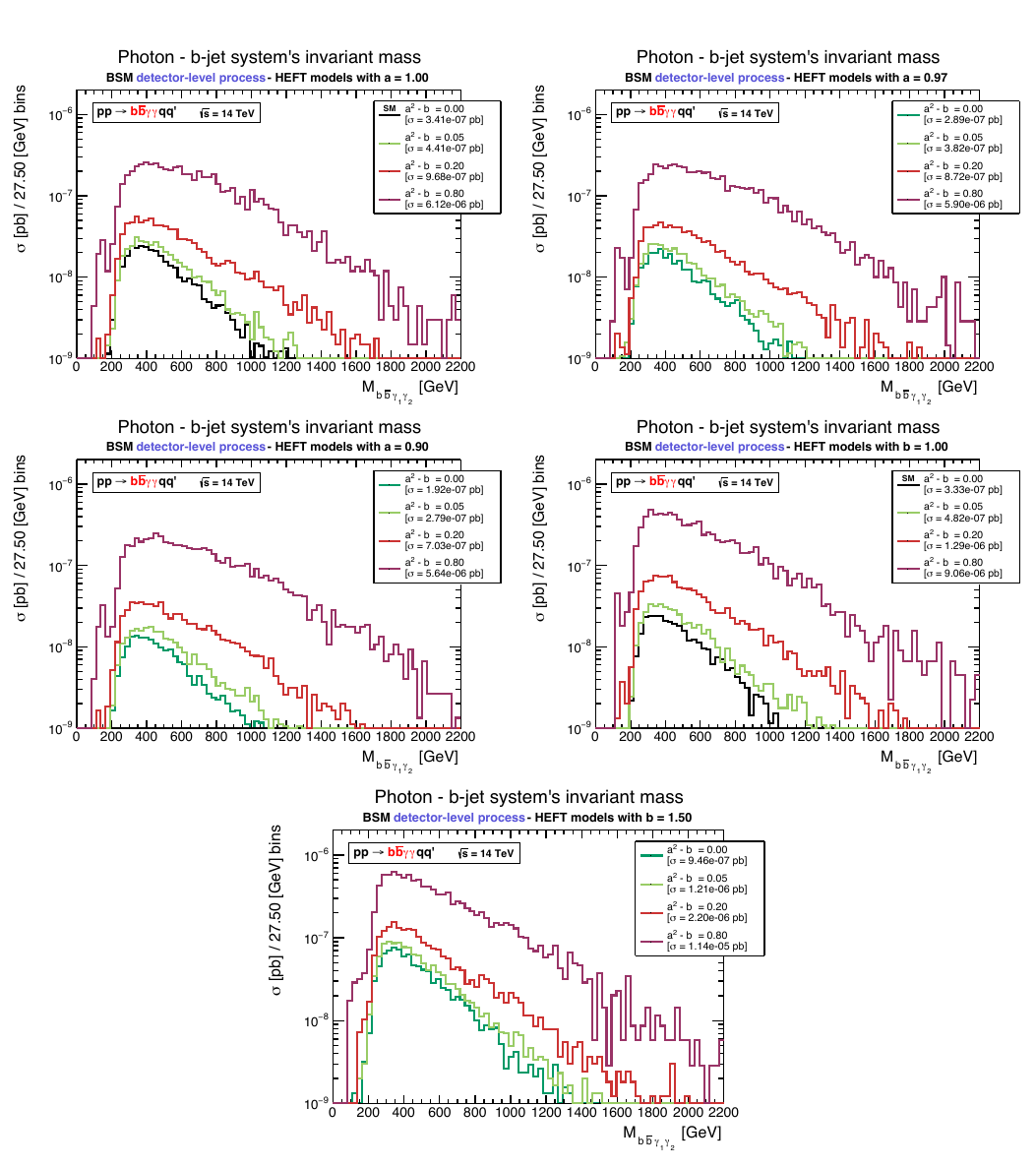}}
\caption{HEFT detector-level predictions for the differential cross-section of the $pp \to HHjj' \to b\overline{b}\gamma\gamma j j'$ process with respect to the invariant mass of the combined $b\overline{b}\gamma\gamma$ system,  $d\sigma(pp\to b\overline{b}\gamma\gamma j j') / dM_{b\overline{b}\gamma\gamma}$, for different values of $a$ and $b$ parameters. Each line's color corresponds to a given value of the combination $(a^2-b)$:  
$(a^2-b) = 0.00$ [\textcolor{ForestGreen}{dark green}], $(a^2-b) = 0.05$ [\textcolor{SpringGreen}{light green}], $(a^2-b) = 0.20$ [\textcolor{BrickRed}{red}] and $(a^2-b) = 0.80$ [\textcolor{RedViolet}{violet}]. The parameter fixed for each graph, either $a=\kappa_V$ or $b=\kappa_{2V}$, is indicated in the title of each plot. Specifically, from top to bottom, left to right,  the choices are  $a$ = 1.00,  $a$ = 0.97,  $a$ = 0.9,  $b$ = 1.00,  $b$ = 1.50.  Distributions in black correspond to SM ($a=b=1$) predictions.} \label{fig:06_31} \end{figure}

In the last part of this subsection,  we analyze the effects of showering,  hadronization,  and detector level on the distributions for some selected  BSM cases  within the HEFT.  This is shown in Figs.\ref{fig:06_21},  \ref{fig:06_22},  and  \ref{fig:06_31}.
Figure \ref{fig:06_21} shows the differential cross-section distributions with respect to the pseudo-rapidity of the detector-level photon pair, for all the selected BSM points. Figure \ref{fig:06_22} shows the differential cross-section distributions with respect to the pseudo-rapidity of the detector-level $b$-jets. Qualitatively, all graphs maintain the same distribution shapes for the pseudo-rapidity of the pair products from the  Higgs boson decay as those found previously from the parton-level predictions with respect to the Higgs boson pseudo-rapidity.  All plots show the central peak at vanishing pseudo-rapidity of the pair, both at $\eta_{\gamma \gamma} =0$ and $\eta_{b \bar b}=0$.  Signals calculated using HEFT models with an identical value for the combination $\kappa_V^2 - \kappa_{2V}$ have nearly  identical pseudo-rapidity distribution shapes, not varying so much with the separate values of $\kappa_{V}$ and $\kappa_{2V}$.  Signals calculated from HEFT models with an increased value of $\kappa_V^2 - \kappa_{2V}$  predict a higher central 
peak,  which is obviously correlated with the high transversality of the final Higgs bosons produced as already indicated.  
Figure \ref{fig:06_31} shows the differential cross-section distributions with respect to $M_{b \bar b \gamma \gamma}$,  i.e. ,  the invariant mass of the photons and $b$-jets combined system at detector-level.   From these plots we see that HEFT  distributions with higher values of $\kappa_V^2- \kappa_{2V}$ have higher rates.  Also the tails at large invariant mass,  $M_{b \bar b \gamma \gamma}$,  are higher and longer as the combination parameter $\kappa_V^2- \kappa_{2V}$ increases.  This shape at detector level  is equivalent to the $M_{HH}$ distribution shape that is predicted at the parton level,   as expected.  Therefore,  in general,  the enhancement of rates at large invariant mass $M_{\gamma \gamma b \bar b}$ can also be used as good discriminants among BSM and SM,  and when compared them to the background rates could provide an increasing sensitivity to the combination parameter $\kappa_V^2- \kappa_{2V}$. 

\clearpage

\section{Sensitivity to $(\kappa_V^2-\kappa_{2V})$ in  $b\overline{b}\gamma\gamma jj'$ events from $HHjj$ production at HL-LHC} 
\label{section:5}
In this section we finally evaluate the expected sensitivity to the combination parameter of our interest $\kappa_V^2-\kappa_{2V}$ in  $b\overline{b}\gamma\gamma jj'$ events from $HHjj$ production at HL-LHC.   For all the predictions of event rates in this section we assume the canonical parameters for HL-LHC,  i.e. ,    energy of 14 TeV and integrated luminosity of 3000 fb$^{-1}$.  
For the evaluation of the mentioned sensitivity,   we first compare the signal rates from the BSM events given by the HEFT for several choices of $a^2-b$ and the background rates from the most relevant SM processes producing the same final state  $b\overline{b}\gamma\gamma jj'$.  This first evaluation is done at the parton-level and the results are shown in table \ref{table:BKGs}.  The three selected background processes to be compared with the BSM signal with $a^2-b \neq 0$ are:  1) SM-EW $\gamma\gamma b\bar{b}jj$ via $HHjj$  (i.e.  for $a=b=1$, tree level) 2) SM-QCD-EW $\gamma\gamma b\bar{b}jj$ non-resonant (tree level,  $\mathcal{O}(\alpha_s^2\alpha^1\hbar^0)$),  3)  SM-QCD-EW $\gamma\gamma b\bar{b}jj$ via $HH$ from gluon-gluon fusion 
(one-loop,  $\mathcal{O}(\alpha_s^2\alpha^3\hbar^1)$),  and 4) SM-EW $\gamma\gamma b\bar{b}jj$ via $ZHjj$ (tree level).  These rates are all generated with MG5 and the specific cuts for each process are also included.  In the background case 3) we have performed the calculation using the large $m_t$ limit that provides a 
rough  estimate for the most important contributions coming from the top loops.  We have used an effective approach to describe the $ggHH$ effective vertex in this large $m_t$ limit,  which is $\mathcal{O}(\hbar^1)$,   following  Refs. \cite{Dawson:1998py,  Dolan:2015zja}.  According to Ref.  \cite{Dolan:2015zja} this rough estimate
provides  larger $HHjj$  rates compared to the exact computation by a factor of up to 2-3 mainly in large invariant mass $M_{HH}$ region,  although other distributions are not so much affected.   Using MG5 with this $ggHH$ effective interaction we have then generated to order $\mathcal{O}(\alpha_s^2\alpha^3\hbar^1)$  the final $b\overline{b}\gamma\gamma jj'$ events.   Our simple estimate of this background being larger than the expected full one is therefore a conservative estimate.  The most important conclusion from table \ref{table:BKGs}   is that the most relevant background among the four selected ones is by far the second one,  i.e.  SM-QCD-EW $\gamma\gamma b\bar{b}jj$ non-resonant,  which  overpasses the others by several orders of magnitude.  The starting parton-level cross section for this process is $8.1 \times 10^{-4} {\rm pb}$ to be compared with our BSM signal from the HEFT models which is bellow $10^{-4} {\rm pb}$.  In the following we will take just this dominant background and we will explore for specific and more optimal strategies to enhance the sensitivity to our BSM signal with $a^2-b \neq 0$.    From now on we will perform our estimates of both the signal and the background at the detector level,  i.e. ,  all the forthcoming predictions are generated  with MG5 and passed through PYTHIA and DELPHES. 


\vspace{.5em} \begin{table}[!h] \footnotesize \centerline{
\begin{tabular}{|c|c|c|c|c|c|}

\hline

\multicolumn{6}{|c|}{$\sigma(\gamma\gamma b \bar{b} j j)(\text{pb})$: BSM signal versus SM background rates (parton level)}\\ 

\hline

\multicolumn{6}{c}{} \\ [-10pt]

\hline

\multicolumn{6}{|c|}{Cuts applied: $p_{T_j}\geq20\text{ GeV},\quad2<|\eta_j|<5,\quad\eta_{j_1}\cdot\eta_{j_2}<0,\quad\Delta R_{jj}>0.4,\quad M_{jj}>500\text{ GeV}$}\\

\multicolumn{6}{|c|}{Additional cuts in SM-QCD-EW (non resonant) background: $M_{b\bar{b}}>100\text{ GeV},\quad M_{\gamma\gamma}>100\text{ GeV} $}\\ 

\hline

\multicolumn{6}{c}{}\\ [-10pt] 

\hline

\multicolumn{2}{|c|}{BSM Signal ($a^2-b\neq 0$)} & \multicolumn{1}{c|}{SM-EW ($\mathcal{O}(\alpha^3\hbar^0)$)}& \multicolumn{1}{c|}{SM-QCD-EW ($\mathcal{O}(\alpha_s^2\alpha^1\hbar^0)$)} & \multicolumn{1}{c|}{SM-QCD-EW ($\mathcal{O}(\alpha_s^2\alpha^3\hbar^1)$)} & \multicolumn{1}{c|}{SM-EW ($\mathcal{O}(\alpha^3\hbar^0)$)}\\ 

\multicolumn{2}{|c|}{$\gamma\gamma b\bar{b}jj$ via $HHjj$} & \multicolumn{1}{c|}{$\gamma\gamma b\bar{b}jj$ via $HHjj$} & \multicolumn{1}{c|}{$\gamma\gamma b\bar{b}jj$ (non-resonant)} & \multicolumn{1}{c|}{$\gamma\gamma b\bar{b}jj$ via $HH$ from $gg$F} & \multicolumn{1}{c|}{$\gamma\gamma b\bar{b}jj$ via $ZHjj$}\\

\hline

{$a^2-b$ $(a=1)$} &  & {\multirow{5}{*}{$9.8\times10^{-7}$}} & {\multirow{5}{*}{$8.1\times10^{-4}$}} & {\multirow{5}{*}{$4.7\times10^{-7}$}} & {\multirow{5}{*}{$3.6\times10^{-8}$}}\\ 

\cline{1-2} 

{0.2} & {$2.2\times10^{-6}$} &&&&\\
{0.8} & {$1.2\times10^{-5}$} &&&&\\
{1} & {$2.2\times10^{-5}$} &&&&\\
{-1} & {$7.4\times10^{-5}$} &&&&\\

\hline

\end{tabular}} \caption{Cross section, $\sigma(\gamma\gamma b \bar{b} j j)(\text{pb})$,  at parton level: BSM rates within  HEFT versus SM backgrounds rates.}
\label{table:BKGs}  
\end{table}

\newpage

In order to find the proper selection of cuts that optimizes the signal versus background rates we have designed the following strategy.   First we select the events with VBF topology and require also some basic cuts for the $HH$ topology.  This is called in short Selection A 
(sometimes also called simply VBF).  Then with this selected events we study the most relevant  differential cross section distributions and compare the HEFT predictions  for various values of the combination parameter  $(\kappa_V^2-\kappa_{2V})$ with the dominant SM-QCD-EW (non-resonant) background.  This will lead us to the proper variables to discriminate among signal and background.  With these optimal variables we will then design the final strategy using Selection A and also additional Selections B,   C and C',  which will be defined later,  to enhance the sensitivity to $(\kappa_V^2-\kappa_{2V})$ which is our main objective in this work.

In table \ref{table:SelectionA} we summarize the cuts applied to the generated $b\overline{b}\gamma\gamma jj'$ events that define 
Selection A.  The corresponding efficiencies for the cases SM-EW and SM-QCD-EW (non-resonant) are also included. The case of HEFT will be discussed later,  after knowing the distributions that are presented next. 

\begin{figure}[!h]
\vspace{-1cm}
\centerline{\includegraphics[width=1.0\textwidth]{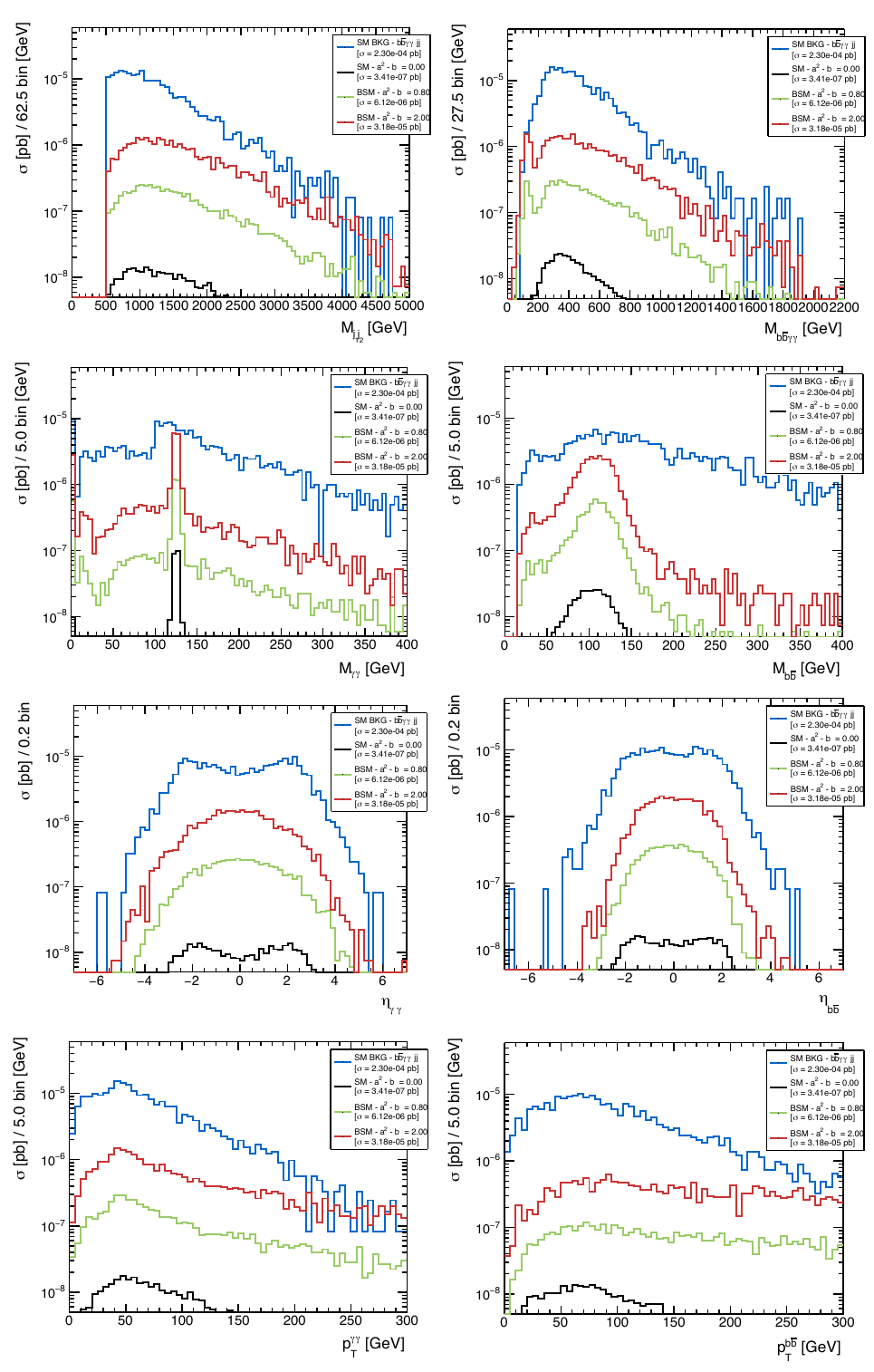}}
\caption{$d\sigma$ distributions for HEFT (red,  green),  SM (black) and main BKG (blue).  Details in the text.}
\label{final-distributions}
\end{figure}

\begin{table}[!h] \centerline{ \renewcommand{\arraystretch}{0.8} \fontsize{10}{16} \selectfont
\begin{tabular}{clccc|c}
\cline{5-6} 
\multicolumn{4}{c|}{}                                                                                                                                                                                                                                     & \multicolumn{1}{c|}{SM $E_i$}  & \multicolumn{1}{c|}{$b\bar{b}\gamma\gamma jj$  BKG $E_i$}          \\  \cline{5-6} 
\multicolumn{6}{c}{}                                                                                                                                                                                                                                                                                     \\ [-10pt] \cline{1-2} \cline{4-6} 
\multicolumn{2}{|c|}{\multirow{4}{*}{\makecell{\\\\VBF topology}}}    & \multicolumn{1}{c|}{\multirow{4}{*}{}} & \multicolumn{1}{c|}{\makecell{At least two jets with: $P_{T, j} >   20$ GeV, $2 < \left| \eta_j \right| < 5$}}
& \multicolumn{1}{c|}{\multirow{2}{*}{\makecell{\\$66\%$}}} & \multicolumn{1}{c|}{\multirow{2}{*}{\makecell{\\$62\%$}}} \\ \cline{4-4}
\multicolumn{2}{|c|}{}                                   & \multicolumn{1}{c|}{}                  & \multicolumn{1}{c|}{\makecell{Selected VBF jet pair $jj'$ maximizes\\the $\Delta R$ between light jets}}                                                                       & \multicolumn{1}{c|}{}      & \multicolumn{1}{c|}{}                  \\ \cline{4-6} 
\multicolumn{2}{|c|}{}                                   & \multicolumn{1}{c|}{}                  & \multicolumn{1}{c|}{\makecell{VBF jets with: $\eta_j\times\eta_{j'} < 0$}}                                                                                              & \multicolumn{1}{c|}{$91\%$} & \multicolumn{1}{c|}{$81\%$}                  \\ \cline{4-6} 
\multicolumn{2}{|c|}{}                                   & \multicolumn{1}{c|}{}                  & \multicolumn{1}{c|}{\makecell{VBF jet pair with: $m_{jj'} > 500$ GeV}}                                                                                           & \multicolumn{1}{c|}{$98\%$}   & \multicolumn{1}{c|}{$93\%$}               \\ \cline{1-2} \cline{4-6} 
\multicolumn{5}{c}{}                                                                                                                                                                                                                                                                                     \\ [-10pt] \cline{1-2} \cline{4-6} 
\multicolumn{2}{|c|}{\multirow{4}{*}{\makecell{\\\\\\$HH$ topology}}} & \multicolumn{1}{c|}{}                  & \multicolumn{1}{c|}{\makecell{At least two loosely identified photons\\ with $P_{T, \gamma} > 18$ GeV, $\left| \eta_\gamma \right|   < 2.5$}}                                            & \multicolumn{1}{c|}{\multirow{2}{*}{\makecell{\\$79\%$}}} & \multicolumn{1}{c|}{\multirow{2}{*}{\makecell{\\$77\%$}}} \\ \cline{4-4}
\multicolumn{2}{|c|}{}                                   & \multicolumn{1}{c|}{\multirow{3}{*}{}} & \multicolumn{1}{c|}{{Take the two photons $\gamma\gamma$ with the greatest $P_T$}}                                               & \multicolumn{1}{c|}{}    & \multicolumn{1}{c|}{}                     \\ \cline{4-6} 
\multicolumn{2}{|c|}{}                                   & \multicolumn{1}{c|}{}                  & \multicolumn{1}{c|}{\makecell{At least two loosely b-tagged jets\\with $P_{T, j_b} > 25$ GeV, $\left| \eta_{j_b} \right| < 2.5$}}                                     & \multicolumn{1}{c|}{\multirow{2}{*}{\makecell{\\$76\%$}}} & \multicolumn{1}{c|}{\multirow{2}{*}{\makecell{\\$79\%$}}} \\ \cline{4-4}
\multicolumn{2}{|c|}{}                                   & \multicolumn{1}{c|}{}                  & \multicolumn{1}{c|}{{Take the two b-jets $b\bar{b}$ with the greatest $P_T$}}                                                      & \multicolumn{1}{c|}{}      &  \multicolumn{1}{c|}{}                 \\ \cline{1-2} \cline{4-6} 
\multicolumn{5}{l}{}                                                                          \\ [-8pt] \hline \multicolumn{4}{|c|}{Total selection efficiency $(\Pi^iE_i)$}                                                                                                                                                                                                          & \multicolumn{1}{c|}{$36\%$} & \multicolumn{1}{c|}{$28\%$}                  \\ \hline
\end{tabular}}   
 \caption{Cuts defining Selection A.  In the last two columns the corresponding efficiencies for the cases SM-EW and SM-QCD-EW (non-resonant) are shown.} 
\label{table:SelectionA}
\end{table} 
In figure \ref{final-distributions} we include the HEFT predictions  for the selected differential cross sections,   for 
$(a^2-b)=2$ (red lines) (obtained with $(a,b) =(1,-1)$),   for $a^2-b=0.8$ (green lines) (obtained with $(a,b)=(1, 0.2)$),  and for $a^2-b=0$ (black lines) (obtained in this case with $(a,b)=(1,1)$).  Notice, that this later prediction is indeed the SM-EW prediction.  The corresponding predictions for the dominant SM-QCD-EW (non-resonant) background (BKG) (blue lines) are also included for comparison.  All predictions are done after applying Selection A,  as defined in table \ref{table:SelectionA}.  The selected variable in each panel corresponds to a particular variable of the $b\overline{b}\gamma\gamma j_1j_2$ final state.  The variables for the plots ordered from left to right and from upper to lower plots  are,  respectively,   $M_{j_1j_2}$,  $M_{b\overline{b}\gamma\gamma}$,  $M_{\gamma \gamma}$,  $M_{b\overline{b}}$,  $\eta_{\gamma \gamma}$,  $\eta_{b\overline{b}}$,   $P_T^{\gamma \gamma}$,  and $P_T^{b\overline{b}}$.  Here,  we follow a similar notation/definition for these variables as stated in previous sections.

\clearpage


Some comments on these distributions of Fig. \ref{final-distributions}  are in order.  First,  we see that in all the panels the predictions  for the dominant SM-QCD-EW (non-resonant) background (blue lines) are above the HEFT signals (red,  green and black lines),  as already manifested for the total parton-level cross sections in table \ref{table:BKGs}.   Second,  the plots showing the largest differences in the shapes between signal and background are those for the variables of the pairs 
$\gamma \gamma$ and $b\overline{b}$.  Specifically,  these are the transverse momentum,  the pseudo-rapidity and the invariant mass, i.e.,   $P_T^{b\overline{b}}$,  $P_T^{\gamma \gamma}$, $\eta_{\gamma \gamma}$,  $\eta_{b\overline{b}}$,   $M_{\gamma \gamma}$,  $M_{b\overline{b}}$.  Also the total invariant mass $M_{b\overline{b}\gamma\gamma}$ shows some differences in the large invariant mass region.   The HEFT predictions for the distributions in $P_T^{b\overline{b}}$,  $P_T^{\gamma \gamma}$, $\eta_{\gamma \gamma}$ and  $\eta_{b\overline{b}}$ confirm the high transversality of these pairs, as expected from our previous comments on the high transversality of the two Higgs bosons produced.  In the $\eta$ distributions we see that central peaks at $\eta=0$ in the red and green lines as opposed to the blue and black lines.  In the case of the $P_T$ distributions we see the long tails with small slopes  at large transverse momentum.  Finally,  The HEFT predictions for the distributions in $M_{\gamma \gamma}$ and $M_{b\overline{b}}$ show clearly the peaks around $m_H$,  as expected,  in contrast to the broad distribution for the background (non-resonant) prediction.  Any of these kinematic variables are therefore good candidates as signal/background discriminants.  
Accordingly,  in the following we design new selections to improve  the $HH$ final state isolation that are based on the above behaviour found.  
First,  we define Selections B,  C and C' in tables \ref{table:SelectionB}, \ref{table:SelectionC}, \ref{table:SelectionC'} and provide  the corresponding  efficiencies for the cases SM-EW and SM-QCD-EW (non-resonant).  After this we provide the predictions for the HEFT in tables \ref{strategy1} and \ref{strategy2} which summarize the signal event rates at HL-LHC within HEFT after applying those Selections A,B,C and C',  in comparison with the predicted main background rates (BKG) which are also included in these tables. 
The cuts defining  Selection B,  based on $M_{\gamma \gamma}$ and $M_{b\overline{b}}$,  are given  in table \ref{table:SelectionB}.
The cuts defining  Selection C,  based on  $\eta_{\gamma \gamma}$, are given  in table \ref{table:SelectionC}.  
The cuts defining  Selection C',  based on  P$^T_{\gamma \gamma}$, are given in table \ref{table:SelectionC'}.

\vspace{-0.4cm}
\begin{table}[!h] \centerline{ \renewcommand{\arraystretch}{0.8} \fontsize{10}{16} \selectfont
\begin{tabular}{clccc|c}
\cline{5-6} 
\multicolumn{4}{c|}{}                                                                                                                                                                                                                                     & \multicolumn{1}{c|}{SM $E_i$}  & \multicolumn{1}{c|}{$b\bar{b}\gamma\gamma jj$  BKG $E_i$}          \\  \cline{5-6} 
\multicolumn{6}{c}{}                                                                                                                                                                                                                                                                                     \\ [-10pt] \cline{1-2} \cline{4-6} 
\multicolumn{2}{|c|}{\multirow{2}{*}{\makecell{$HH$ final state isolation}}}    & \multicolumn{1}{c|}{\multirow{4}{*}{}} & \multicolumn{1}{c|}{$120 < M_{\gamma\gamma} < 130$ GeV}& \multicolumn{1}{c|}{\multirow{1}{*}{$55\%$}} & \multicolumn{1}{c|}{\multirow{1}{*}{$6\%$}} \\ \cline{4-6} 
\multicolumn{2}{|c|}{}                                   & \multicolumn{1}{c|}{}                  & \multicolumn{1}{c|}{$50 < M_{b\bar{b}} < 150$ GeV}                                                                                              & \multicolumn{1}{c|}{$89\%$} & \multicolumn{1}{c|}{$45\%$}                                 \\ \cline{1-2} \cline{4-6} 
\multicolumn{5}{l}{}                                                                          \\ [-8pt] \hline \multicolumn{4}{|c|}{Total selection efficiency 
(A+B) $(\Pi^iE_i)$}                                                                                                                                                                                                          & \multicolumn{1}{c|}{$18\%$} & \multicolumn{1}{c|}{$1\%$}                  \\ \hline
\end{tabular}} 
\caption{Cuts defining Selection B.  In the last two columns the corresponding efficiencies for the cases SM-EW and SM-QCD-EW (non-resonant) are shown.}
\label{table:SelectionB}
\end{table}
\begin{table}[!h] \centerline{ \renewcommand{\arraystretch}{0.8} \fontsize{10}{16} \selectfont
\begin{tabular}{clccc|c}
\cline{5-6} 
\multicolumn{4}{c|}{}                                                                                                                                                                                                                                     & \multicolumn{1}{c|}{SM $E_i$}  & \multicolumn{1}{c|}{$b\bar{b}\gamma\gamma jj$  BKG $E_i$}          \\  \cline{5-6} 
\multicolumn{6}{c}{}                                                                                                                                                                                                                                                                                     \\ [-10pt] \cline{1-2} \cline{4-6} 
\multicolumn{2}{|c|}{\makecell{$HH$ final state isolation}}    & \multicolumn{1}{c|}{\multirow{4}{*}{}} & \multicolumn{1}{c|}{$-2 < \eta_{\gamma\gamma} < 2$}& \multicolumn{1}{c|}{\multirow{1}{*}{$60\%$}} & \multicolumn{1}{c|}{\multirow{1}{*}{$59\%$}}                                 \\ \cline{1-2} \cline{4-6} 
\multicolumn{5}{l}{}                                                                          \\ [-8pt] \hline \multicolumn{4}{|c|}{Total selection efficiency 
(A+C) $(\Pi^iE_i)$}                                                                                                                                                                                                          & \multicolumn{1}{c|}{$22\%$} & \multicolumn{1}{c|}{$16\%$}                  \\ \hline
\end{tabular}}
\caption{Cuts defining Selection C.  In the last two columns the corresponding efficiencies for the cases SM-EW and SM-QCD-EW (non-resonant) are shown.} 
\label{table:SelectionC}
\end{table} 
\begin{table}[!h] \centerline{ \renewcommand{\arraystretch}{0.8} \fontsize{10}{16} \selectfont
\begin{tabular}{clccc|c}
\cline{5-6} 
\multicolumn{4}{c|}{}                                                                                                                                                                                                                                     & \multicolumn{1}{c|}{SM $E_i$}  & \multicolumn{1}{c|}{$b\bar{b}\gamma\gamma jj$  BKG $E_i$}          \\  \cline{5-6} 
\multicolumn{6}{c}{}                                                                                                                                                                                                                                                                                     \\ [-10pt] \cline{1-2} \cline{4-6} 
\multicolumn{2}{|c|}{\makecell{$HH$ final state isolation}}    & \multicolumn{1}{c|}{\multirow{4}{*}{}} & \multicolumn{1}{c|}{P$^t_{\gamma\gamma} > 200$ GeV}& \multicolumn{1}{c|}{\multirow{1}{*}{$8\%$}} & \multicolumn{1}{c|}{\multirow{1}{*}{$3\%$}}                                 \\ \cline{1-2} \cline{4-6} 
\multicolumn{5}{l}{}                                                                          \\ [-8pt] \hline \multicolumn{4}{|c|}{Total selection efficiency 
(A+C')  $(\Pi^iE_i)$}                                                                                                                                                                                                          & \multicolumn{1}{c|}{$3\%$} & \multicolumn{1}{c|}{$1\%$}                  \\ \hline
\end{tabular}} 
\caption{Cuts defining Selection C'.  In the last two columns the corresponding efficiencies for the cases SM-EW and SM-QCD-EW (non-resonant) are shown.} 
\label{table:SelectionC'}
\end{table}  

Next we propose two strategies,  I and II,  optimized for $\eta_{\gamma \gamma}$ and $P^T_{\gamma\gamma}$ respectively. 

{\bf{STRATEGY I}} (using $\eta_{\gamma \gamma}$) :  Applies sequentially Selection A,  Selection B and Selection C. 

{\bf{STRATEGY II}} (using $P^T_{\gamma\gamma}$) : Applies sequentially  Selection A,  Selection B and Selection C'. 
 
The total number of ($b\bar{b}\gamma\gamma jj$) events expected at  HL-LHC that are predicted for the HEFT models (with $-2 \leq  a^2-b \leq 2$),  the SM $(a=b=1)$  and the dominant background (BKG),  are summarized in the two tables \ref{strategy1} and \ref{strategy2}, 
 using Strategy I and Strategy II,  respectively.

\begin{table}[b!] \fontsize{9}{16} \selectfont \renewcommand{\arraystretch}{0.8}
\begin{tabular}{ccccccccccc}
\hline
\multicolumn{11}{c}{}                                                                                                                                                                                                                                                                                                                                                  
Number of predicted events,  $N(b\bar{b}\gamma\gamma jj)$,  with  \bf{STRATEGY I}
\\ \hline
\hline
\multicolumn{2}{|c|}{\multirow{3}{*}{Selection}}                       & \multicolumn{1}{c|}{\multirow{3}{*}{BKG}}           & \multicolumn{1}{c|}{a =}               & \multicolumn{7}{c|}{1.10}                                                                                                                                                                              \\ \cline{4-11} 
\multicolumn{2}{|c|}{}                                         & \multicolumn{1}{c|}{}                               & \multicolumn{1}{c|}{b =}               & \multicolumn{1}{c|}{3.21}  & \multicolumn{1}{c|}{2.00}  & \multicolumn{1}{c|}{1.20} & \multicolumn{1}{c|}{1.00} & \multicolumn{1}{c|}{0.40}  & \multicolumn{1}{c|}{-0.30} & \multicolumn{1}{c|}{-0.79} \\ \cline{4-11} 
\multicolumn{2}{|c|}{}                                         & \multicolumn{1}{c|}{}                               & \multicolumn{1}{c|}{a$^2$ -b =}           & \multicolumn{1}{c|}{-2.00} & \multicolumn{1}{c|}{-0.80} & \multicolumn{1}{c|}{0.00} & \multicolumn{1}{c|}{0.20} & \multicolumn{1}{c|}{0.80}  & \multicolumn{1}{c|}{1.50}  & \multicolumn{1}{c|}{2.00}  \\ \hline
\multicolumn{11}{c}{}                                                                                                                                                                                                                                                                                                                                                  \\ \hline
\multicolumn{1}{|c|}{1)}   & \multicolumn{1}{c|}{A}           & \multicolumn{1}{c|}{690.70}                         & \multicolumn{1}{c|}{\multirow{4}{*}{}} & \multicolumn{1}{c|}{69.07} & \multicolumn{1}{c|}{9.43}  & \multicolumn{1}{c|}{1.58} & \multicolumn{1}{c|}{3.87} & \multicolumn{1}{c|}{20.29} & \multicolumn{1}{c|}{58.65} & \multicolumn{1}{c|}{97.67} \\ \cline{1-3} \cline{5-11} 
\multicolumn{1}{|c|}{2)}  & \multicolumn{1}{c|}{A+C}     & \multicolumn{1}{c|}{409.63}                         & \multicolumn{1}{c|}{}                  & \multicolumn{1}{c|}{52.84} & \multicolumn{1}{c|}{7.18}  & \multicolumn{1}{c|}{0.99} & \multicolumn{1}{c|}{2.62} & \multicolumn{1}{c|}{14.89} & \multicolumn{1}{c|}{43.40} & \multicolumn{1}{c|}{72.59} \\ \cline{1-3} \cline{5-11} 
\multicolumn{1}{|c|}{3)} & \multicolumn{1}{c|}{A+B}       & \multicolumn{1}{c|}{19.80}                          & \multicolumn{1}{c|}{}                  & \multicolumn{1}{c|}{18.96} & \multicolumn{1}{c|}{2.49}  & \multicolumn{1}{c|}{0.75} & \multicolumn{1}{c|}{1.58} & \multicolumn{1}{c|}{6.71}  & \multicolumn{1}{c|}{18.30} & \multicolumn{1}{c|}{31.63} \\ \cline{1-3} \cline{5-11} 
\multicolumn{1}{|c|}{4)}  & \multicolumn{1}{c|}{A+B+C} & \multicolumn{1}{c|}{13.20}                          & \multicolumn{1}{c|}{}                  & \multicolumn{1}{c|}{17.30} & \multicolumn{1}{c|}{2.28}  & \multicolumn{1}{c|}{0.53} & \multicolumn{1}{c|}{1.25} & \multicolumn{1}{c|}{5.69}  & \multicolumn{1}{c|}{15.97} & \multicolumn{1}{c|}{27.57} \\ \hline
\multicolumn{11}{c}{}                                                                                                                                                                                                                                                                                                                                                  \\ \hline
\multicolumn{2}{|c|}{\multirow{3}{*}{Selection}}                       & \multicolumn{1}{c|}{\multirow{3}{*}{SM (a=b=1)}} & \multicolumn{1}{c|}{a =}               & \multicolumn{7}{c|}{0.95}                                                                                                                                                                              \\ \cline{4-11} 
\multicolumn{2}{|c|}{}                                         & \multicolumn{1}{c|}{}                               & \multicolumn{1}{c|}{b =}               & \multicolumn{1}{c|}{2.90}  & \multicolumn{1}{c|}{1.70}  & \multicolumn{1}{c|}{0.90} & \multicolumn{1}{c|}{0.70} & \multicolumn{1}{c|}{0.10}  & \multicolumn{1}{c|}{-0.60} & \multicolumn{1}{c|}{-1.10} \\ \cline{4-11} 
\multicolumn{2}{|c|}{}                                         & \multicolumn{1}{c|}{}                               & \multicolumn{1}{c|}{a$^2$ -b =}           & \multicolumn{1}{c|}{-2.00} & \multicolumn{1}{c|}{-0.80} & \multicolumn{1}{c|}{0.00} & \multicolumn{1}{c|}{0.20} & \multicolumn{1}{c|}{0.80}  & \multicolumn{1}{c|}{1.50}  & \multicolumn{1}{c|}{2.00}  \\ \hline
\multicolumn{11}{c}{}                                                                                                                                                                                                                                                                                                                                                  \\ \hline
\multicolumn{1}{|c|}{1)}   & \multicolumn{1}{c|}{A}           & \multicolumn{1}{c|}{1.02}                           & \multicolumn{1}{c|}{\multirow{4}{*}{}} & \multicolumn{1}{c|}{76.30} & \multicolumn{1}{c|}{10.56} & \multicolumn{1}{c|}{0.78} & \multicolumn{1}{c|}{2.50} & \multicolumn{1}{c|}{17.55} & \multicolumn{1}{c|}{55.10} & \multicolumn{1}{c|}{92.93} \\ \cline{1-3} \cline{5-11} 
\multicolumn{1}{|c|}{2)}  & \multicolumn{1}{c|}{A+C}     & \multicolumn{1}{c|}{0.62}                           & \multicolumn{1}{c|}{}                  & \multicolumn{1}{c|}{58.12} & \multicolumn{1}{c|}{8.01}  & \multicolumn{1}{c|}{0.48} & \multicolumn{1}{c|}{1.74} & \multicolumn{1}{c|}{13.10} & \multicolumn{1}{c|}{41.37} & \multicolumn{1}{c|}{70.40} \\ \cline{1-3} \cline{5-11} 
\multicolumn{1}{|c|}{3)} & \multicolumn{1}{c|}{A+B}       & \multicolumn{1}{c|}{0.50}                           & \multicolumn{1}{c|}{}                  & \multicolumn{1}{c|}{21.85} & \multicolumn{1}{c|}{2.74}  & \multicolumn{1}{c|}{0.37} & \multicolumn{1}{c|}{0.96} & \multicolumn{1}{c|}{5.99}  & \multicolumn{1}{c|}{16.94} & \multicolumn{1}{c|}{28.88} \\ \cline{1-3} \cline{5-11} 
\multicolumn{1}{|c|}{4)}  & \multicolumn{1}{c|}{A+B+C} & \multicolumn{1}{c|}{0.34}                           & \multicolumn{1}{c|}{}                  & \multicolumn{1}{c|}{20.14} & \multicolumn{1}{c|}{2.52}  & \multicolumn{1}{c|}{0.27} & \multicolumn{1}{c|}{0.79} & \multicolumn{1}{c|}{5.20}  & \multicolumn{1}{c|}{14.97} & \multicolumn{1}{c|}{25.92} \\ \hline
\end{tabular}
\caption{Total number of ($b\bar{b}\gamma\gamma jj$) events expected at  HL-LHC  for the HEFT models (with $-2 \leq  a^2-b \leq 2$) (in the last 7 columns),  the SM  ($a=b=1$)  and the dominant background, (BKG)  following Strategy I  (i.e applying cuts on  $\eta_{\gamma \gamma}$).  The sequence in the selection of cuts is specified in the first columns. }
\label{strategy1}
\end{table}
We learn from these tables,  that both strategies are very efficient to discriminate between the HEFT signals and the background.  
The best one seems to be using the $P^T_{\gamma\gamma}$ cut at large transverse momentum (here,  $P^T_{\gamma\gamma}>200$ GeV).   However this strategy if used with a too stringent cut may produce a rather poor signal statistics. 

Finally,  to complete our study we explore more values of the combination $a^2-b=\kappa_V^2-\kappa_{2V}$ and compute the expected sensitivity to this combination at HL-LHC for the two strategies I and II.  For this computation we define the sensitivity  $S$ in terms of the number of signal events,  $N_S$ and background events $N_B$,  as follows:
\begin{eqnarray}
S&=&\sqrt{-2 \Big( (N_S+N_B)\log(\frac{N_B}{N_S+N_B})+N_S \Big)}
\end{eqnarray}
In Figs.  \ref{S-final-eta} and  \ref{S-final-PT} we show the predictions for $S$ as a function of $\kappa_V^2-\kappa_{2V}$ using Strategy I and Strategy II,  respectively.  In these (interpolated) predictions we have also explored the variations when moving $\kappa_V$ and $\kappa_{2V}$ separately within the ranges $0.95 < \kappa_V <1.10$ and $-1 < \kappa_{2V}< 3$,    
 for each fixed $\kappa_V^2-\kappa_{2V}$ value within the range $-2< \kappa_V^2-\kappa_{2V}< 2$.  These variations spread the predicted $S$ parabolas a little bit producing the narrow colored bands.


\begin{table}[h] \fontsize{9}{16} \selectfont \renewcommand{\arraystretch}{0.8}
\begin{tabular}{ccccccccccc}
\hline
\multicolumn{11}{c}{}                                                                                                                                                                                                                                                                                                                                                  
Number of predicted events,  $N(b\bar{b}\gamma\gamma jj)$,  with \bf{STRATEGY II}
\\ \hline
\hline
\multicolumn{2}{|c|}{\multirow{3}{*}{Selection}}                      & \multicolumn{1}{c|}{\multirow{3}{*}{BKG}}           & \multicolumn{1}{c|}{a =}               & \multicolumn{7}{c|}{1.10}                                                                                                                                                                              \\ \cline{4-11} 
\multicolumn{2}{|c|}{}                                        & \multicolumn{1}{c|}{}                               & \multicolumn{1}{c|}{b =}               & \multicolumn{1}{c|}{3.21}  & \multicolumn{1}{c|}{2.00}  & \multicolumn{1}{c|}{1.20} & \multicolumn{1}{c|}{1.00} & \multicolumn{1}{c|}{0.40}  & \multicolumn{1}{c|}{-0.30} & \multicolumn{1}{c|}{-0.79} \\ \cline{4-11} 
\multicolumn{2}{|c|}{}                                        & \multicolumn{1}{c|}{}                               & \multicolumn{1}{c|}{a$^2$ -b =}           & \multicolumn{1}{c|}{-2.00} & \multicolumn{1}{c|}{-0.80} & \multicolumn{1}{c|}{0.00} & \multicolumn{1}{c|}{0.20} & \multicolumn{1}{c|}{0.80}  & \multicolumn{1}{c|}{1.50}  & \multicolumn{1}{c|}{2.00}  \\ \hline
\multicolumn{11}{c}{}                                                                                                                                                                                                                                                                                                                                                 \\ \hline
\multicolumn{1}{|c|}{1)}   & \multicolumn{1}{c|}{A}          & \multicolumn{1}{c|}{690.70}                         & \multicolumn{1}{c|}{\multirow{4}{*}{}} & \multicolumn{1}{c|}{69.07} & \multicolumn{1}{c|}{9.43}  & \multicolumn{1}{c|}{1.58} & \multicolumn{1}{c|}{3.87} & \multicolumn{1}{c|}{20.29} & \multicolumn{1}{c|}{58.65} & \multicolumn{1}{c|}{97.67} \\ \cline{1-3} \cline{5-11} 
\multicolumn{1}{|c|}{2)}  & \multicolumn{1}{c|}{A+B}      & \multicolumn{1}{c|}{19.80}                          & \multicolumn{1}{c|}{}                  & \multicolumn{1}{c|}{18.96} & \multicolumn{1}{c|}{2.49}  & \multicolumn{1}{c|}{0.75} & \multicolumn{1}{c|}{1.58} & \multicolumn{1}{c|}{6.71}  & \multicolumn{1}{c|}{18.30} & \multicolumn{1}{c|}{31.63} \\ \cline{1-3} \cline{5-11} 
\multicolumn{1}{|c|}{3)} & \multicolumn{1}{c|}{A+C'}     & \multicolumn{1}{c|}{17.60}                          & \multicolumn{1}{c|}{}                  & \multicolumn{1}{c|}{18.41} & \multicolumn{1}{c|}{2.56}  & \multicolumn{1}{c|}{0.13} & \multicolumn{1}{c|}{0.57} & \multicolumn{1}{c|}{4.13}  & \multicolumn{1}{c|}{13.01} & \multicolumn{1}{c|}{23.27} \\ \cline{1-3} \cline{5-11} 
\multicolumn{1}{|c|}{4)}  & \multicolumn{1}{c|}{A+B+C'} & \multicolumn{1}{c|}{0.24}                           & \multicolumn{1}{c|}{}                  & \multicolumn{1}{c|}{11.86} & \multicolumn{1}{c|}{1.70}  & \multicolumn{1}{c|}{0.09} & \multicolumn{1}{c|}{0.38} & \multicolumn{1}{c|}{2.80}  & \multicolumn{1}{c|}{8.42}  & \multicolumn{1}{c|}{15.95} \\ \hline
\multicolumn{11}{c}{}                                                                                                                                                                                                                                                                                                                                                 \\ \hline
\multicolumn{2}{|c|}{\multirow{3}{*}{Selection}}                      & \multicolumn{1}{c|}{\multirow{3}{*}{SM (a=b=1)}} & \multicolumn{1}{c|}{a =}               & \multicolumn{7}{c|}{0.95}                                                                                                                                                                              \\ \cline{4-11} 
\multicolumn{2}{|c|}{}                                        & \multicolumn{1}{c|}{}                               & \multicolumn{1}{c|}{b =}               & \multicolumn{1}{c|}{2.90}  & \multicolumn{1}{c|}{1.70}  & \multicolumn{1}{c|}{0.90} & \multicolumn{1}{c|}{0.70} & \multicolumn{1}{c|}{0.10}  & \multicolumn{1}{c|}{-0.60} & \multicolumn{1}{c|}{-1.10} \\ \cline{4-11} 
\multicolumn{2}{|c|}{}                                        & \multicolumn{1}{c|}{}                               & \multicolumn{1}{c|}{a$^2$ -b =}           & \multicolumn{1}{c|}{-2.00} & \multicolumn{1}{c|}{-0.80} & \multicolumn{1}{c|}{0.00} & \multicolumn{1}{c|}{0.20} & \multicolumn{1}{c|}{0.80}  & \multicolumn{1}{c|}{1.50}  & \multicolumn{1}{c|}{2.00}  \\ \hline
\multicolumn{11}{c}{}                                                                                                                                                                                                                                                                                                                                                 \\ \hline
\multicolumn{1}{|c|}{1)}   & \multicolumn{1}{c|}{A}          & \multicolumn{1}{c|}{1.02}                           & \multicolumn{1}{c|}{\multirow{4}{*}{}} & \multicolumn{1}{c|}{76.30} & \multicolumn{1}{c|}{10.56} & \multicolumn{1}{c|}{0.78} & \multicolumn{1}{c|}{2.50} & \multicolumn{1}{c|}{17.55} & \multicolumn{1}{c|}{55.10} & \multicolumn{1}{c|}{92.93} \\ \cline{1-3} \cline{5-11} 
\multicolumn{1}{|c|}{2)}  & \multicolumn{1}{c|}{A+B}      & \multicolumn{1}{c|}{0.50}                           & \multicolumn{1}{c|}{}                  & \multicolumn{1}{c|}{21.85} & \multicolumn{1}{c|}{2.74}  & \multicolumn{1}{c|}{0.37} & \multicolumn{1}{c|}{0.96} & \multicolumn{1}{c|}{5.99}  & \multicolumn{1}{c|}{16.94} & \multicolumn{1}{c|}{28.88} \\ \cline{1-3} \cline{5-11} 
\multicolumn{1}{|c|}{3)} & \multicolumn{1}{c|}{A+C'}     & \multicolumn{1}{c|}{0.08}                           & \multicolumn{1}{c|}{}                  & \multicolumn{1}{c|}{20.05} & \multicolumn{1}{c|}{2.77}  & \multicolumn{1}{c|}{0.06} & \multicolumn{1}{c|}{0.42} & \multicolumn{1}{c|}{4.30}  & \multicolumn{1}{c|}{12.72} & \multicolumn{1}{c|}{23.29} \\ \cline{1-3} \cline{5-11} 
\multicolumn{1}{|c|}{4)}  & \multicolumn{1}{c|}{A+B+C'} & \multicolumn{1}{c|}{0.06}                           & \multicolumn{1}{c|}{}                  & \multicolumn{1}{c|}{12.84} & \multicolumn{1}{c|}{1.79}  & \multicolumn{1}{c|}{0.04} & \multicolumn{1}{c|}{0.29} & \multicolumn{1}{c|}{2.85}  & \multicolumn{1}{c|}{8.43}  & \multicolumn{1}{c|}{15.60} \\ \hline
\end{tabular}
\caption{Total number of ($b\bar{b}\gamma\gamma jj$) events expected at  HL-LHC  for the HEFT models (with $-2 \leq  a^2-b \leq 2$) (in the last 7 columns) ,  the SM ($a=b=1$)  and the dominant background (BKG),   following Strategy II  (i.e applying cuts on $P^T_{\gamma\gamma}$).  The sequence in the selection of cuts is specified in the first columns }
\label{strategy2}
\end{table}
 From these two figures,  \ref{S-final-eta} and \ref{S-final-PT},  we learn that the expected sensitivity at HL-LHC  to  $\kappa_V^2-\kappa_{2V}$ grows with the absolute  value of this combination and reaches high values at the larger considered values in the interval (-2,2),   for both strategies.   In the case of Strategy I the maximum sensitivity reached is when applying the three selected cuts A+B+C,  i.e.  for VBF+$M_{\gamma \gamma,  b \bar b}$+$\eta_{\gamma \gamma}$.  Concretely, we find $S=4.9$ at  $\kappa_V^2-\kappa_{2V}=-2$ and $S=6.4$ at $\kappa_V^2-\kappa_{2V}=2$.  In the case of Strategy II we find even higher sensitivities than  for Strategy I.  The maximum sensitivity is reached when applying the three selected cuts A+B+C', i.e. for VBF+$M_{\gamma \gamma,  b \bar b}$+ $P^T_{\gamma\gamma}$.  Concretely, we find $S=9.6$ at $\kappa_V^2-\kappa_{2V}=-2$ and $S=10.4$ at $\kappa_V^2-\kappa_{2V}=2$.  The spread in the prediction of $S$ due to varying separately $\kappa_V$ and $\kappa_{2V}$ for each value of the combination produces the colored bands but they are very narrow , indicating that for this study of $HHjj$ production at LHC the relevant parameter is not the separated $\kappa$'s but their combination $\kappa_V^2-\kappa_{2V}=2$,  as we have said through all this work.  Finally,  these sensitivity plots also demonstrate that the decay products of the two final Higgs bosons,  i.e. the $\gamma \gamma$ and $j_bj_b$ pairs inherit the high transversality of the  two H's, and as a consequence,  the use of differential distributions with variables like $\eta_{\gamma \gamma}$ and $P^T_{\gamma\gamma}$ as discriminants between the BSM signal and the background turns out to be very convenient.  
 
Last,  but not least,  it is also important to comment on the results for $S$ 
in the narrower interval $(-0.6, 0.6)$.  Although,  as already said,  there is not a direct constraint on the combination parameter  
$(\kappa_{V}^2-\kappa_{2V})$ from one single process,  this allowed  interval can be approximately  inferred from the separate $(95\%CL)$ constraints from LHC data  on $\kappa_V$ and $\kappa_{2V}$ (see Eqs.  \ref{ATLASconstraints} and \ref{CMSconstraints} and references quoted there) by combining several channels and processes (single Higgs,  double Higgs (inclusive),  various final states from Higgs decays, etc).  It  is clearly visible from our figures \ref{S-final-eta} and \ref{S-final-PT} that within the interval $(\kappa_V^2-\kappa_{2V}) \in (-0.6, 0.6)$ the maximum value of $S$ that can be obtained from the single process studied in this section,  $pp \to HHjj \to b \bar b \gamma \gamma jj$ is around 2.  Once the real data from HL-LHC be available in the future,  an improved study,  similar to the one done in here, but using more sophisticated tools like Neural Networks,   Multivariate Analysis,  etc,  will lead to either a determination of 
$(\kappa_V^2-\kappa_{2V})$ or to an improved constraint on this combination parameter.  On the other hand,  the particular scenarios within the HEFT with $(\kappa_V^2-\kappa_{2V})=0$,   including  the SM point itself with $\kappa_V=\kappa_{2V}=1$,  in contrast,  will not produce any visible signal at the HL-LHC in the process studied here,   i.e.,  in $pp \to HHjj \to b \bar b \gamma \gamma jj$.   
 \begin{minipage}{\linewidth}
      \centering
          \begin{figure}[H]
	\centerline{ \includegraphics[width=0.5\textwidth]{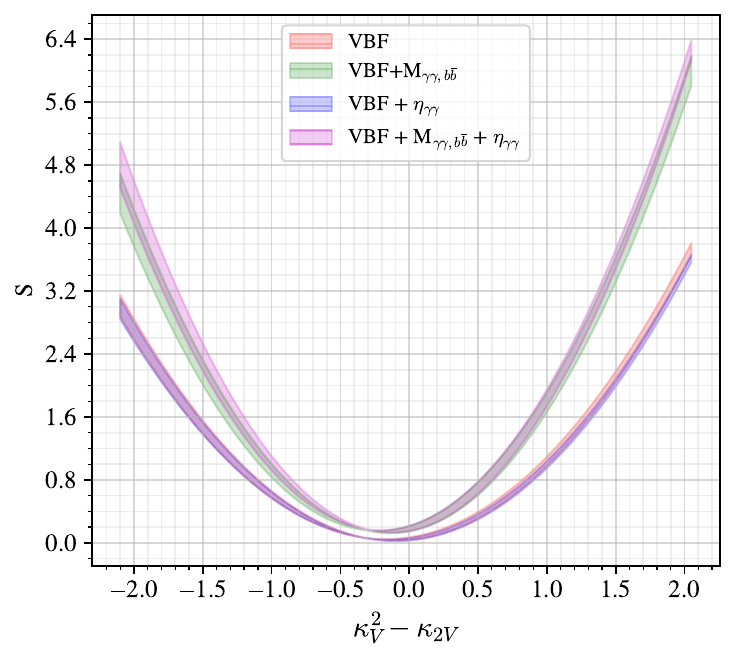}}
        \caption{Predicted sensitivity to $\kappa_V^2-\kappa_{2V}$ using Strategy I, based on $\eta_{\gamma\gamma}$.  The results for the Selections,  A (salmon)  A+B (green),  A+C (blue) and A+B+C (pink),  are shown in different colors.  The narrow colored bands along each parabola are produced when moving separately $\kappa_{V}$ and $\kappa_{2V}$ for each $\kappa_V^2-\kappa_{2V}$ value. }.  
\label{S-final-eta}        
        	\end{figure}
  \end{minipage}
   \begin{minipage}{\linewidth}
      \centering
          \begin{figure}[H]
	\centerline{ \includegraphics[width=0.5\textwidth]{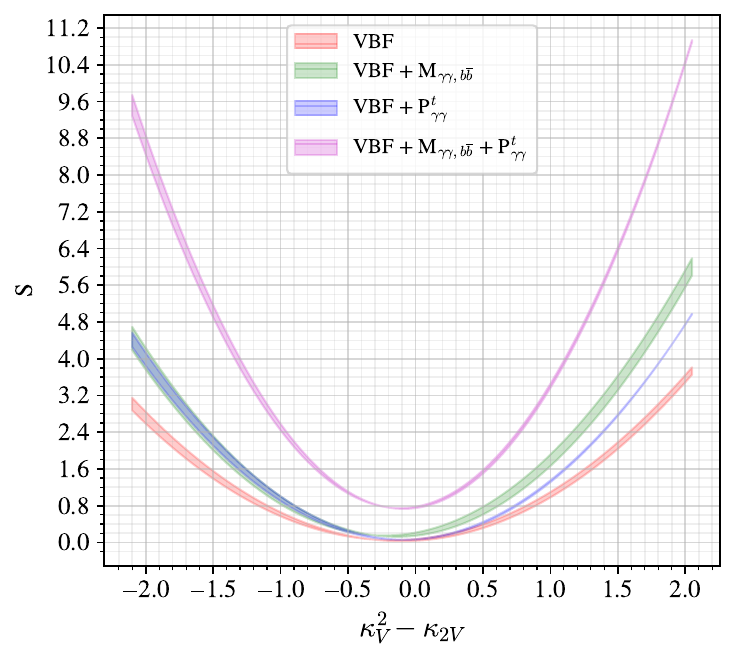}}
        \caption{Predicted sensitivity to $\kappa_V^2-\kappa_{2V}$ using Strategy II,  based on $P^T_{\gamma\gamma}$.  The results for the Selections,   A (salmon)  A+B (green),  A+C' (blue) and A+B+C' (pink),  are shown in different colors.  The narrow colored bands along each parabola are produced when moving separately 
        $\kappa_{V}$ and $\kappa_{2V}$ for each $\kappa_V^2-\kappa_{2V}$ value. }.  
\label{S-final-PT}        
        \end{figure}
  \end{minipage}

\section*{Conclusions} \addcontentsline{toc}{section}{Conclusions} \label{section:7} 
In this work,  we have studied the effective interactions $HVV$ and $HHVV$ within the HEFT context and their potential correlations described by the combination parameter $a^2-b=\kappa_V^2-\kappa_{2V}$.  We have selected  the process $pp \to HH jj'$ at LHC,  which is sensitive to this particular combination $\kappa_V^2-\kappa_{2V}$,  in contrast to single Higgs production,  which is only sensitive to $\kappa_V$.  The interest in this particular combination of $\kappa$'s parameters is because it can give access to potential correlations between the $HVV$ and $HHVV$ effective interactions that may occur in the low energy limit of the particular high energy fundamental theory,  which is  underlying the Higgs sector of the Electroweak Theory and the spontaneous symmetry breaking of $SU(2) \times U(1)$.  The particular setting $\kappa_V=\kappa_{2V}=1$ corresponds to the SM case,  and therefore any deviation from this setting implies BSM Higgs Physics.  We have used here as low energy theory the HEFT, which is a gauge invariant $SU(2) \times U(1)$ theory,  allowing for a fully gauge invariant computation of all scattering amplitudes and cross sections of interest.   Within the HEFT,  these two effective coefficients,  $a=\kappa_V$ and $b=\kappa_{2V}$,  are treated as independent parameters and are generically uncorrelated parameters.  It is in particular settings for the UV underlying theory,  where,  after integration of the heavy modes,  the predictions for $a$,  $b$ and $a^2-b$ could give rise to the mentioned correlations between the effective interactions $HVV$ and $HHVV$ at low energies.  Therefore,  studying the potential  correlations via the combination $\kappa_V^2-\kappa_{2V}$ which we propose here could provide interesting signals of BSM Higgs physics.

Here we have first described,  both analytically and numerically,  how this combination  $\kappa_V^2-\kappa_{2V}$ appears at the subprocess level.  This particular combination appears explicitly at large energies compared to the masses,  in the scattering amplitude of double Higgs production from two longitudinal EW gauge bosons: $V_LV_L \to HH$  with $VV=WW$ and $VV=ZZ$.  The particular behaviour at $\sqrt{s} \gg m_H, m_V$ of this dominant helicity amplitude growing with energy as $ {\cal A} \sim  (\kappa_V^2-\kappa_{2V}) s/v^2$ is the clue to understand the relevance of studying this combination.

We have then focused our study on the production of $HHjj$ at LHC which,   with a properly designed strategy,  may access directly to this combination.  
The strategy proposed here is to first make a selection of events having dominantly the so-called Vector Boson Fusion (VBF) topology,  precisely because in this topology is where the $\kappa_V^2-\kappa_{2V}$ combination governs the $HHjj$ event rates.  Our study of $HHjj$ event rates with VBF topology within the HEFT models indicates that not only the total cross section $\sigma (pp \to HHjj)$ but also the differential cross sections like $d \sigma (pp \to HHjj) /d \eta_H$ are very sensitive to this  $\kappa_V^2-\kappa_{2V}$ combination.  The total rates clearly grow with $\kappa_V^2-\kappa_{2V}$,  and the distributions with respect to the final Higgs boson pseudo-rapidity show a clear peak at central rapidity $\eta_H =0$.  This demonstrates that within these HEFT scenarios with $(\kappa_V^2 -\kappa_{2V}) \neq 0$ a good discriminant of BSM signals with respect to the SM rates will be to search for these central peaks indicating the high transversality of the Higgs produced, and in consequence also of its decays. 

We have then fully studied the complete process,  $pp \to HHjj' \to b \bar b \gamma \gamma j j'$,  that includes the particular Higgs decays,  $H \to b \bar b$ and $H \to \gamma \gamma $ which offer good prospects for the detection of BSM Higgs physics.  We have analyzed  the final state  events,  $b\overline{b}\gamma\gamma jj'$,  with two light jets,  two b-jets and two photons for both the HEFT signals,  considering various values of the combination parameter,  and for the dominant background,  SM-QCD-EW (non-resonant),  and we have taken into account  in all cases  the realistic effects from 
showering,  clustering,  hadronization and detector.  To implement these effects,  the parton-level $pp \rightarrow b\overline{b}\gamma\gamma jj'$ simulated events with MG5  are filtered to incorporate \textsc{Pythia} and \textsc{Delphes} simulations,  which account for showering and hadronization effects, as well as the jet clustering, and for a parametric simulation of the detector, in particular the future upgraded Phase-2 CMS detector.  As a result of the excellent capabilities for photon and b-jet identification of the detector in the phase space regions of our interest,  the $HH$ relative sensitivities that allow us to discriminate between HEFT models with different values for the combination $\kappa_V^2 - \kappa_{2V}$ are preserved almost unchanged.  In particular,  the shapes of the distributions with respect to the pseudo-rapidity of $H$,   translated to the 
pseudo-rapidity of the $\gamma \gamma$ pairs remain unchanged and show clearly the centrality and high transversality of these $\gamma \gamma$ pairs in the HEFT events as compared to the SM events.  The Detector-level predictions deviate from parton-level predictions almost exclusively by a general scale-down of the total rates,  resulting from a prediction of the experimental acceptance and resolution of the process.  We have shown that higher combined selection efficiency values are obtained from HEFT $HHjj$ signals with a greater $\kappa_V^2 - \kappa_{2V}$ combination value.  We have also shown that HEFT $HHjj$ production involving $a$ and $b$ parameters with higher values for the $\kappa_V^2 - \kappa_{2V}$ combination leads to signals with greater detector-level efficiencies, i.e.  smaller event loss with respect to the parton-level estimations, and a greater number of predicted events at the $pp$ LHC experiments.  

Finally,  we have presented specific strategies focused on the choice of the proper variables and  the proper cuts that lead to enhancing the sensitivity to the $\kappa_V^2 - \kappa_{2V}$ combination for the future HL-LHC operation.  These proper variables have been extracted by studying the various relevant differential cross sections for signal and background.  To enhance the signal versus background sensitivity,  we have proposed two main routes.  The two of them use first the  
$b\overline{b}\gamma\gamma jj'$ events selection by means of the important requirement of VBF topology.  The two of them also use the $HH$ final state isolation criteria based on 
$M_{\gamma \gamma}$,  and $M_{b \bar b}$ being close to $m_H$.   The main difference between the two strategies is the variable chosen for the study of the transversality of the final $\gamma \gamma$ and $b \bar b$ pairs.  Strategy I uses mainly $\eta_{\gamma \gamma}$,  and Strategy II uses mainly $P^T_{\gamma \gamma}$.  Our final results in figures   \ref{S-final-eta} and \ref{S-final-PT} show that both Strategies will provide efficient access to this  $\kappa_V^2 - \kappa_{2V}$ combination  and  will lead to notable sensitivities to the largest values of this parameter,  once the proper cuts for VBF+$M_{\gamma \gamma,  b \bar b}$+
$\eta_{\gamma \gamma}$  or for VBF+$M_{\gamma \gamma,  b \bar b}$+
$P^T_{\gamma \gamma}$, respectively,  have been applied. 

The general conclusion drawn from our work is that the BSM $HHjj$ signal strengths grow explicitly with the value of the $\kappa_V^2 - \kappa_{2V}$ combination within the considered HEFT models. This conclusion is preserved when considering the decay of both Higgs bosons, in particular for the $b\overline{b}\gamma\gamma$ decay channel,  which offers good experimental prospects.  It is also (nearly) independent of whether we include the showering, hadronization, and clustering effects and the parametric simulation of the detector. This is so  because the event distributions with respect to the relevant variables,  like $\eta_{\gamma \gamma}$ and $P^T_{\gamma \gamma}$,  do not change appreciably when moving from parton-level to detector-level predictions.  This leads to clear theoretical predictions and kinematic properties that can be translated into experimentally testable event characteristics within the projected HL-LHC operation and Phase-2 CMS detector, for which we expect to obtain the first evidence of the $HH$ production.   In the most pessimistic case that this future experimental search of BSM $HHjj$ production at HL-LHC does not find any deviation with respect to the SM predictions,  our study indicates that improved constraints on BSM Higgs via the particular combination $\kappa_V^2 - \kappa_{2V}$ will be definitely set. 
\section*{Acknowledgments}
We wish to warmly thank Maria Cepeda for her valuable help in the evaluation of the efficiencies loss due to showering, hadronization, and detector effects,  which have been simulated in the present work with PYTHIA and DELPHES.  The results presented here would not have been possible without her guide and advice. 
We would also like to thank Michele Selvaggi for kindly providing us with the Snowmass21 Delphes card.  Its use has been fundamental and has greatly contributed to the results of this work.
We also warmly acknowledge  the valuable help from Roberto Morales who participated in the building of  the HEFT model file for MG5 which is used in this work.  
We acknowledge partial financial support by the Spanish Research Agency (Agencia Estatal de Investigación) through the grant  with reference number PID2022-137127NB-I00 
funded by MCIN/AEI/10.13039/501100011033/ FEDER, UE.  We also thank financial support from the  grant IFT Centro de Excelencia Severo Ochoa with reference number CEX2020-001007-S
 funded by MCIN/AEI/10.13039/501100011033, from the previous AEI project PID2019-108892RB-I00 funded by MCIN/AEI/10.13039/501100011033, and from the European Union’s Horizon 2020 research and innovation programme under the Marie Sklodowska-Curie grant agreement No 860881-HIDDeN.  
The work of D.D. is also supported by the Spanish Ministry of Science and Innovation via an FPU grant No FPU22/03485. 
\newpage


\begin{appendices} 

   \section{Feynman diagrams for double Higgs production in hadronic collisions} \label{anexo}
   
The present appendix summarizes the diagrams involved in calculating the Higgs boson pair production
and two jets  at hadron colliders,  $pp \rightarrow HHjj'$. For brevity, only the partonic $q_1\overline{q}_2 \rightarrow HH q_3\overline{q}_4$ case is displayed here, but all the partonic cases with initial quarks and anti-quarks are included in the final computation.  Notice also that only the diagrams with asterisk are of VBF topologies, whereas all the others are of associated production (double Higgs-strahlung) topologies, whose respective generic representations have been summarized in Figure \ref{fig:04_00}.  All the diagrams are obtained using \textsc{Madgraph5} (\textsc{MG5}) \cite{Alwall_2014} and the particular HEFT model defined in Refs.  \cite{PhysRevD.102.075040, PhysRevD.104.075013, PhysRevD.106.073008, PhysRevD.106.115027}

\vspace{0em} \begin{figure}[!h] 
\centerline{ \includegraphics[width=1.05\textwidth]{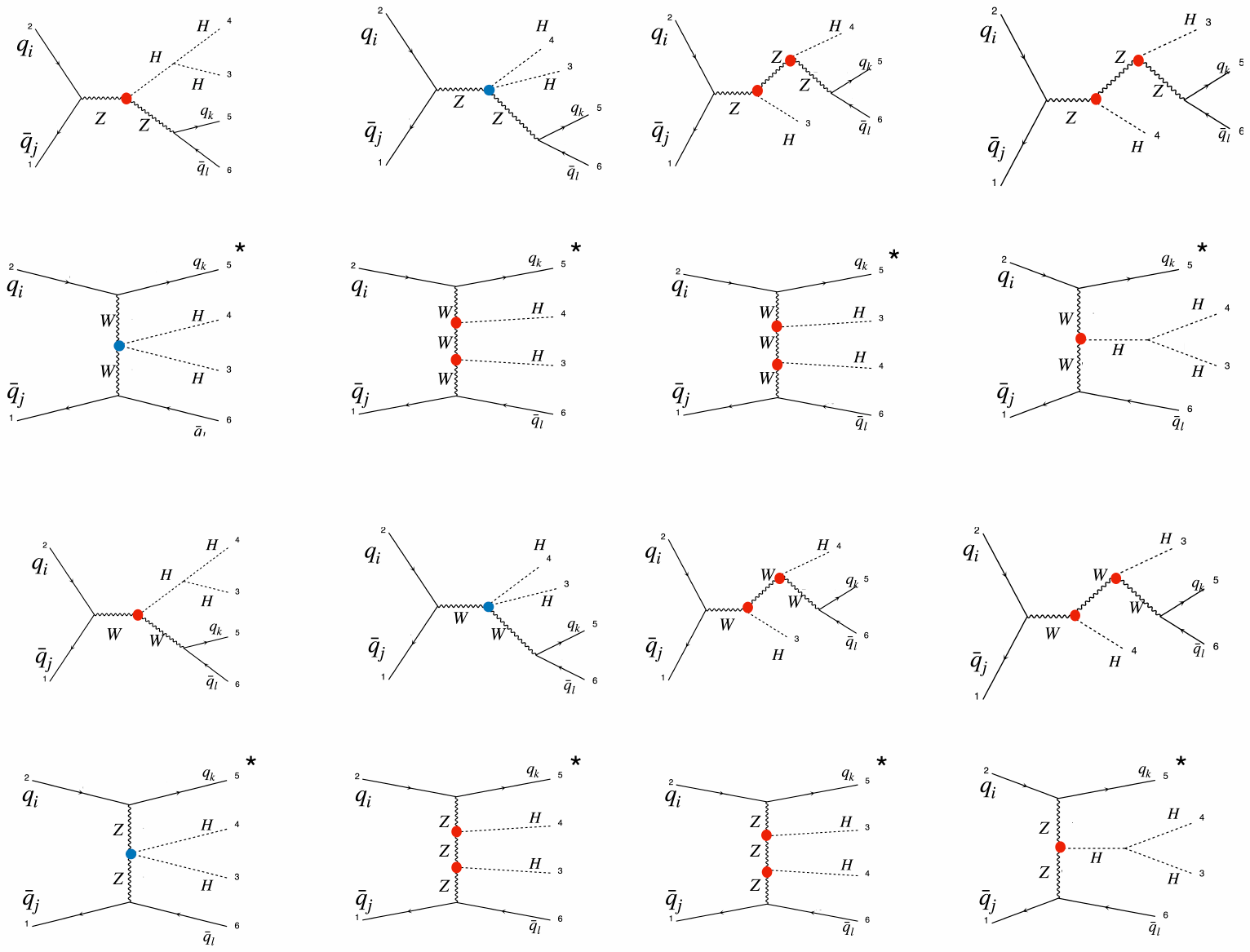} }
\caption{Feynman diagrams for double Higgs production in proton-proton collisions. Diagram topologies marked with a * are also present in the MG5 calculus considering an $q$-$q$ and $\overline{q}$-$\overline{q}$ initial states. Remind that $q \equiv u, d , c, s$ and $\overline{q} \equiv \overline{u}, \overline{d}, \overline{c}, \overline{s}.$}
\label{fig:A} \end{figure}   
 
\end{appendices}

\clearpage

\bibliography{DGH-revised-final}

\begin{thebibliography}{42}%
\makeatletter
\providecommand \@ifxundefined [1]{%
 \@ifx{#1\undefined}
}%
\providecommand \@ifnum [1]{%
 \ifnum #1\expandafter \@firstoftwo
 \else \expandafter \@secondoftwo
 \fi
}%
\providecommand \@ifx [1]{%
 \ifx #1\expandafter \@firstoftwo
 \else \expandafter \@secondoftwo
 \fi
}%
\providecommand \natexlab [1]{#1}%
\providecommand \enquote  [1]{``#1''}%
\providecommand \bibnamefont  [1]{#1}%
\providecommand \bibfnamefont [1]{#1}%
\providecommand \citenamefont [1]{#1}%
\providecommand \href@noop [0]{\@secondoftwo}%
\providecommand \href [0]{\begingroup \@sanitize@url \@href}%
\providecommand \@href[1]{\@@startlink{#1}\@@href}%
\providecommand \@@href[1]{\endgroup#1\@@endlink}%
\providecommand \@sanitize@url [0]{\catcode `\\12\catcode `\$12\catcode
  `\&12\catcode `\#12\catcode `\^12\catcode `\_12\catcode `\%12\relax}%
\providecommand \@@startlink[1]{}%
\providecommand \@@endlink[0]{}%
\providecommand \url  [0]{\begingroup\@sanitize@url \@url }%
\providecommand \@url [1]{\endgroup\@href {#1}{\urlprefix }}%
\providecommand \urlprefix  [0]{URL }%
\providecommand \Eprint [0]{\href }%
\providecommand \doibase [0]{http://dx.doi.org/}%
\providecommand \selectlanguage [0]{\@gobble}%
\providecommand \bibinfo  [0]{\@secondoftwo}%
\providecommand \bibfield  [0]{\@secondoftwo}%
\providecommand \translation [1]{[#1]}%
\providecommand \BibitemOpen [0]{}%
\providecommand \bibitemStop [0]{}%
\providecommand \bibitemNoStop [0]{.\EOS\space}%
\providecommand \EOS [0]{\spacefactor3000\relax}%
\providecommand \BibitemShut  [1]{\csname bibitem#1\endcsname}%
\let\auto@bib@innerbib\@empty
\bibitem [{\citenamefont {Andersen}\ \emph {et~al.}(2013)\citenamefont
  {Andersen} \emph {et~al.}}]{LHCHiggsCrossSectionWorkingGroup:2013rie}%
  \BibitemOpen
  \bibfield  {author} {\bibinfo {author} {\bibfnamefont {J.~R.}\ \bibnamefont
  {Andersen}} \emph {et~al.} (\bibinfo {collaboration} {LHC Higgs Cross Section
  Working Group}),\ }\href {\doibase 10.5170/CERN-2013-004} {\  (\bibinfo
  {year} {2013}),\ 10.5170/CERN-2013-004},\ \Eprint
  {http://arxiv.org/abs/1307.1347} {arXiv:1307.1347 [hep-ph]} \BibitemShut
  {NoStop}%
\bibitem [{\citenamefont {Dainese}\ \emph {et~al.}(2019)\citenamefont
  {Dainese}, \citenamefont {Mangano}, \citenamefont {Meyer}, \citenamefont
  {Nisati}, \citenamefont {Salam},\ and\ \citenamefont
  {Vesterinen}}]{Dainese:2019rgk}%
  \BibitemOpen
  \bibinfo {editor} {\bibfnamefont {A.}~\bibnamefont {Dainese}}, \bibinfo
  {editor} {\bibfnamefont {M.}~\bibnamefont {Mangano}}, \bibinfo {editor}
  {\bibfnamefont {A.~B.}\ \bibnamefont {Meyer}}, \bibinfo {editor}
  {\bibfnamefont {A.}~\bibnamefont {Nisati}}, \bibinfo {editor} {\bibfnamefont
  {G.}~\bibnamefont {Salam}}, \ and\ \bibinfo {editor} {\bibfnamefont {M.~A.}\
  \bibnamefont {Vesterinen}},\ eds.,\ \href {\doibase 10.23731/CYRM-2019-007}
  {\emph {\bibinfo {title} {{Report on the Physics at the HL-LHC,and
  Perspectives for the HE-LHC}}}},\ \bibinfo {series} {CERN Yellow Reports:
  Monographs}, Vol.\ \bibinfo {volume} {7/2019}\ (\bibinfo  {publisher}
  {CERN},\ \bibinfo {address} {Geneva, Switzerland},\ \bibinfo {year}
  {2019})\BibitemShut {NoStop}%
\bibitem [{ATL(2019)}]{ATLAS:2019mfr}%
  \BibitemOpen
  \href {\doibase 10.23731/CYRM-2019-007.Addendum} {\bibfield  {journal}
  {\bibinfo  {journal} {CERN Yellow Rep. Monogr.}\ }\textbf {\bibinfo {volume}
  {7}},\ \bibinfo {pages} {Addendum} (\bibinfo {year} {2019})},\ \Eprint
  {http://arxiv.org/abs/1902.10229} {arXiv:1902.10229 [hep-ex]} \BibitemShut
  {NoStop}%
\bibitem [{\citenamefont {Cepeda}\ \emph {et~al.}(2019)\citenamefont {Cepeda}
  \emph {et~al.}}]{Cepeda:2019klc}%
  \BibitemOpen
  \bibfield  {author} {\bibinfo {author} {\bibfnamefont {M.}~\bibnamefont
  {Cepeda}} \emph {et~al.},\ }\href {\doibase 10.23731/CYRM-2019-007.221}
  {\bibfield  {journal} {\bibinfo  {journal} {CERN Yellow Rep. Monogr.}\
  }\textbf {\bibinfo {volume} {7}},\ \bibinfo {pages} {221} (\bibinfo {year}
  {2019})},\ \Eprint {http://arxiv.org/abs/1902.00134} {arXiv:1902.00134
  [hep-ph]} \BibitemShut {NoStop}%
\bibitem [{\citenamefont {Mlynarikova}(2023)}]{Mlynarikova:2023bvx}%
  \BibitemOpen
  \bibfield  {author} {\bibinfo {author} {\bibfnamefont {M.}~\bibnamefont
  {Mlynarikova}} (\bibinfo {collaboration} {ATLAS, CMS}),\ }in\ \href@noop {}
  {\emph {\bibinfo {booktitle} {{30th International Workshop on Deep-Inelastic
  Scattering and Related Subjects}}}}\ (\bibinfo {year} {2023})\ \Eprint
  {http://arxiv.org/abs/2307.07772} {arXiv:2307.07772 [hep-ex]} \BibitemShut
  {NoStop}%
\bibitem [{\citenamefont {ATLAS-CMS}(2025)}]{ATL-PHYS-PUB-2025-018}%
  \BibitemOpen
  \bibfield  {author} {\bibinfo {author} {\bibnamefont {ATLAS-CMS}},\ }\href
  {https://arxiv.org/abs/2504.00672} {\enquote {\bibinfo {title} {Highlights of
  the hl-lhc physics projections by atlas and cms},}\ } (\bibinfo {year}
  {2025}),\ \Eprint {http://arxiv.org/abs/2504.00672} {arXiv:2504.00672
  [hep-ex]} \BibitemShut {NoStop}%
\bibitem [{\citenamefont {Brivio}\ and\ \citenamefont
  {Trott}(2019)}]{Brivio:2017vri}%
  \BibitemOpen
  \bibfield  {author} {\bibinfo {author} {\bibfnamefont {I.}~\bibnamefont
  {Brivio}}\ and\ \bibinfo {author} {\bibfnamefont {M.}~\bibnamefont {Trott}},\
  }\href {\doibase 10.1016/j.physrep.2018.11.002} {\bibfield  {journal}
  {\bibinfo  {journal} {Phys. Rept.}\ }\textbf {\bibinfo {volume} {793}},\
  \bibinfo {pages} {1} (\bibinfo {year} {2019})},\ \Eprint
  {http://arxiv.org/abs/1706.08945} {arXiv:1706.08945 [hep-ph]} \BibitemShut
  {NoStop}%
\bibitem [{\citenamefont {Dobado}\ and\ \citenamefont
  {Espriu}(2019)}]{Dobado:2019fxe}%
  \BibitemOpen
  \bibfield  {author} {\bibinfo {author} {\bibfnamefont {A.}~\bibnamefont
  {Dobado}}\ and\ \bibinfo {author} {\bibfnamefont {D.}~\bibnamefont
  {Espriu}},\ }\href@noop {} {\  (\bibinfo {year} {2019})},\ \Eprint
  {http://arxiv.org/abs/1911.06844} {arXiv:1911.06844 [hep-ph]} \BibitemShut
  {NoStop}%
\bibitem [{\citenamefont {Herrero}\ and\ \citenamefont
  {Morales}(2020)}]{PhysRevD.102.075040}%
  \BibitemOpen
  \bibfield  {author} {\bibinfo {author} {\bibfnamefont {M.}~\bibnamefont
  {Herrero}}\ and\ \bibinfo {author} {\bibfnamefont {R.~A.}\ \bibnamefont
  {Morales}},\ }\href {\doibase 10.1103/PhysRevD.102.075040} {\bibfield
  {journal} {\bibinfo  {journal} {Phys. Rev. D}\ }\textbf {\bibinfo {volume}
  {102}},\ \bibinfo {pages} {075040} (\bibinfo {year} {2020})}\BibitemShut
  {NoStop}%
\bibitem [{\citenamefont {Herrero}\ and\ \citenamefont
  {Morales}(2021)}]{PhysRevD.104.075013}%
  \BibitemOpen
  \bibfield  {author} {\bibinfo {author} {\bibfnamefont {M.~J.}\ \bibnamefont
  {Herrero}}\ and\ \bibinfo {author} {\bibfnamefont {R.~A.}\ \bibnamefont
  {Morales}},\ }\href {\doibase 10.1103/PhysRevD.104.075013} {\bibfield
  {journal} {\bibinfo  {journal} {Phys. Rev. D}\ }\textbf {\bibinfo {volume}
  {104}},\ \bibinfo {pages} {075013} (\bibinfo {year} {2021})}\BibitemShut
  {NoStop}%
\bibitem [{\citenamefont {Herrero}\ and\ \citenamefont
  {Morales}(2022)}]{PhysRevD.106.073008}%
  \BibitemOpen
  \bibfield  {author} {\bibinfo {author} {\bibfnamefont {M.~J.}\ \bibnamefont
  {Herrero}}\ and\ \bibinfo {author} {\bibfnamefont {R.~A.}\ \bibnamefont
  {Morales}},\ }\href {\doibase 10.1103/PhysRevD.106.073008} {\bibfield
  {journal} {\bibinfo  {journal} {Phys. Rev. D}\ }\textbf {\bibinfo {volume}
  {106}},\ \bibinfo {pages} {073008} (\bibinfo {year} {2022})}\BibitemShut
  {NoStop}%
\bibitem [{\citenamefont {Domenech}\ \emph {et~al.}(2022)\citenamefont
  {Domenech}, \citenamefont {Herrero}, \citenamefont {Morales},\ and\
  \citenamefont {Ramos}}]{PhysRevD.106.115027}%
  \BibitemOpen
  \bibfield  {author} {\bibinfo {author} {\bibfnamefont {D.}~\bibnamefont
  {Domenech}}, \bibinfo {author} {\bibfnamefont {M.~J.}\ \bibnamefont
  {Herrero}}, \bibinfo {author} {\bibfnamefont {R.~A.}\ \bibnamefont
  {Morales}}, \ and\ \bibinfo {author} {\bibfnamefont {M.}~\bibnamefont
  {Ramos}},\ }\href {\doibase 10.1103/PhysRevD.106.115027} {\bibfield
  {journal} {\bibinfo  {journal} {Phys. Rev. D}\ }\textbf {\bibinfo {volume}
  {106}},\ \bibinfo {pages} {115027} (\bibinfo {year} {2022})}\BibitemShut
  {NoStop}%
\bibitem [{\citenamefont {Domenech}\ \emph {et~al.}(2025)\citenamefont
  {Domenech}, \citenamefont {Herrero}, \citenamefont {Morales},\ and\
  \citenamefont {Salas-Bern{\'a}rdez}}]{Domenech:2025gmn}%
  \BibitemOpen
  \bibfield  {author} {\bibinfo {author} {\bibfnamefont {D.}~\bibnamefont
  {Domenech}}, \bibinfo {author} {\bibfnamefont {M.}~\bibnamefont {Herrero}},
  \bibinfo {author} {\bibfnamefont {R.~A.}\ \bibnamefont {Morales}}, \ and\
  \bibinfo {author} {\bibfnamefont {A.}~\bibnamefont {Salas-Bern{\'a}rdez}},\
  }\href@noop {} {\  (\bibinfo {year} {2025})},\ \Eprint
  {http://arxiv.org/abs/2506.21716} {arXiv:2506.21716 [hep-ph]} \BibitemShut
  {NoStop}%
\bibitem [{\citenamefont {Arco}\ \emph {et~al.}(2023)\citenamefont {Arco},
  \citenamefont {Domenech}, \citenamefont {Herrero},\ and\ \citenamefont
  {Morales}}]{Arco:2023sac}%
  \BibitemOpen
  \bibfield  {author} {\bibinfo {author} {\bibfnamefont {F.}~\bibnamefont
  {Arco}}, \bibinfo {author} {\bibfnamefont {D.}~\bibnamefont {Domenech}},
  \bibinfo {author} {\bibfnamefont {M.~J.}\ \bibnamefont {Herrero}}, \ and\
  \bibinfo {author} {\bibfnamefont {R.~A.}\ \bibnamefont {Morales}},\ }\href
  {\doibase 10.1103/PhysRevD.108.095013} {\bibfield  {journal} {\bibinfo
  {journal} {Phys. Rev. D}\ }\textbf {\bibinfo {volume} {108}},\ \bibinfo
  {pages} {095013} (\bibinfo {year} {2023})},\ \Eprint
  {http://arxiv.org/abs/2307.15693} {arXiv:2307.15693 [hep-ph]} \BibitemShut
  {NoStop}%
\bibitem [{\citenamefont {D{\'a}vila}\ \emph {et~al.}(2024)\citenamefont
  {D{\'a}vila}, \citenamefont {Domenech}, \citenamefont {Herrero},\ and\
  \citenamefont {Morales}}]{D_vila_2024}%
  \BibitemOpen
  \bibfield  {author} {\bibinfo {author} {\bibfnamefont {J.~M.}\ \bibnamefont
  {D{\'a}vila}}, \bibinfo {author} {\bibfnamefont {D.}~\bibnamefont
  {Domenech}}, \bibinfo {author} {\bibfnamefont {M.~J.}\ \bibnamefont
  {Herrero}}, \ and\ \bibinfo {author} {\bibfnamefont {R.~A.}\ \bibnamefont
  {Morales}},\ }\href {\doibase 10.1140/epjc/s10052-024-12815-5} {\bibfield
  {journal} {\bibinfo  {journal} {The European Physical Journal C}\ }\textbf
  {\bibinfo {volume} {84}} (\bibinfo {year} {2024}),\
  10.1140/epjc/s10052-024-12815-5}\BibitemShut {NoStop}%
\bibitem [{\citenamefont {Gonzalez-Lopez}\ \emph {et~al.}(2021)\citenamefont
  {Gonzalez-Lopez}, \citenamefont {Herrero},\ and\ \citenamefont
  {Martinez-Suarez}}]{Gonzalez-Lopez:2021aa}%
  \BibitemOpen
  \bibfield  {author} {\bibinfo {author} {\bibfnamefont {M.}~\bibnamefont
  {Gonzalez-Lopez}}, \bibinfo {author} {\bibfnamefont {M.~J.}\ \bibnamefont
  {Herrero}}, \ and\ \bibinfo {author} {\bibfnamefont {P.}~\bibnamefont
  {Martinez-Suarez}},\ }\href {\doibase 10.1140/epjc/s10052-021-09048-1}
  {\bibfield  {journal} {\bibinfo  {journal} {The European Physical Journal C}\
  }\textbf {\bibinfo {volume} {81}},\ \bibinfo {pages} {260} (\bibinfo {year}
  {2021})}\BibitemShut {NoStop}%
\bibitem [{\citenamefont {Contino}\ \emph {et~al.}(2010)\citenamefont
  {Contino}, \citenamefont {Grojean}, \citenamefont {Moretti}, \citenamefont
  {Piccinini},\ and\ \citenamefont {Rattazzi}}]{Contino_2010}%
  \BibitemOpen
  \bibfield  {author} {\bibinfo {author} {\bibfnamefont {R.}~\bibnamefont
  {Contino}}, \bibinfo {author} {\bibfnamefont {C.}~\bibnamefont {Grojean}},
  \bibinfo {author} {\bibfnamefont {M.}~\bibnamefont {Moretti}}, \bibinfo
  {author} {\bibfnamefont {F.}~\bibnamefont {Piccinini}}, \ and\ \bibinfo
  {author} {\bibfnamefont {R.}~\bibnamefont {Rattazzi}},\ }\href {\doibase
  10.1007/jhep05(2010)089} {\bibfield  {journal} {\bibinfo  {journal} {Journal
  of High Energy Physics}\ }\textbf {\bibinfo {volume} {2010}} (\bibinfo {year}
  {2010}),\ 10.1007/jhep05(2010)089}\BibitemShut {NoStop}%
\bibitem [{\citenamefont {Alwall}\ \emph {et~al.}(2014)\citenamefont {Alwall},
  \citenamefont {Frederix}, \citenamefont {Frixione}, \citenamefont {Hirschi},
  \citenamefont {Maltoni}, \citenamefont {Mattelaer}, \citenamefont {Shao},
  \citenamefont {Stelzer}, \citenamefont {Torrielli},\ and\ \citenamefont
  {Zaro}}]{Alwall_2014}%
  \BibitemOpen
  \bibfield  {author} {\bibinfo {author} {\bibfnamefont {J.}~\bibnamefont
  {Alwall}}, \bibinfo {author} {\bibfnamefont {R.}~\bibnamefont {Frederix}},
  \bibinfo {author} {\bibfnamefont {S.}~\bibnamefont {Frixione}}, \bibinfo
  {author} {\bibfnamefont {V.}~\bibnamefont {Hirschi}}, \bibinfo {author}
  {\bibfnamefont {F.}~\bibnamefont {Maltoni}}, \bibinfo {author} {\bibfnamefont
  {O.}~\bibnamefont {Mattelaer}}, \bibinfo {author} {\bibfnamefont {H.-S.}\
  \bibnamefont {Shao}}, \bibinfo {author} {\bibfnamefont {T.}~\bibnamefont
  {Stelzer}}, \bibinfo {author} {\bibfnamefont {P.}~\bibnamefont {Torrielli}},
  \ and\ \bibinfo {author} {\bibfnamefont {M.}~\bibnamefont {Zaro}},\ }\href
  {\doibase 10.1007/jhep07(2014)079} {\bibfield  {journal} {\bibinfo  {journal}
  {Journal of High Energy Physics}\ }\textbf {\bibinfo {volume} {2014}}
  (\bibinfo {year} {2014}),\ 10.1007/jhep07(2014)079}\BibitemShut {NoStop}%
\bibitem [{\citenamefont {Sj{\"o}strand}\ \emph {et~al.}(2015)\citenamefont
  {Sj{\"o}strand}, \citenamefont {Ask}, \citenamefont {Christiansen},
  \citenamefont {Corke}, \citenamefont {Desai}, \citenamefont {Ilten},
  \citenamefont {Mrenna}, \citenamefont {Prestel}, \citenamefont {Rasmussen},\
  and\ \citenamefont {Skands}}]{Sj_strand_2015}%
  \BibitemOpen
  \bibfield  {author} {\bibinfo {author} {\bibfnamefont {T.}~\bibnamefont
  {Sj{\"o}strand}}, \bibinfo {author} {\bibfnamefont {S.}~\bibnamefont {Ask}},
  \bibinfo {author} {\bibfnamefont {J.~R.}\ \bibnamefont {Christiansen}},
  \bibinfo {author} {\bibfnamefont {R.}~\bibnamefont {Corke}}, \bibinfo
  {author} {\bibfnamefont {N.}~\bibnamefont {Desai}}, \bibinfo {author}
  {\bibfnamefont {P.}~\bibnamefont {Ilten}}, \bibinfo {author} {\bibfnamefont
  {S.}~\bibnamefont {Mrenna}}, \bibinfo {author} {\bibfnamefont
  {S.}~\bibnamefont {Prestel}}, \bibinfo {author} {\bibfnamefont {C.~O.}\
  \bibnamefont {Rasmussen}}, \ and\ \bibinfo {author} {\bibfnamefont {P.~Z.}\
  \bibnamefont {Skands}},\ }\href {\doibase 10.1016/j.cpc.2015.01.024}
  {\bibfield  {journal} {\bibinfo  {journal} {Computer Physics Communications}\
  }\textbf {\bibinfo {volume} {191}},\ \bibinfo {pages} {159} (\bibinfo {year}
  {2015})}\BibitemShut {NoStop}%
\bibitem [{\citenamefont {de~Favereau}\ \emph {et~al.}(2014)\citenamefont
  {de~Favereau}, \citenamefont {Delaere}, \citenamefont {Demin}, \citenamefont
  {Giammanco}, \citenamefont {Lema{\^\i}tre}, \citenamefont {Mertens},
  \citenamefont {Selvaggi},\ and\ \citenamefont
  {collaboration}}]{Favereau:2014aa}%
  \BibitemOpen
  \bibfield  {author} {\bibinfo {author} {\bibfnamefont {J.}~\bibnamefont
  {de~Favereau}}, \bibinfo {author} {\bibfnamefont {C.}~\bibnamefont
  {Delaere}}, \bibinfo {author} {\bibfnamefont {P.}~\bibnamefont {Demin}},
  \bibinfo {author} {\bibfnamefont {A.}~\bibnamefont {Giammanco}}, \bibinfo
  {author} {\bibfnamefont {V.}~\bibnamefont {Lema{\^\i}tre}}, \bibinfo {author}
  {\bibfnamefont {A.}~\bibnamefont {Mertens}}, \bibinfo {author} {\bibfnamefont
  {M.}~\bibnamefont {Selvaggi}}, \ and\ \bibinfo {author} {\bibfnamefont
  {T.~D.~.}\ \bibnamefont {collaboration}},\ }\href {\doibase
  10.1007/JHEP02(2014)057} {\bibfield  {journal} {\bibinfo  {journal} {Journal
  of High Energy Physics}\ }\textbf {\bibinfo {volume} {2014}},\ \bibinfo
  {pages} {57} (\bibinfo {year} {2014})}\BibitemShut {NoStop}%
\bibitem [{\citenamefont {Selvaggi}()}]{delphes_card_idea}%
  \BibitemOpen
  \bibfield  {author} {\bibinfo {author} {\bibfnamefont {M.}~\bibnamefont
  {Selvaggi}},\ }\href@noop {} {\enquote {\bibinfo {title} {Fcc-ee idea
  detector delphes card},}\ }\bibinfo {howpublished}
  {\url{https://github.com/delphes/delphes/blob/master/cards/delphes_card_IDEA.tcl}}\BibitemShut
  {NoStop}%
\bibitem [{\citenamefont {Ball}\ \emph {et~al.}(2013)\citenamefont {Ball},
  \citenamefont {Bertone}, \citenamefont {Carrazza}, \citenamefont
  {Del~Debbio}, \citenamefont {Forte}, \citenamefont {Guffanti}, \citenamefont
  {Hartland},\ and\ \citenamefont {Rojo}}]{Ball2013290}%
  \BibitemOpen
  \bibfield  {author} {\bibinfo {author} {\bibfnamefont {R.~D.}\ \bibnamefont
  {Ball}}, \bibinfo {author} {\bibfnamefont {V.}~\bibnamefont {Bertone}},
  \bibinfo {author} {\bibfnamefont {S.}~\bibnamefont {Carrazza}}, \bibinfo
  {author} {\bibfnamefont {L.}~\bibnamefont {Del~Debbio}}, \bibinfo {author}
  {\bibfnamefont {S.}~\bibnamefont {Forte}}, \bibinfo {author} {\bibfnamefont
  {A.}~\bibnamefont {Guffanti}}, \bibinfo {author} {\bibfnamefont {N.~P.}\
  \bibnamefont {Hartland}}, \ and\ \bibinfo {author} {\bibfnamefont
  {J.}~\bibnamefont {Rojo}},\ }\href {\doibase 10.1016/j.nuclphysb.2013.10.010}
  {\bibfield  {journal} {\bibinfo  {journal} {Nuclear Physics B}\ }\textbf
  {\bibinfo {volume} {877}},\ \bibinfo {pages} {290 } (\bibinfo {year}
  {2013})},\ \bibinfo {note} {cited by: 440; All Open Access, Green Open
  Access}\BibitemShut {NoStop}%
\bibitem [{\citenamefont {Aad}\ \emph {et~al.}(2022)\citenamefont {Aad} \emph
  {et~al.}}]{ATLAS_main}%
  \BibitemOpen
  \bibfield  {author} {\bibinfo {author} {\bibfnamefont {G.}~\bibnamefont
  {Aad}} \emph {et~al.} (\bibinfo {collaboration} {ATLAS Collaboration}),\
  }\href {\doibase 10.1038/s41586-022-04893-w} {\bibfield  {journal} {\bibinfo
  {journal} {Nature}\ }\textbf {\bibinfo {volume} {607}},\ \bibinfo {pages}
  {52} (\bibinfo {year} {2022})}\BibitemShut {NoStop}%
\bibitem [{\citenamefont {Aad}\ \emph {et~al.}(2024)\citenamefont {Aad} \emph
  {et~al.}}]{PhysRevLett.133.101801}%
  \BibitemOpen
  \bibfield  {author} {\bibinfo {author} {\bibfnamefont {G.}~\bibnamefont
  {Aad}} \emph {et~al.} (\bibinfo {collaboration} {ATLAS Collaboration}),\
  }\href {\doibase 10.1103/PhysRevLett.133.101801} {\bibfield  {journal}
  {\bibinfo  {journal} {Phys. Rev. Lett.}\ }\textbf {\bibinfo {volume} {133}},\
  \bibinfo {pages} {101801} (\bibinfo {year} {2024})}\BibitemShut {NoStop}%
\bibitem [{\citenamefont {Tumasyan}\ \emph {et~al.}(2022)\citenamefont
  {Tumasyan} \emph {et~al.}}]{CMS_main}%
  \BibitemOpen
  \bibfield  {author} {\bibinfo {author} {\bibfnamefont {A.}~\bibnamefont
  {Tumasyan}} \emph {et~al.} (\bibinfo {collaboration} {CMS Collaboration}),\
  }\href {\doibase 10.1038/s41586-022-04892-x} {\bibfield  {journal} {\bibinfo
  {journal} {Nature}\ }\textbf {\bibinfo {volume} {607}},\ \bibinfo {pages}
  {60} (\bibinfo {year} {2022})}\BibitemShut {NoStop}%
\bibitem [{\citenamefont {Hayrapetyan}\ \emph {et~al.}(2025)\citenamefont
  {Hayrapetyan} \emph {et~al.}}]{CMS:2024awa}%
  \BibitemOpen
  \bibfield  {author} {\bibinfo {author} {\bibfnamefont {A.}~\bibnamefont
  {Hayrapetyan}} \emph {et~al.} (\bibinfo {collaboration} {CMS}),\ }\href
  {\doibase 10.1016/j.physletb.2024.139210} {\bibfield  {journal} {\bibinfo
  {journal} {Phys. Lett. B}\ }\textbf {\bibinfo {volume} {861}},\ \bibinfo
  {pages} {139210} (\bibinfo {year} {2025})},\ \Eprint
  {http://arxiv.org/abs/2407.13554} {arXiv:2407.13554 [hep-ex]} \BibitemShut
  {NoStop}%
\bibitem [{\citenamefont {Arganda}\ \emph {et~al.}(2019)\citenamefont
  {Arganda}, \citenamefont {Garcia-Garcia},\ and\ \citenamefont
  {Herrero}}]{ARGANDA2019114687}%
  \BibitemOpen
  \bibfield  {author} {\bibinfo {author} {\bibfnamefont {E.}~\bibnamefont
  {Arganda}}, \bibinfo {author} {\bibfnamefont {C.}~\bibnamefont
  {Garcia-Garcia}}, \ and\ \bibinfo {author} {\bibfnamefont {M.~J.}\
  \bibnamefont {Herrero}},\ }\href {\doibase
  https://doi.org/10.1016/j.nuclphysb.2019.114687} {\bibfield  {journal}
  {\bibinfo  {journal} {Nuclear Physics B}\ }\textbf {\bibinfo {volume}
  {945}},\ \bibinfo {pages} {114687} (\bibinfo {year} {2019})}\BibitemShut
  {NoStop}%
\bibitem [{\citenamefont {Cornwall}\ \emph {et~al.}(1974)\citenamefont
  {Cornwall}, \citenamefont {Levin},\ and\ \citenamefont
  {Tiktopoulos}}]{Cornwall:1974km}%
  \BibitemOpen
  \bibfield  {author} {\bibinfo {author} {\bibfnamefont {J.~M.}\ \bibnamefont
  {Cornwall}}, \bibinfo {author} {\bibfnamefont {D.~N.}\ \bibnamefont {Levin}},
  \ and\ \bibinfo {author} {\bibfnamefont {G.}~\bibnamefont {Tiktopoulos}},\
  }\href {\doibase 10.1103/PhysRevD.10.1145} {\bibfield  {journal} {\bibinfo
  {journal} {Phys. Rev. D}\ }\textbf {\bibinfo {volume} {10}},\ \bibinfo
  {pages} {1145} (\bibinfo {year} {1974})},\ \bibinfo {note} {[Erratum:
  Phys.Rev.D 11, 972 (1975)]}\BibitemShut {NoStop}%
\bibitem [{\citenamefont {Vayonakis}(1976)}]{Vayonakis:1976vz}%
  \BibitemOpen
  \bibfield  {author} {\bibinfo {author} {\bibfnamefont {C.~E.}\ \bibnamefont
  {Vayonakis}},\ }\href {\doibase 10.1007/BF02746538} {\bibfield  {journal}
  {\bibinfo  {journal} {Lett. Nuovo Cim.}\ }\textbf {\bibinfo {volume} {17}},\
  \bibinfo {pages} {383} (\bibinfo {year} {1976})}\BibitemShut {NoStop}%
\bibitem [{\citenamefont {Lee}\ \emph {et~al.}(1977)\citenamefont {Lee},
  \citenamefont {Quigg},\ and\ \citenamefont {Thacker}}]{Lee:1977eg}%
  \BibitemOpen
  \bibfield  {author} {\bibinfo {author} {\bibfnamefont {B.~W.}\ \bibnamefont
  {Lee}}, \bibinfo {author} {\bibfnamefont {C.}~\bibnamefont {Quigg}}, \ and\
  \bibinfo {author} {\bibfnamefont {H.~B.}\ \bibnamefont {Thacker}},\ }\href
  {\doibase 10.1103/PhysRevD.16.1519} {\bibfield  {journal} {\bibinfo
  {journal} {Phys. Rev. D}\ }\textbf {\bibinfo {volume} {16}},\ \bibinfo
  {pages} {1519} (\bibinfo {year} {1977})}\BibitemShut {NoStop}%
\bibitem [{\citenamefont {Delgado}\ \emph {et~al.}(2014)\citenamefont
  {Delgado}, \citenamefont {Dobado},\ and\ \citenamefont
  {Llanes-Estrada}}]{Delgado_2014}%
  \BibitemOpen
  \bibfield  {author} {\bibinfo {author} {\bibfnamefont {R.~L.}\ \bibnamefont
  {Delgado}}, \bibinfo {author} {\bibfnamefont {A.}~\bibnamefont {Dobado}}, \
  and\ \bibinfo {author} {\bibfnamefont {F.~J.}\ \bibnamefont
  {Llanes-Estrada}},\ }\href {\doibase 10.1007/jhep02(2014)121} {\bibfield
  {journal} {\bibinfo  {journal} {Journal of High Energy Physics}\ }\textbf
  {\bibinfo {volume} {2014}} (\bibinfo {year} {2014}),\
  10.1007/jhep02(2014)121}\BibitemShut {NoStop}%
\bibitem [{\citenamefont {Doroba}\ \emph {et~al.}(2012)\citenamefont {Doroba},
  \citenamefont {Kalinowski}, \citenamefont {Kuczmarski}, \citenamefont
  {Pokorski}, \citenamefont {Rosiek}, \citenamefont {Szleper},\ and\
  \citenamefont {Tkaczyk}}]{PhysRevD.86.036011}%
  \BibitemOpen
  \bibfield  {author} {\bibinfo {author} {\bibfnamefont {K.}~\bibnamefont
  {Doroba}}, \bibinfo {author} {\bibfnamefont {J.}~\bibnamefont {Kalinowski}},
  \bibinfo {author} {\bibfnamefont {J.}~\bibnamefont {Kuczmarski}}, \bibinfo
  {author} {\bibfnamefont {S.}~\bibnamefont {Pokorski}}, \bibinfo {author}
  {\bibfnamefont {J.}~\bibnamefont {Rosiek}}, \bibinfo {author} {\bibfnamefont
  {M.}~\bibnamefont {Szleper}}, \ and\ \bibinfo {author} {\bibfnamefont
  {S.}~\bibnamefont {Tkaczyk}},\ }\href {\doibase 10.1103/PhysRevD.86.036011}
  {\bibfield  {journal} {\bibinfo  {journal} {Phys. Rev. D}\ }\textbf {\bibinfo
  {volume} {86}},\ \bibinfo {pages} {036011} (\bibinfo {year}
  {2012})}\BibitemShut {NoStop}%
\bibitem [{\citenamefont {Szleper}(2015)}]{szleper2015higgsbosonphysicsww}%
  \BibitemOpen
  \bibfield  {author} {\bibinfo {author} {\bibfnamefont {M.}~\bibnamefont
  {Szleper}},\ }\href {https://arxiv.org/abs/1412.8367} {\enquote {\bibinfo
  {title} {The higgs boson and the physics of $ww$ scattering before and after
  higgs discovery},}\ } (\bibinfo {year} {2015}),\ \Eprint
  {http://arxiv.org/abs/1412.8367} {arXiv:1412.8367 [hep-ph]} \BibitemShut
  {NoStop}%
\bibitem [{\citenamefont {Delgado}\ \emph {et~al.}(2017)\citenamefont
  {Delgado}, \citenamefont {Dobado}, \citenamefont {Espriu}, \citenamefont
  {Garcia-Garcia}, \citenamefont {Herrero}, \citenamefont {Marcano},\ and\
  \citenamefont {Sanz-Cillero}}]{Delgado:2017aa}%
  \BibitemOpen
  \bibfield  {author} {\bibinfo {author} {\bibfnamefont {R.~L.}\ \bibnamefont
  {Delgado}}, \bibinfo {author} {\bibfnamefont {A.}~\bibnamefont {Dobado}},
  \bibinfo {author} {\bibfnamefont {D.}~\bibnamefont {Espriu}}, \bibinfo
  {author} {\bibfnamefont {C.}~\bibnamefont {Garcia-Garcia}}, \bibinfo {author}
  {\bibfnamefont {M.~J.}\ \bibnamefont {Herrero}}, \bibinfo {author}
  {\bibfnamefont {X.}~\bibnamefont {Marcano}}, \ and\ \bibinfo {author}
  {\bibfnamefont {J.~J.}\ \bibnamefont {Sanz-Cillero}},\ }\href {\doibase
  10.1007/JHEP11(2017)098} {\bibfield  {journal} {\bibinfo  {journal} {Journal
  of High Energy Physics}\ }\textbf {\bibinfo {volume} {2017}},\ \bibinfo
  {pages} {98} (\bibinfo {year} {2017})}\BibitemShut {NoStop}%
\bibitem [{\citenamefont {Gon{\c c}alves}\ \emph {et~al.}(2018)\citenamefont
  {Gon{\c c}alves}, \citenamefont {Han}, \citenamefont {Kling}, \citenamefont
  {Plehn},\ and\ \citenamefont {Takeuchi}}]{Gon_alves_2018}%
  \BibitemOpen
  \bibfield  {author} {\bibinfo {author} {\bibfnamefont {D.}~\bibnamefont
  {Gon{\c c}alves}}, \bibinfo {author} {\bibfnamefont {T.}~\bibnamefont {Han}},
  \bibinfo {author} {\bibfnamefont {F.}~\bibnamefont {Kling}}, \bibinfo
  {author} {\bibfnamefont {T.}~\bibnamefont {Plehn}}, \ and\ \bibinfo {author}
  {\bibfnamefont {M.}~\bibnamefont {Takeuchi}},\ }\href {\doibase
  10.1103/physrevd.97.113004} {\bibfield  {journal} {\bibinfo  {journal}
  {Physical Review D}\ }\textbf {\bibinfo {volume} {97}} (\bibinfo {year}
  {2018}),\ 10.1103/physrevd.97.113004}\BibitemShut {NoStop}%
\bibitem [{\citenamefont {Dreyer}\ and\ \citenamefont
  {Karlberg}(2018)}]{PhysRevD.98.114016}%
  \BibitemOpen
  \bibfield  {author} {\bibinfo {author} {\bibfnamefont {F.~A.}\ \bibnamefont
  {Dreyer}}\ and\ \bibinfo {author} {\bibfnamefont {A.}~\bibnamefont
  {Karlberg}},\ }\href {\doibase 10.1103/PhysRevD.98.114016} {\bibfield
  {journal} {\bibinfo  {journal} {Phys. Rev. D}\ }\textbf {\bibinfo {volume}
  {98}},\ \bibinfo {pages} {114016} (\bibinfo {year} {2018})}\BibitemShut
  {NoStop}%
\bibitem [{\citenamefont {Cacciari}\ \emph {et~al.}(2008)\citenamefont
  {Cacciari}, \citenamefont {Salam},\ and\ \citenamefont
  {Soyez}}]{Cacciari_2008}%
  \BibitemOpen
  \bibfield  {author} {\bibinfo {author} {\bibfnamefont {M.}~\bibnamefont
  {Cacciari}}, \bibinfo {author} {\bibfnamefont {G.~P.}\ \bibnamefont {Salam}},
  \ and\ \bibinfo {author} {\bibfnamefont {G.}~\bibnamefont {Soyez}},\ }\href
  {\doibase 10.1088/1126-6708/2008/04/063} {\bibfield  {journal} {\bibinfo
  {journal} {Journal of High Energy Physics}\ }\textbf {\bibinfo {volume}
  {2008}},\ \bibinfo {pages} {063} (\bibinfo {year} {2008})}\BibitemShut
  {NoStop}%
\bibitem [{\citenamefont {Cacciari}\ \emph {et~al.}(2012)\citenamefont
  {Cacciari}, \citenamefont {Salam},\ and\ \citenamefont
  {Soyez}}]{Cacciari_2012}%
  \BibitemOpen
  \bibfield  {author} {\bibinfo {author} {\bibfnamefont {M.}~\bibnamefont
  {Cacciari}}, \bibinfo {author} {\bibfnamefont {G.~P.}\ \bibnamefont {Salam}},
  \ and\ \bibinfo {author} {\bibfnamefont {G.}~\bibnamefont {Soyez}},\ }\href
  {\doibase 10.1140/epjc/s10052-012-1896-2} {\bibfield  {journal} {\bibinfo
  {journal} {The European Physical Journal C}\ }\textbf {\bibinfo {volume}
  {72}} (\bibinfo {year} {2012}),\ 10.1140/epjc/s10052-012-1896-2}\BibitemShut
  {NoStop}%
\bibitem [{\citenamefont {{CMS-PAS-FTR-18-019}}(2018)}]{CMS:2018ccd}%
  \BibitemOpen
  \bibfield  {author} {\bibinfo {author} {\bibnamefont {{CMS-PAS-FTR-18-019}}}
  (\bibinfo {collaboration} {{CMS Collaboration}}),\ }\href@noop {} {\
  (\bibinfo {year} {{2018}})}\BibitemShut {NoStop}%
\bibitem [{\citenamefont {Collaboration}(2022)}]{CMS-PAS-FTR-21-004}%
  \BibitemOpen
  \bibfield  {author} {\bibinfo {author} {\bibfnamefont {C.}~\bibnamefont
  {Collaboration}} (\bibinfo {collaboration} {CMS}),\ }\href
  {https://cds.cern.ch/record/2803918} {\emph {\bibinfo {title} {Prospects for
  non-resonant Higgs boson pair production measurement in bb$\gamma\gamma$
  final states in proton-proton collisions at $\sqrt{s}=14$ TeV at the
  High-Luminosity LHC}}},\ \bibinfo {type} {Tech. Rep.}\ (\bibinfo
  {institution} {CERN},\ \bibinfo {address} {Geneva},\ \bibinfo {year}
  {2022})\BibitemShut {NoStop}%
\bibitem [{\citenamefont {Dawson}\ \emph {et~al.}(1998)\citenamefont {Dawson},
  \citenamefont {Dittmaier},\ and\ \citenamefont {Spira}}]{Dawson:1998py}%
  \BibitemOpen
  \bibfield  {author} {\bibinfo {author} {\bibfnamefont {S.}~\bibnamefont
  {Dawson}}, \bibinfo {author} {\bibfnamefont {S.}~\bibnamefont {Dittmaier}}, \
  and\ \bibinfo {author} {\bibfnamefont {M.}~\bibnamefont {Spira}},\ }\href
  {\doibase 10.1103/PhysRevD.58.115012} {\bibfield  {journal} {\bibinfo
  {journal} {Phys. Rev. D}\ }\textbf {\bibinfo {volume} {58}},\ \bibinfo
  {pages} {115012} (\bibinfo {year} {1998})},\ \Eprint
  {http://arxiv.org/abs/hep-ph/9805244} {arXiv:hep-ph/9805244} \BibitemShut
  {NoStop}%
\bibitem [{\citenamefont {Dolan}\ \emph {et~al.}(2015)\citenamefont {Dolan},
  \citenamefont {Englert}, \citenamefont {Greiner}, \citenamefont {Nordstrom},\
  and\ \citenamefont {Spannowsky}}]{Dolan:2015zja}%
  \BibitemOpen
  \bibfield  {author} {\bibinfo {author} {\bibfnamefont {M.~J.}\ \bibnamefont
  {Dolan}}, \bibinfo {author} {\bibfnamefont {C.}~\bibnamefont {Englert}},
  \bibinfo {author} {\bibfnamefont {N.}~\bibnamefont {Greiner}}, \bibinfo
  {author} {\bibfnamefont {K.}~\bibnamefont {Nordstrom}}, \ and\ \bibinfo
  {author} {\bibfnamefont {M.}~\bibnamefont {Spannowsky}},\ }\href {\doibase
  10.1140/epjc/s10052-015-3622-3} {\bibfield  {journal} {\bibinfo  {journal}
  {Eur. Phys. J. C}\ }\textbf {\bibinfo {volume} {75}},\ \bibinfo {pages} {387}
  (\bibinfo {year} {2015})},\ \Eprint {http://arxiv.org/abs/1506.08008}
  {arXiv:1506.08008 [hep-ph]} \BibitemShut {NoStop}%
\end{thebibliography}%

\end{document}